\documentclass[cernpreprint, texlive=2023, UKenglish, texmf, orcidlogo]{atlasdoc}
\usepackage{atlaspackage}
\usepackage{atlasbiblatex}
 
\usepackage{atlasphysics}
\usepackage{atlasjetetmiss}
\usepackage{atlasprocess}

\graphicspath{{logos/}{figures/}}
 
\usepackage{ANA-EXOT-2021-01-PAPER-defs}
 
\usepackage{nicefrac}

\AtlasTitle{Search for dark matter produced in association with a single top quark and an energetic $W$ boson in $\sqrt{s}=$ 13 TeV \(pp\) collisions with the ATLAS detector}

\AtlasAbstract{
This paper presents a search for dark matter, $\chi$, using events with a
single top quark and an energetic $W$ boson. The analysis is based on proton--proton collision data collected with the ATLAS experiment at
$\sqrt{s}=$ 13 TeV during LHC Run 2 (2015--2018), corresponding to an integrated luminosity of 139 fb$^{-1}$. The search
considers final states with zero or one charged lepton (electron or muon), at least one $b$-jet and large missing transverse momentum.
In addition, a result from a previous search considering two-charged-lepton final states is included in the interpretation of the results.
The data are found to be in good agreement with the Standard Model predictions and the results are interpreted in terms of
95\% confidence-level exclusion limits in the context of a class of dark matter models involving an extended two-Higgs-doublet sector together with a pseudoscalar mediator particle.
The search is particularly sensitive to on-shell production of the charged Higgs boson state, $H^{\pm}$, arising from the two-Higgs-doublet mixing, and its
semi-invisible decays via the mediator particle, $a$: $H^{\pm} \rightarrow W^\pm a (\rightarrow \chi\chi)$.
Signal models with $H^{\pm}$ masses up to 1.5 TeV and $a$ masses up to 350 GeV are excluded assuming a $\tan\beta$ value of 1. For masses of $a$ of 150 (250) GeV, $\tan\beta$ values up to 2 are excluded for $H^{\pm}$ masses between 200 (400) GeV and 1.5 TeV.
Signals with $\tan\beta$ values between 20 and 30 are excluded for $H^{\pm}$ masses between 500 and 800 GeV.
}
 
\author{The ATLAS Collaboration}
\PreprintIdNumber{CERN-EP-2022-146}
\AtlasRefCode{EXOT-2021-01}
\AtlasDate{\today}

\AtlasJournal{Eur. Phys. J. C}
\AtlasJournalRef{Eur. Phys. J. C 83 (2023) 603}
\AtlasDOI{10.1140/epjc/s10052-023-11582-z}

\hypersetup{pdftitle={ATLAS document},pdfauthor={The ATLAS Collaboration}}
 
\begin{document}
 
\maketitle

\section{Introduction}
\label{sec:intro}
 
The existence of non-luminous matter, referred to as dark matter (DM),
is strongly suggested by a wide variety of astrophysical
and cosmological measurements~\cite{Zwicky:1933gu, DMreview}. Despite
the strong evidence supporting the presence of DM, which accounts for 26\%
of the energy content of the universe~\cite{Hinshaw:2012aka,Planck:2018nkj},  
its nature and properties remain largely unknown and constitute one of
the most important unanswered questions in modern physics.
Assuming that its main component is a weakly interacting massive
particle (WIMP or $\chi$)~\cite{Steigman:1984ac}, DM produced in proton--proton collisions does not interact with the
ATLAS detector and it can be detected only if produced in association with Standard Model (SM) particles. This leads
to signatures with missing transverse momentum (\ptmiss, its modulus denoted by \etmiss).
\begin{figure}[b]
\centering
\subfloat[][]{
\includegraphics[width=0.3\textwidth,trim={0cm 5cm 0cm 5cm},clip]{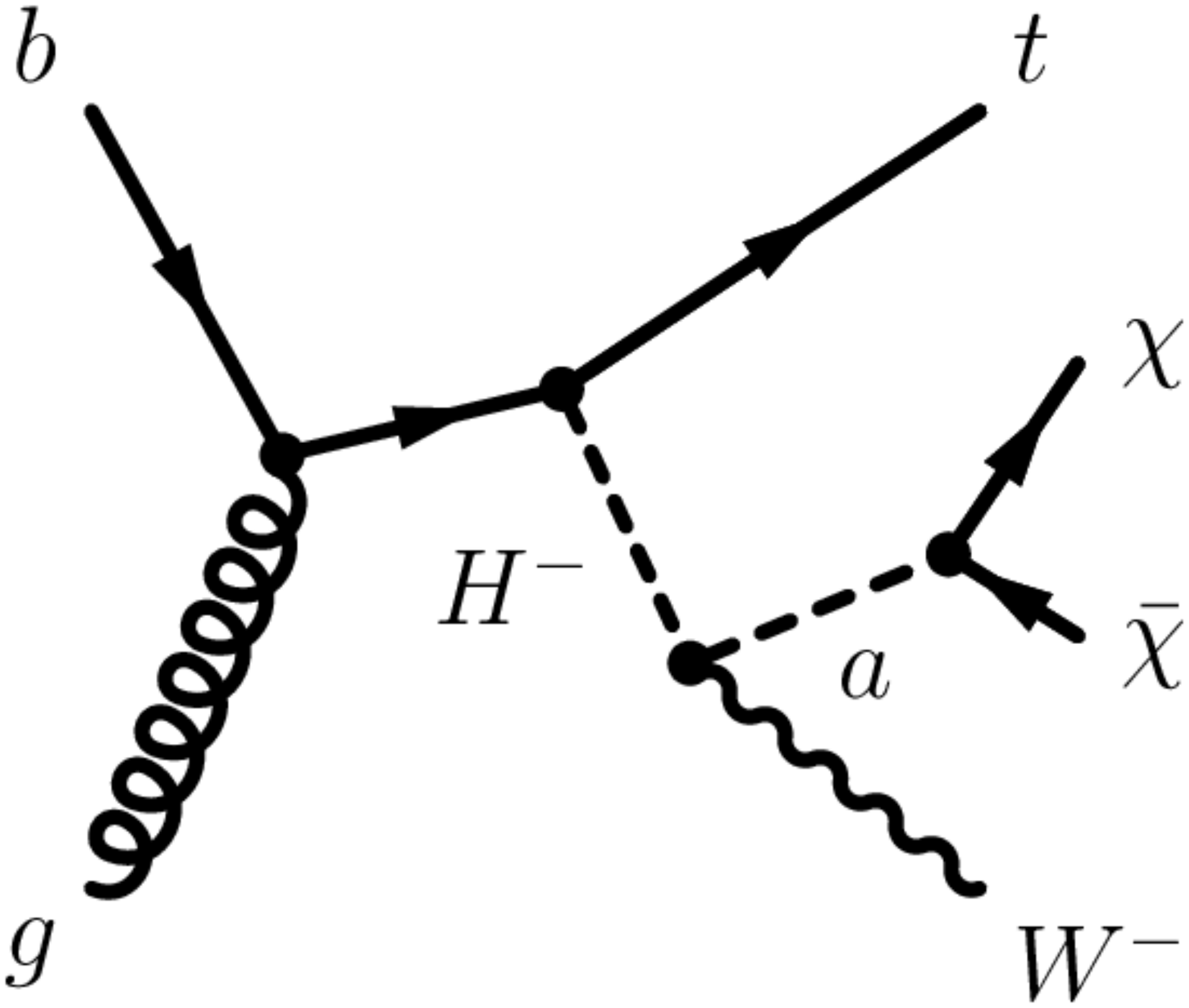}
}
\subfloat[][]{
\includegraphics[width=0.3\textwidth,trim={0cm 5cm 0cm 5cm},clip]{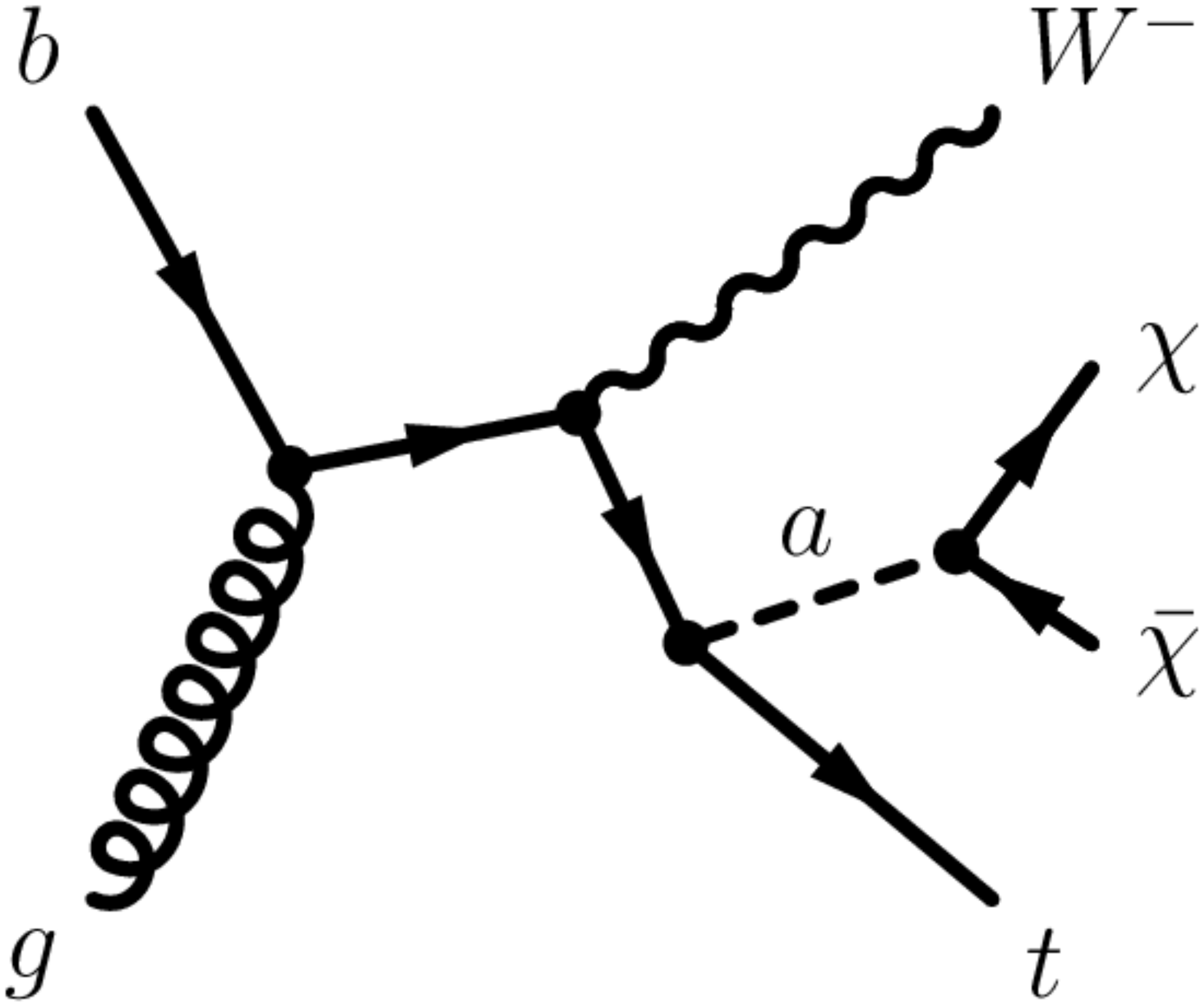}
}
\caption{
Representative diagrams of \twdm\ production from the 2HDM$+a$ model
considered in this analysis. Charge-conjugate diagrams are considered as well.
}
\label{fig:THDMpa_feync}
\end{figure}
 
The signal model considered in this search belongs to a class of
simplified models for DM searches at the Large Hadron Collider (LHC).
It involves an extended two-Higgs-doublet sector (2HDM)~\cite{Ipek:2014gua,No:2015xqa,Goncalves:2016iyg,Bell:2016ekl,Bauer:2017ota,Bauer:2017fsw,Haisch:2017gql,Arcadi:2017wqi,Bell:2017rgi},
together with an additional pseudoscalar mediator ($a$) that couples to a fermionic DM candidate.
This 2HDM$+a$
model~\cite{Bauer:2017ota,Abe:2018bpo} represents the simplest ultraviolet-complete
and renormalisable framework for investigating the broad phenomenology predicted by
spin-0 mediator-based DM
models~\cite{Abdallah:2015ter,Abercrombie:2015wmb,Abe:2018bpo,Cheung:2010zf,Haisch:2013fla,Buckley:2014fba,Buckley:2015ctj,Haisch:2015ioa,Backovic:2015soa,Arina:2016cqj,Haisch:2016gry,Banerjee:2017wxi,Haisch:2018bby}.
 
The 2HDM$+a$ model offers a rich
phenomenology~\cite{Pani:2017qyd,CidVidal:2018eel,ATL-PHYS-PUB-2018-027,Haisch:2018djm,Butterworth:2020vnb,Robens:2021lov},
with a variety of final states that might arise
depending on the production and decay modes of the various bosons composing the Higgs sector,
as investigated in Refs.~\cite{EXOT-2017-32,Abe:2018bpo,CMS-EXO-16-050,EXOT-2018-46,CMS-EXO-19-003,HIGG-2019-02}. 
A recent analysis performed by the ATLAS Collaboration
\cite{EXOT-2018-43} has considered topologies
characterised by the presence of \etmiss and a single top quark in the context of 2HDM$+a$ models. That search allowed masses of the additional
charged Higgs bosons, $H^{\pm}$, from 400~\GeV\ to 1.1~\TeV\ to be excluded at a 95\% confidence level (CL) for different values of the $a$-boson mass and for low values ($ < 2 $) of $\tan\beta$ (the ratio of the vacuum expectation values of the two Higgs doublets), which significantly affects
the phenomenology of the 2HDM$+a$ model. Values of the $a$-boson mass up to 330~\GeV\ are also excluded at 95\% CL for $\tan\beta = 1 $ and an
$H^{\pm}$ mass of 800~\GeV. CMS has also performed a search for these topologies~\cite{CMS-EXO-18-010}, where the results are interpreted in the context of a different set of simplified models.
 
As in the case of the SM single top-quark production, the associated production of DM with a single
top quark has three production modes at leading order (LO): $t$-channel production, $s$-channel production and
associated production with a \Wboson\ boson ($tW$). In the 2HDM$+a$ model, the dominant production mode for single-top-quark final states
is the \twdm\ channel, through the diagrams depicted in Figure~\ref{fig:THDMpa_feync}.
On-shell production of charged Higgs bosons dominates the \twdm\ production mode
when $H^{\pm}\rightarrow W^{\pm} a$ decays are kinematically allowed
and the $H^{\pm}$ mass is a few hundred~\GeV.  Furthermore, the cross-section for this inclusive \twdm\ production mode has a local minimum
at $\tan\beta \approx 5$ and two local maxima at low $\tan\beta$ ($< 2$) and $\tan\beta \in [20,30]$~\cite{Pani:2017qyd}.
The aim of the search presented in this paper is to extend the current results obtained by ATLAS for the 2HDM$+a$ model by improving
the sensitivity to single top-quark production in association with dark matter in the \twdm\ process. The focus is to improve upon the current ATLAS limits at low $\tan\beta$
and to provide, for the first time, sensitivity to signal benchmarks probing the \twdm\ cross-section maximum at high $\tan\beta$ values.
 
This paper presents
a dedicated search for associated production of a single top quark, a
\Wboson\ boson and DM particles, based on 139~fb$^{-1}$ of
proton--proton ($pp$) collisions at a centre-of-mass energy $\sqrt{s}= 13$~\TeV\
produced at the LHC and collected by the ATLAS detector (see Section \ref{sec:detector}) between $2015$ and $2018$.
Due to the similarity of the experimental signature to \ttbar\ production, the analysis is also sensitive to DM
produced in association with two top quarks (\ttdm).
This final state is not considered in the optimization of the
analysis, but its contribution is added to the \twdm\ signal, according to the prediction of the 2HDM$+a$ model, when interpreting the final result.
The analysis relies on Monte Carlo (MC) simulations, described in Section \ref{sec:datamc},
which aid in the estimation of the SM background and DM signals.
This search improves upon previous results~\cite{EXOT-2018-43}
by targeting final states with an energetic $W$ boson decaying hadronically or leptonically and
characterised by the presence of exactly zero or one lepton ($\ell =e, \mu$).
The hadronic decays of the \Wboson\ boson are identified by requiring the presence of at least one high-\pt\ large-radius jet
consistent with originating from the hadronization of a resonant di-quark pair. In addition, at least one jet arising from the fragmentation of $b$-hadrons ($b$-jet)
is required as a signature of the additional presence of a top quark, while large \etmiss\ is required as a sign of the production of DM particles.
The identification of these objects, as well as the event reconstruction,
is described in Section~\ref{sec:objsel}.
Section~\ref{sec:analysis} presents the selection
of events in the one- or zero-lepton analysis channels, and also
the method used to combine these two channels
with the dilepton analysis described in Ref.~\cite{EXOT-2018-43}. Their combination maximises the sensitivity to \twdm\ processes and
further tightens previous constraints for 2HDM$+a$ models using the \twdm\ channel.
Systematic uncertainties are described in Section
\ref{sec:unc}, followed by the experimental results
and their interpretation in the context of the 2HDM$+a$ model
in Section \ref{sec:results}. Finally, Section \ref{sec:conclusion}
is devoted to the conclusions.


\section{The ATLAS detector}
\label{sec:detector}
 
The ATLAS detector~\cite{PERF-2007-01} is a multipurpose particle detector with a forward--backward symmetric cylindrical
geometry and nearly 4$\pi$ coverage in solid angle.\footnote{ATLAS uses a right-handed coordinate system with its origin at the nominal
interaction point in the centre of the detector. The positive $x$-axis is defined by the direction from the interaction point to the centre of the LHC ring, with the positive $y$-axis pointing upwards, while the beam direction defines the $z$-axis. Cylindrical coordinates
$(r,\phi)$ are used in the transverse plane, $\phi$ being the azimuthal angle around the $z$-axis. The pseudorapidity $\eta$ is
defined in terms of the polar angle $\theta$ by $\eta=-\ln\tan(\theta/2)$. Rapidity is defined as $y$ = $0.5 \ln[(E + p_z)/(E-p_z)]$ where $E$
denotes the energy and $p_z$ is the component of the momentum along the beam direction. The angular distance $\DeltaR$ is defined as $\sqrt{(\Delta \eta)^2 + (\Delta \phi)^2}$.}
The inner tracking detector (ID) consists of pixel and microstrip silicon detectors
covering the pseudorapidity region $|\eta|<2.5$, surrounded by a transition radiation tracker
which enhances electron identification in the region $|\eta|<2.0$.
An inner pixel layer, the insertable B-layer~\cite{ATLAS-TDR-2010-19, PIX-2018-001}, was added at a mean radius of 3.3~cm during the period between Run~1 and Run~2 of the LHC.
The inner detector is surrounded by a thin superconducting solenoid providing an axial 2~T magnetic field and by a fine-granularity lead/liquid-argon
(LAr) electromagnetic calorimeter covering $|\eta|<3.2$. A steel/scintillator-tile calorimeter provides hadronic coverage in the central pseudorapidity
range ($|\eta|<1.7$).
The endcap ($1.5<|\eta|<3.2$) and forward ($3.1<|\eta|<4.9$) regions of the hadron calorimeter are made of LAr active layers with either copper or tungsten
as the absorber material.
A muon spectrometer with an air-core toroid magnet system surrounds the calorimeters.
Three layers of high-precision tracking chambers provide coverage in the range $|\eta|<2.7$, while dedicated fast chambers allow triggering
in the region $|\eta|<2.4$. The ATLAS trigger system consists of a hardware-based level-1 trigger followed by a software-based high-level trigger~\cite{TRIG-2016-01}.
An extensive software suite~\cite{ATL-SOFT-PUB-2021-001} is used in the reconstruction and analysis of real and simulated data, in detector operations, and in the trigger and data acquisition systems of the experiment.


\section{Data and simulated events}
\label{sec:datamc}

The dataset used in the analysis corresponds to an integrated luminosity
of 139~fb$^{-1}$ of $pp$ collisions at a centre-of-mass energy of 13~\TeV\ recorded
by the ATLAS detector with stable beam conditions.
The uncertainty in the integrated luminosity is
1.7\%~\cite{ATLAS-CONF-2019-021},
obtained using the LUCID-2 detector~\cite{Avoni:2633501} for the primary luminosity measurements.
The number of interactions in the same and temporally adjacent bunch crossings (pile-up)
is 33.7 on average across all data-taking years.
All detector subsystems are required to be operational for this dataset~\cite{DAPR-2018-01}. Candidate events were recorded using a combined set of triggers based on the presence of \etmiss\ or charged leptons ($\ell = e, \mu$).
The $\etmiss$\ trigger~\cite{TRIG-2019-01} is fully efficient for events with reconstructed $\etmiss > 250~\GeV$
and is used for the zero-lepton and one-lepton analysis channels. Triggers based on single
leptons~\cite{TRIG-2018-05,TRIG-2018-01} are used to define auxiliary selections that aid in the
estimation of the SM background processes.
These selections require the presence of a muon or electron with
transverse momentum, \pt (or transverse energy \et for electrons), above certain thresholds, and impose data quality
and lepton identification and isolation requirements.
 
Dedicated MC simulated samples are used to model the SM and
signal processes and to estimate their expected yields.
The 2HDM$+a$ model considered in this paper assumes a
type-II~\cite{Gunion:2002zf,Branco:2011iw} coupling structure of the
Higgs sector, and has a CP-conserving potential and a softly broken $Z_2$ symmetry~\cite{Gunion:2002zf}.
The additional pseudoscalar mediator of the
model couples to DM particles and to the SM, and mixes with the
pseudoscalar partner of the SM Higgs boson.
The most important parameters that determine the phenomenology of the model are
the masses of the CP-even ($h$ and $H$), CP-odd ($a$ and $A$)
and charged ($H^\pm$) bosons; the mass of the DM particle ($\chi$);
the three quartic couplings between the scalar doublets and the $a$ boson ($\lambda_{P1}, \lambda_{P2}$ and $\lambda_3$)
and the coupling between the $a$ boson and the DM particle ($g_\chi$);
the ratio of the vacuum expectation values (VEVs) of the
two Higgs doublets ($\tan\beta$); and the mixing angles of the
CP-even and CP-odd weak eigenstates, denoted by $\alpha$ and $\theta$, respectively.
The alignment limit ($\cos(\beta-\alpha) = 0$) and the
decoupling limit are assumed. Thus the lightest CP-even mass eigenstate, $h$, can be identified as
the SM Higgs boson with couplings predicted by the SM. The electroweak vacuum expectation value is set to $246~\GeV$.
The mixing angle $\theta$ is fixed at $\sin\theta=1/\sqrt{2}$, yielding full mixing between the $a$ and $A$ bosons
and the largest cross-section for the processes of interest.
To further reduce the parameter space, unitary couplings between the $a$-boson mediator and the DM particle $\chi$ ($g_\chi = 1$) are considered,
with the DM particle mass set to $m_\chi=10\;\GeV$. This has a negligible effect on the kinematic properties in the final states of interest, as long as
$a\rightarrow \chi \chi$ is kinematically allowed.
Following the prescriptions in Ref.~\cite{Abe:2018bpo}, the masses of the heavy CP-even Higgs boson, $H$,
and charged bosons, $H^\pm$, are set equal to the mass of the heavy
CP-odd partner, $A$, and the three quartic couplings are set to a
value of $3$ for compatibility with
constraints from electroweak precision
measurements~\cite{Bauer:2017ota} and to ensure the stability of the Higgs
potential for most of the parameter space of interest.
\footnote{As pointed out in Ref.~\cite{Abe:2018bpo}, for the parameter space considered in this paper, only values
$m_{H^{\pm}} < 600$~\GeV\ provide a bounded-from-below scalar potential~\cite{Gunion:2002zf} for the 2HDM$+a$ model.
This constraint can be relaxed by up to a factor of 2 if the quartic coupling $\lambda_3$ assumes a value closer
to the perturbativity limit, and it can be relaxed further in more general 2HDMs containing
additional quartic couplings~\cite{Bauer:2017fsw}, as discussed in Ref.~\cite{Haisch:2018djm}.}
 
The signal MC samples include $tW$ production in association with DM particles.
They were generated using LO matrix elements calculated by the \MGNLO[2.7.3]~\cite{Alwall:2014hca} generator interfaced
to \PYTHIA[8.244]~\cite{Sjostrand:2014zea}, which used parameter values set to the A14 tune~\cite{ATL-PHYS-PUB-2014-021} to model parton showering (PS),
hadronization and the underlying event.
The five-flavour scheme \NNPDF[3.0nlo]~\cite{NNPDF:2014otw} set of parton distribution functions (PDFs) was used.
Signal cross-sections are calculated at LO accuracy in QCD.
Additional simulated samples are used for \ttdm\ processes.
They were generated using LO matrix elements with up to one
extra parton calculated by \MGNLO[2.6.7] interfaced to \PYTHIA[8.244], with the same PDF set and tune as used
for the $tW$ processes. In this case, signal cross-sections are calculated at next-to-leading-order (NLO) accuracy using the same version of
\MGNLO as suggested in Ref.~\cite{Backovic:2015soa}.
The top-quark decays in all signal samples were simulated using \MADSPIN~\cite{Artoisenet:2012st}.
The final results are presented as a function of the $(m_a, m_{H^{\pm}})$ parameters while setting $\tan\beta$ to unity,
or varying the $(m_{H^{\pm}}, \tan\beta)$ parameters while setting $m_a$ to 250~\GeV\ or 150~\GeV.
 
The SM background processes were simulated using various MC event generators, accurate to NLO in perturbation theory, depending on the process.
All background processes are normalised to the best available theoretical calculation of their respective cross-sections.
The event generators, the accuracy of theoretical cross-sections, the underlying-event set of tuned parameters, and the PDF sets
used in simulating the SM background processes most relevant for this analysis are summarised in \Tab{\ref{tab-MCsamples}}. Note that
the \NNPDF[2.3lo] PDF sets~\cite{Ball:2012cx} were used for the parton-shower and hadronization steps in samples using \PYTHIA 8.
 
For all samples, except those generated using \SHERPA~\cite{Gleisberg:2008ta,Gleisberg:2008fv,Cascioli:2011va,Schumann:2007mg,Hoeche:2012yf,Bothmann:2019yzt},
the \EVTGEN[1.2.0]~\cite{Lange:2001uf} program was used to simulate the properties of the $b$- and $c$-hadron decays.
All generated events were then processed
using the ATLAS simulation infrastructure~\cite{SOFT-2010-01} and $\GEANT$~\cite{Agostinelli:2002hh}, which models
the response of the various ATLAS subdetectors with high precision.
In some cases, a faster simulation based on a parameterization of the calorimeter response, and on $\GEANT$ for the other detector
subsystems~\cite{SOFT-2010-01}, was used.
Subsequently, simulated events are reconstructed after including a realistic modelling
of pile-up interactions, with pile-up profiles matching the ones in
data.
These profiles were obtained
by overlaying each hard-scatter event with minimum-bias events simulated using the soft QCD processes of \PYTHIA[8.186]~\cite{Sjostrand:2007gs}
with the \NNPDF[2.3lo] set of PDFs~\cite{Ball:2012cx} and the
A3 tune~\cite{ATL-PHYS-PUB-2016-017}.
 
\begin{table}[t!]
\caption{List of MC generators and the PDF sets used to simulate the different SM background processes. Diboson includes $WW$, $WZ$ and $ZZ$ production.
Information is also given on the generator used to simulate the parton-shower and hadronization step and the underlying-event set of tuned parameters. The last column
presents the perturbative QCD highest-order accuracy  (NLO, next-to-next-to-leading order (NNLO), and next-to-next-to-leading logarithm (NNLL)) achieved
in the calculation of the cross-section later used to normalizate  of the different samples.
Samples using \PYTHIA 8 used \NNPDF[2.3lo] set of PDFs~\cite{Ball:2012cx} for the parton-shower and hadronization step. Diboson cross-sections are directly taken from \SHERPA.}
\label{tab-MCsamples}
\centering
\scalebox{0.69}{
\begin{tabular}{lllllll}
\toprule
\multicolumn{2}{l}{\multirow{2}{*}{Process}} & \multirow{2}{*}{Generator} & \multirow{2}{*}{PDF set}  & PS and & \multirow{2}{*}{UE tune}   & Cross-section\\
& & & & frag./hadr. &  & accuracy \\
\midrule
\multicolumn{2}{l}{Top pair ($t\bar{t}$)} & \POWHEGBOX[v2]~\cite{Frixione:2007nw,Nason:2004rx,Frixione:2007vw,Alioli:2010xd} &  \NNPDF[3.0nlo] & \PYTHIA 8 & A14 & NNLO+NNLL~\cite{Czakon:2011xx} \\
\multirow{2}{*}{Single top$\bigg\{$} & $t$-channel  & \POWHEGBOX[v2]~\cite{Frederix:2012dh} & \NNPDF[3.0nlo] & \PYTHIA 8 & A14 &NNLO+NNLL~\cite{Kidonakis:2011wy} \\
& $s$- and $tW$-channel & \POWHEGBOX[v2]~\cite{Re:2010bp} & \NNPDF[3.0nlo] & \PYTHIA 8  & A14 & NNLO+NNLL~\cite{Kidonakis:2010ux,Kidonakis:2010tc} \\
\multicolumn{2}{l}{$V$+jets $(V=W/Z)$} & \SHERPA~2.2.1~\cite{Gleisberg:2008ta,Gleisberg:2008fv,Cascioli:2011va,Schumann:2007mg,Hoeche:2012yf,Bothmann:2019yzt} & \NNPDF[3.0nnlo] & \SHERPA & Default & NNLO~\cite{Catani:2009sm}\\
\multicolumn{2}{l}{Diboson} & \SHERPA~2.2.1 or 2.2.2~\cite{Bothmann:2019yzt} & \NNPDF[3.0nnlo] & \SHERPA & Default & NLO\\
\multicolumn{2}{l}{$tZ$; $tWZ(\rightarrow \ell\ell)$; $t\bar{t}+V$, $V=W,Z,h$} & \MGNLO~2.3.3~\cite{Alwall:2014hca} & \NNPDF[3.0nlo] & \PYTHIA 8 & A14 & NLO~\cite{Alwall:2014hca,LHCHiggsCrossSectionWorkingGroup:2016ypw}\\
\multicolumn{2}{l}{$ttt$; $t\bar{t}t\bar{t}$} & \MGNLO~2.3.3~\cite{Alwall:2014hca} & \NNPDF[3.1nlo] & \PYTHIA 8  & A14 & NLO~\cite{Alwall:2014hca,LHCHiggsCrossSectionWorkingGroup:2016ypw}\\
\multicolumn{2}{l}{$tWZ(\rightarrow \nu\nu)$} & \MGNLO~2.6.7~\cite{Alwall:2014hca} & \NNPDF[3.0nlo] & \PYTHIA 8 & A14 & NLO~\cite{Alwall:2014hca}\\
\bottomrule
\end{tabular}
}
\end{table}


\section{Object identification and event reconstruction}
\label{sec:objsel}
All collision events considered in this paper are required to have at
least one reconstructed interaction vertex with a minimum of two
associated tracks each having $\pt>500$~\MeV.
In events with multiple vertices, the one with the highest sum of
squared transverse momenta of associated tracks is
chosen as the primary vertex~\cite{ATL-PHYS-PUB-2015-026}.
Minimal quality criteria are applied to reject events with detector noise~\cite{DAPR-2018-01},
non-collision backgrounds or events with jets failing basic cleaning requirements~\cite{ATLAS-CONF-2015-029}.
 
Electrons ($e$), muons ($\mu$) and jets are considered
with two levels of quality requirements: baseline and signal.
The baseline requirements have looser identification criteria.
For each event, the missing transverse momentum~\cite{ATLAS-CONF-2018-023} \ptmiss, with magnitude \met,
is calculated as the negative vector sum of the transverse momenta of all
baseline reconstructed objects and the `soft term'. The soft term includes all tracks associated with the primary vertex but not matched to any reconstructed lepton or jet.
Tracks not associated with the primary vertex are not considered in the \ptmiss\ calculation,
improving the \ptmiss\ resolution by reducing the effect of
pile-up. A quality criterion for the matching of topological cell clusters~\cite{PERF-2014-07} in the electromagnetic calorimeter
to electrons is also imposed in events containing electrons with $|\eta| \in [1.37,1.52]$
in data recorded during 2015 and 2016. Baseline reconstructed physics objects are also used when resolving
possible reconstruction ambiguities (overlap removal).
The details of the object prioritization and requirements in this
procedure can be found in Ref.~\cite{SUSY-2018-08}.
 
Electron candidates are reconstructed from energy deposits in the electromagnetic calorimeter that are
matched to charged-particle tracks in the inner detector~\cite{PERF-2017-01}.
Baseline quality criteria include
$\pt>4.5$~\GeV, $|\eta|<2.47$ and satisfying the `LooseAndBLayer' likelihood identification\footnote{The electron identification criteria are based on a likelihood built by combining information from the electron track and shower shapes of the electron energy deposit in the calorimeter. Four operating are defined - VeryLoose, Loose, Medium and Tight - corresponding to increasing cuts on the likelihood discriminant. Track quality criteria are required on top of this selection. The `LooseAndBLayer' uses the Loose working point likelihood requirement, but requires in addition that the electron track has a hit in the innermost pixel layer.} operating point~\cite{PERF-2017-01}.
The longitudinal impact parameter, $z_0$, relative to the primary vertex is required to satisfy $|z_0\sin\theta| <0.5$ mm.
Signal-quality electrons are required to also satisfy $\pt>20$~\GeV\ and the `Medium' likelihood identification criterion.
The significance of the transverse impact parameter, $d_0$, must satisfy $ |d_0/ \sigma(d_0)|<5$ for these electrons.
Signal  electrons with $\pt < 200$~\GeV\ are also required to be isolated\footnote{Isolation quality cuts are assessed by studying the energy deposits within a cone in $\Delta R$ around the physics object. Energy deposits in the calorimeter and the tracker are considered to define the `Loose' working points while only energy deposits in the calorimeter are used to define the `HighPtCaloOnly'.} from other activity in the detector
by satisfying the `Loose' isolation working point, while those with larger $\pt$ are required to pass the `HighPtCaloOnly' isolation working point, as described in Ref.~\cite{EGAM-2018-01}.
 
Muon candidates are reconstructed from matching tracks in the inner detector and muon spectrometer~\cite{MUON-2018-03}.
Requirements for baseline-quality muons include $\pt>4$~\GeV, $|\eta|<2.7$ and satisfying the `Medium' identification criterion\footnote{The `Medium' muon identification criteria requires well-reconstructed muon tracks the inner detector and muon spectrometer (only muon spectrometer tracks in the region 2.5 < $|\eta|$ < 2.7) and a good $q/p$ compatiblity between both.}~\cite{MUON-2018-03}.
Like the electrons, their longitudinal impact parameter relative to the primary vertex is required to satisfy $|z_0\sin\theta| <0.5$~mm.
Signal-quality muons must satisfy \pt$>20$~\GeV\ and a requirement on their transverse impact parameter significance of $\left|d_0/ \sigma(d_0)\right| <3$.
Furthermore, they are required to be isolated based on the `Loose' isolation criterion~\cite{MUON-2018-03}, which relies on variables calculated from energy deposits
within a cone around the muon. The angular width of this cone depends on the \pt\ of the muon, decreasing at higher \pt.
 
Jets are reconstructed using the anti-$k_t$ algorithm~\cite{Cacciari:2008gp,Cacciari:2011ma} with a radius parameter $R=0.4$
and particle-flow objects (PFO) as inputs. PFOs combine information from the inner detector and calorimeter to reconstruct the
energy and path of charged particles and neutral particles as described in Ref.~\cite{PERF-2015-09}.
Jet energy scale corrections, derived from MC simulation
and data, are used to calibrate the average energies of jet candidates to the scale of their
constituent particles~\cite{JETM-2018-05}.
To further reduce the effect of pile-up interactions, a jet-vertex-tagger (JVT) algorithm is used to identify jets originating
from the primary vertex using track information~\cite{PERF-2014-03}. Jets with $|\eta|<2.4$ and $\pt<60$~\GeV\ are
required to satisfy the `Tight' working point of this tagger, which corresponds to a JVT score of at least 0.5. In addition, jets
with $|\eta|> 2.5$ and $\pt<50$~\GeV\ are required to pass a `Tight' forward-JVT requirement~\cite{PERF-2016-06}, which corresponds
to a forward-JVT score of less than 0.4 and a jet-timing requirement of less than 10~ns.
Baseline-quality jets are selected in the region $|\eta|<4.5$ and must have a $\pt>20$~\GeV.
Signal-quality jets are required to fulfil $|\eta|<2.5$ and $\pt>30$~\GeV.
 
Jets containing $b$-hadrons are identified as arising from $b$-quarks
(`$b$-tagged' jets or \bjets)
using a multivariate algorithm (DL1r)~\cite{FTAG-2018-01}.
These $b$-tagged jets are reconstructed in the region $|\eta|<2.5$ and require $\pt>30$~\GeV.
The $b$-tagging working point used in this analysis provides an efficiency of 77\% for \bjets in simulated \ttbar events.
 
A second category of jets is reconstructed by applying the anti-$k_t$ algorithm with radius parameter $R=1.0$ to a collection
of noise-suppressed topological calorimeter-cell clusters calibrated using `local hadronic cell weighting'~\cite{PERF-2014-07} to
correct for the non-compensating response of the ATLAS calorimeter. These jets are referred to as large-$R$ jets to
distinguish them from the $R=0.4$ version, also called small-$R$ jets or simply `jets'. Large-$R$ jets~\cite{JETM-2018-06} are trimmed to remove pile-up and underlying-event effects. This trimming, extensively described in Ref.~\cite{Krohn:2009th}, is a grooming technique in which the original
constituents of the jets are reclustered using the $k_t$ algorithm~\cite{Ellis:1993tq} with a radius parameter, $R_{\mathrm{sub}}$,
to produce a collection of subjets. These subjets are then discarded if they have less than a specific fraction, $f_{\mathrm{cut}}$,
of the \pt\ of the original jet. The trimming parameters used are $R_{\mathrm{sub}} = 0.2$ and $f_{\mathrm{cut}}=0.05$.
The jet energy scale and resolution and the mass scale and resolution of these large-$R$ jets are then corrected via a calibration procedure described in Refs.~\cite{JETM-2018-02,ATLAS-CONF-2020-022}.
Large-$R$ jets are required to have a $\pt\ > 200$~\GeV\ and
$|\eta|<2.0$. To avoid reconstruction ambiguities between large-$R$
jets and leptons, those large-$R$ jets overlapping with signal leptons
are removed. Ambiguities between large-$R$ and baseline jets are not
removed, as large-$R$ jets are only used to construct higher-level
quantities in order to identify hadronically decaying $W$ bosons in
the event. A set of \Wboson-tagging identification criteria~\cite{JETM-2018-03}
are applied to these large-$R$ jets to identify those with topologies
consistent with the decay of energetic hadronically decaying \Wboson\ bosons.
These identification criteria are only used for jets with a
mass between 40~\GeV\ and 600~\GeV\ and a $\pt < 2.5$~\TeV\ and are based on the mass of the large $R$-jet,
the number of inner-detector tracks associated with the jet and the
$D_2$ variable~\cite{Larkoski:2014gra}. This latter variable uses jet substructure energy correlations to
identify deposits consistent with 2-prong particle decays against QCD quark and gluon initiated jets.
 
To compensate for remaining differences between data and simulation in trigger,
particle identification and reconstruction efficiencies,
correction factors are derived and applied to the samples of simulated events~\cite{EGAM-2018-01,MUON-2018-03,PERF-2016-05,ATL-PHYS-PUB-2020-017}.


\section{Analysis strategy}
\label{sec:analysis}

This analysis complements and extends a previous search performed by
the ATLAS Collaboration~\cite{EXOT-2018-43},
by targeting final states with an energetic $W$ boson and
characterised by the presence of exactly zero or one lepton, referred to as the \tWZeroL\ and \tWOneL\ channels, respectively.
 
Top-quark decays contain a \Wboson\ boson, and hence \twdm\ signals contain two \Wboson\ bosons in the decay chain.
The \tWZeroL\ channel selects \twdm\ events where both bosons
in the event decay hadronically, while the \tWOneL\ channel selects
events where one of them decays hadronically and the other decays leptonically. Both selections
require high jet multiplicity and significant
\met\ from two DM particles escaping detection.
In both channels, the \Wboson\ boson arising from the decay of the massive $H^\pm$ boson
is often produced with relatively
high \pt, thus being significantly boosted. When this \Wboson\ boson decays hadronically, it is reconstructed as a single large-$R$ jet and $W$-tagged
using the procedure described in Section~\ref{sec:objsel}.
The one-lepton channel
described in Ref.~\cite{EXOT-2018-43} is extended to
include such boosted \Wboson-boson events. It is constructed to be statistically
independent of the \tWZeroL\ channel so that
all signal regions (SRs) in this paper can be statistically combined.
The \tWTwoL\ analysis channel in Ref.~\cite{EXOT-2018-43} targets \twdm\ events with two opposite-sign leptons
and is statistically independent of the SRs presented in this paper. As a consequence, this channel can
be combined with \tWZeroL\ and \tWOneL\ to derive the final results.

The relative importance of SM background processes varies
across the different SRs. However, the most important can be broadly classified by
the presence of genuine \met produced by non-interacting particles,
e.g.\ neutrinos, or \met\ associated with the presence of particles that are either misidentified, mismeasured or outside the kinematic acceptance of the detector.
Examples of backgrounds containing genuine \met, which constitute a significant part of the SM background yields in their respective channels, are
the \zjets\ background in the \tWZeroL\ channel, where the \Zboson\ boson
decays into two neutrinos; and \wjets\ production in the \tWOneL\ channel, where a lepton and neutrino are present in the decay.
Other backgrounds such as \ttbar\ or \wjets\ (in the \tWZeroL\ channel) are examples of backgrounds that have high \met\ due to leptons in the event which
either escape detection or are misidentified as jets. Due to this, both make a large contribution in the \tWZeroL\ and \tWOneL\ channels.
Contributions from \ttZ and single top-quark production, in particular the
associated production of a top quark with a $W$ boson, are also significant.
The estimation of these five dominant SM backgrounds (\zjets, \wjets, \ttbar, \ttZ and single top quark) is aided by the use of
six dedicated control regions (CRs), which are designed to be orthogonal to the SRs and are
used to constrain six background normalization parameters in a phase
space as close as possible to that of the SRs. The background normalizations are
derived in common regions for the two analysis channels, with the exception of the
\ttbar\ background. Because \ttbar\ has different compositions in
the two channels, separate control regions and normalization parameters are used for \tWZeroL\ and \tWOneL\ channels. 
The validity of the background estimation strategy
is confirmed in specific validation regions (VRs) adapted for each
defined SR. The potential signal contamination in the CRs and  VRs is
found to be small: $<$~2.5\% and $<$~10\% of the total SM expectation for
all analysis channels, respectively.
 
The strategy for the statistical analysis and combinations performed in this paper
closely follows the one used in
Ref.~\cite{EXOT-2018-43}, and relies on a profile likelihood
fit~\cite{Baak:2014wma}, with the systematic
uncertainties, described in \Sec\ref{sec:unc}, introduced as nuisance
parameters constrained by a Gaussian distribution.
Following the definition of Ref.~\cite{EXOT-2018-43}, the fit is
performed using two configurations: background-only and exclusion fit set-ups.
In the background-only configuration the fit is used to estimate the reliability of the
background prediction in the VRs.
It is performed using all \tWZeroL\ and \tWOneL\ CRs in a simultanous fit and assuming
no contribution from `beyond-the-SM' (BSM) physics processes.
The six
normalization factors of the SM backgrounds are hence determined in all the control regions
simultaneously.
The
normalization factors determined in this set-up are applied to the
VRs in order to verify that the background predictions agree with the data.
The background-only fit configuration is also used
to estimate the model-independent limits in Section~\ref{sec:results}, by extrapolating the
background prediction of this fit to the SRs and estimating upper limits on the event yields of a general BSM signal in
inclusive (i.e. single-bin) SRs. In this way, exact knowledge of BSM signal correlations across bins is
not needed to estimate the result. Additionally, this configuration is also used to quantify the significance
of possible data deviations from SM predictions.
In the exclusion fit set-up, all CRs and SRs are fit simultaneously in
order to test a BSM signal plus SM background hypothesis against a SM-only
hypothesis. Unlike the model-independent configuration, these SRs are multi-bin regions that profit of the shape of benchmark signals
to enhance the sensitivity to the 2HDM+$a$ model in different areas of the parameter space.
All correlations between CRs and SRs are taken into account by the common background
normalization parameters and systematic uncertainty nuisance parameters.
This configuration is used to place limits on the production cross-section
at a given point in the parameter space of the 2HDM$+a$ model.
\FloatBarrier

\subsection{Signal regions}
\label{ssec:srs}

An optimization procedure is followed to derive the event selection criteria for the
\tWZeroL\ and \tWOneL\ channels. It
follows a two-step process, using a varying set of kinematic
variables.
First, a manual, coarse optimization is carried out, seeking to
maximise the sensitivity of the event selection to a set of benchmark signal
models. Then a random grid search algorithm
\cite{liashchynskyi2019grid} is used to fine-tune the coarse
selection criteria.
 
The \tWZeroL\ channel selection criteria are summarised in
Table~\ref{tab:SRs}. Following the signal topology, this channel selects events with exactly zero
leptons, at least four jets and at least one large-$R$ jet which is consistent with the hadronic decay
of a \Wboson\ boson (\Wboson-tagged).
Exactly one jet with
$\pt >50\;\GeV$ is required to be $b$-tagged.
Further requirements are placed on the \Wboson-boson candidate and the
$b$-jet to suppress events where they both originate from the
decay of the same top quark, as it is assumed that the boosted
$W$ boson in the signal topology arises from the decay of the charged
Higgs boson. These requirements involve a large angular separation
between the \Wboson-tagged large-$R$ jet and the leading (highest-\pt)
$b$-tagged jet (\dRWb) and an invariant mass of their combined
four-vector (\mWb) larger than the top-quark mass.

\def\arraystretch{1.3}
\begin{table}
\centering
\caption{Summary of the \tWZeroL\ and \tWOneL\ signal region selections. The \tWZeroL\ signal region and the \tWOneL\ region searching for hadronic top decays (\hadTop) are further split into bins of \met to increase the sensitivity for different signal model parameters as described in the text. The signal region seeking leptonic top decays (\lepTop) is used inclusively.}
\label{tab:SRs}
\begin{tabular}{l|c|c|c}
\hline{Variable} & \srzero & {\lepTop} & {\hadTop} \\
\addlinespace[-2ex] & & & \\
\hline
Trigger & \met\  & \met\ & \met \\
\hline
\met\ [\GeV]& $\geq 250$ & $\geq 250$ & $\geq 250$ \\
\metsigObj & $\geq 14$ & $\geq 15$ & - \\
\dphiminfour & $\geq 0.9$ & $\geq 0.5$ & $\geq 0.5$\\
\hline
Number of baseline leptons & 0 & 1 & 1 \\
Number of signal leptons & 0 & 1 & 1 \\
\ptLone[\GeV] & - & $\geq 30$ & $\geq 30$ \\
\hline
Number of signal jets & $\geq 4$ & $\geq 2$ & $\geq 3$ \\
\ptjone[\GeV] & $\geq 100$~~ & $\geq 50$ & $\geq 50$ \\
\ptjtwo[\GeV] & $\geq 60$ & $\geq 30$ & $\geq 30$ \\
\ptjthree[\GeV] & $\geq 60$ & - & $\geq 30$ \\
\ptjfour[\GeV] & $\geq 40$ & - & - \\
\hline
Number of $b$-tagged jets & $\geq 1$ & $\geq 1$ & $\geq 1$\\
\ptbone[\GeV] & $\geq 50$ & $\geq 50$ & $\geq 50$ \\
\ptbtwo[\GeV]& $\leq 50$ & $\leq 50$ & $\leq 50$ \\
\hline
Number of $W$-tagged jets (\nwtagged) & $\geq 1$ & $\geq 1$ & - \\
$\pt^{J;R=1.0}$ [\GeV] & $\geq 200$ & $\geq 200$ & - \\
\dRWb\ & $\geq$ 1.0 & - & -\\
\mWb\ [\GeV] & $\geq 220$ & - & -\\
\hline
\mtbmet\ [\GeV] & $\geq 180$ & - & - \\
\mbj[\GeV] & - & $\geq 200$ & $\leq 200$ \\
\mtw[\GeV] & - & $\geq 130$ & $\geq 200$ \\
\amt[\GeV] & - & $\geq 180$ & $\geq 180$ \\
\hline
\mWHad[\GeV] & - & - & ~~$\geq 60$ \\
\hline
\end{tabular}
\end{table}
\def\arraystretch{1.}

Requirements on \met and
its object-based significance, \metsigObj~\cite{ATLAS-CONF-2018-038},
are used to enhance the selection of events with invisible particles in the final
state. As the momentum of the DM particles in the signal strongly
depends on the mass difference between the $a$-boson mediator and the
$H^{\pm}$ boson, the signal region is further split into five bins in \met
to maximise the sensivity of this analysis throughout the full
considered parameter space. These five bins are defined with \met intervals [250,\,330]~\GeV,
[330,\,400]~\GeV, [400,\,500]~\GeV, [500,\,600]~\GeV\ and $\geq 600$~\GeV,
referred to, respectively, as \srzerobinOne--\srzerobinFive.
Inclusive signal regions, defined with \met\ $\geq $ 250, 330, 400, 500 and 600~\GeV,
are also defined in this analysis as `discovery regions'.
These single-bin overlapping SRs can be used to estimate either the significance of
an excess or an upper limit on the signal yield with less stringent assumptions about the
kinematic properties of the signal.
The minimum azimuthal angle between \met and the leading
four jets, \dphiminfour, is used to suppress fake \met arising
from mismeasured jets. The transverse mass variable constructed
from the leading \bjet\ of the event and the \ptmiss, \mtbmet,
\cite{SUSY-2018-12} is used to suppress events from semileptonic
\ttbar\ decays, which exhibit an endpoint in \mtbmet\ when the
\met\ in the event arises entirely from a missed $W$ boson.
 
The \tWOneL\ channel, also summarised in
Table~\ref{tab:SRs},  selects events with exactly one lepton
and exactly one $b$-tagged jet with $\pt >50\;\GeV$.
As in the \tWZeroL\ channel, requirements in \met, \metsigObj and
\dphiminfour\ are used to enhance the selection of events with
invisible particles and suppress events with fake \met.
The one-lepton channel was explored previously in
Ref.~\cite{EXOT-2018-43} and
strategies used in the previous paper are now extended with ideas
presented in Refs.~\cite{Pani:2017qyd,Brooijmans:2018xbu} and further
enhanced by the use of $W$-tagging techniques.
 
In the \tWOneL channel, events are selected with a boosted hadronically decaying
$W$ boson from the $H^{\pm}$ boson decays and a leptonically decaying $W$-boson from the top
quark decays. These events are selected for the \lepTop\ region. The complementary set of events, where the $W$ boson from the
$H^{\pm}$ boson decays leptonically and the $W$ boson from the top
quark decays hadronically, are selected for the \hadTop\ region.
A new variable, called \mbj, is used to guarantee that \lepTop\ and \hadTop\ are
statistically independent. It is constructed as the invariant mass of the leading
$b$-jet ($\mathrm{b}_{1}$) and the highest-\pt jet that is not $b$-tagged ($\cancel{\mathrm{b}}_{1}$).
Signal events with the top quark decaying hadronically exhibit an
endpoint in \mbj\ slightly below the top-quark mass, while
events with a leptonically decaying top quark extend beyond this
endpoint~\cite{Pani:2017qyd}.
 
Following the previous analysis~\cite{EXOT-2018-43}, the \lepTop\ and \hadTop\ regions exploit both the transverse mass
of the lepton and the \met, \mtw, and the asymmetric stranverse
mass~\cite{Konar:2009qr,Lester:2014yga,Lester:1999tx,Barr:2003rg,SUSY-2016-16}, \amt,
to suppress the background from semileptonic and dileptonic \ttbar decays, respectively.
The latter is constructed to have an
endpoint at the top-quark mass for dileptonic \ttbar\ events where
one of the leptons is outside the acceptance or misidentified.
In addition, the \lepTop\ region requires at least one \Wboson-tagged
large-$R$ jet, while \hadTop uses the variable \mWHad~\cite{EXOT-2018-43,SUSY-2016-16}.
Here, \mWHad\ uses a variable-radius jet reconstruction algorithm with standard
jet inputs to identify the hadronically decaying $W$ bosons in the
event even when their momentum is not high enough to be reconstructed
within a large-$R$ jet.
As in the \tWZeroL\ channel, binning the SRs in \met\ is the optimal
strategy to maximise the sensitivity throughout the full model
parameter space. However, due to the low event yield in \lepTop,
this strategy is implemented only in the \hadTop\ region. Five different bins are
then defined with \met intervals [250,\,300]~\GeV,
[300,\,350]~\GeV, [350,\,400]~\GeV, [400,\,450]~\GeV\ and $\geq 450$~\GeV,  referred to as
\hadTopBinOne--\hadTopBinFive. Similarly, inclusive `discovery' SRs are defined with
\met\ $\geq $ 250, 300, 350, 400 and 450~\GeV.
\FloatBarrier


\subsection{Background estimation and validation}
\label{ssec:crs}
 
\begin{figure}
\centering
\includegraphics[trim={20cm 23cm 37cm 10cm},clip,width=0.44\textwidth]{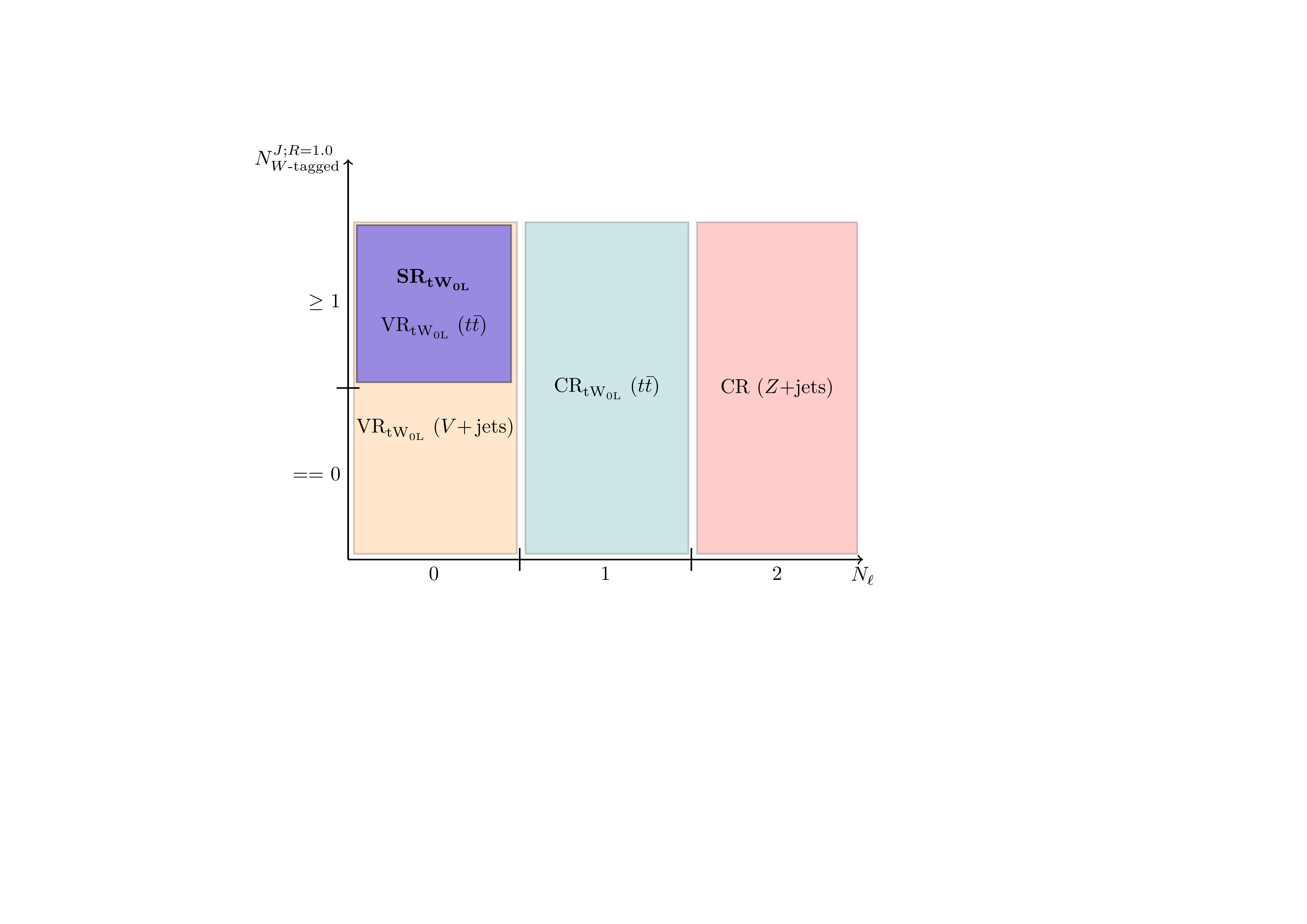}
\includegraphics[trim={20cm 33cm 39cm 13cm},clip,width=0.51\textwidth]{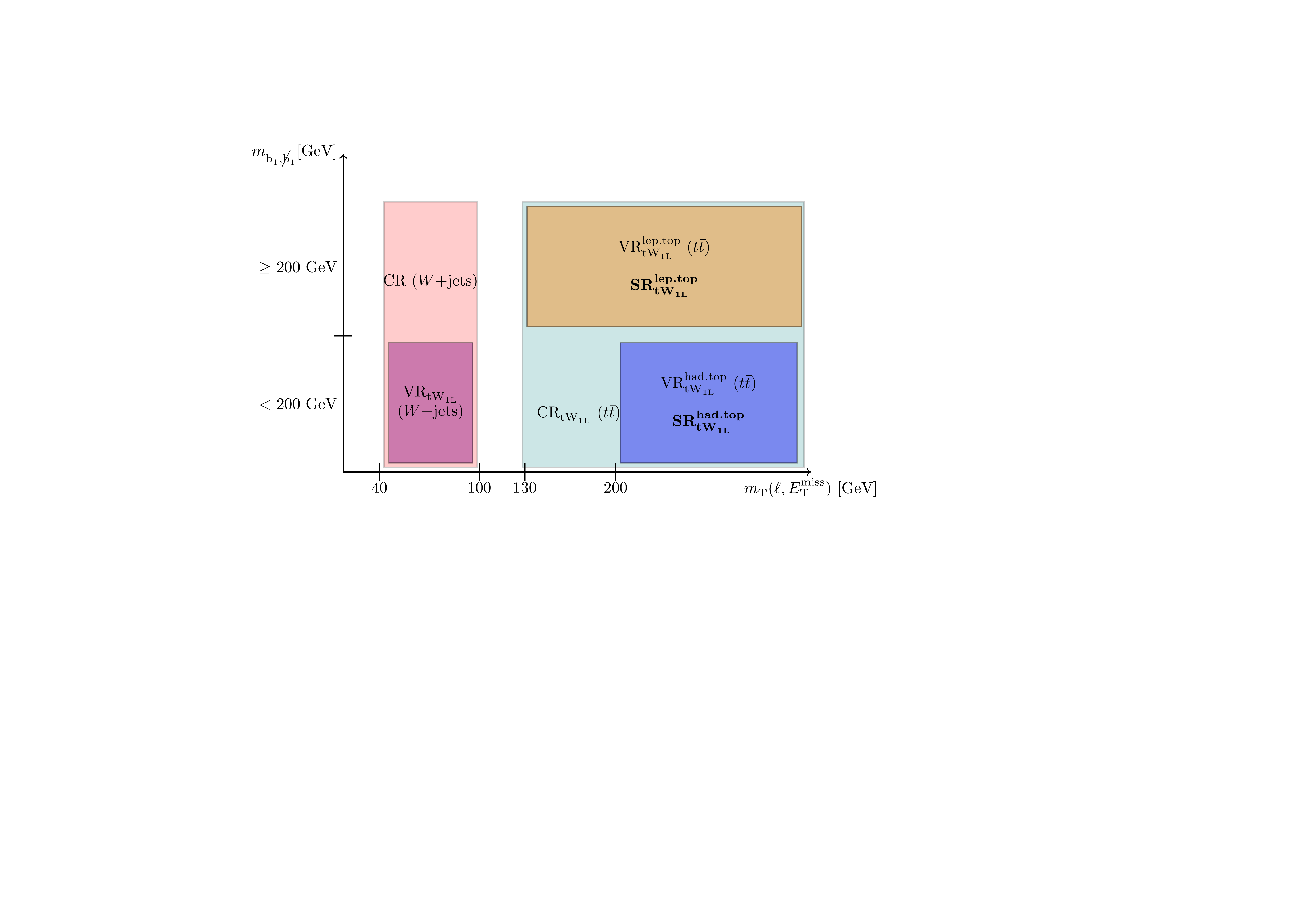}
\caption{Schema of the definitions of the different regions corresponding to (left) the \tWZeroL{} channel in \nwtagged\ and the number of leptons ($N_{\ell}$) and (right) the \tWOneL{} channel in \mt\ and \mbj. Orthogonality between the different regions observed to overlap is ensured by the inversion of some selection cuts in variables that aren't depicted in this figure. \mWb\ and \dRWb\ are inverted to ensure the orthogonality of the \tWZeroL\ signal and validation regions.  The \ttbar\ validation regions and the \tWOneL\ regions are independent thanks to the inversion of the selection cuts on \amt. Finally, \mWHad\ ensures the orthogonality of the \Wjets\ control and validation regions.}
\label{fig:scheme_0L1Lchan}
\end{figure}

Control regions are designed to support the estimation of the dominant backgrounds.
In the \tWZeroL\ channel, the three most important backgrounds are \Zjets, \ttbar and \Wjets. In the
\tWOneL\ channel, the most important backgrounds are \ttbar and \ttZ in
\lepTop\ and \ttbar and \Wjets in \hadTop.
All background processes, with the exception of \ttbar, are estimated
in common CRs and with common normalization parameters for the \tWZeroL\ and \tWOneL\ channels.
Figure~\ref{fig:scheme_0L1Lchan} schematically depicts the requirements imposed on the main analysis observables
in the CRs (and VRs) in order to ensure orthogonality to the SRs
and low signal contamination, as well as high purity in the targeted background, in each region.

The composition of the \ttbar\ background in the \tWZeroL\ and \tWOneL\ channels is very different.
In the former, the background is dominated by
semileptonic \ttbar\ decays, while in the latter, dileptonic
\ttbar\ decays dominate. In both cases, these
backgrounds satisfy the selection criteria because
one lepton is misidentified as a jet or falls outside of the detector
fiducial area.
Due to the difference in composition, two control regions are defined
in order to normalise the \ttbar
background for the \tWZeroL\ and \tWOneL\ channels.
The \tWZeroL\ \ttbar CR is enriched in
semileptonic \ttbar events by requiring exactly one lepton, low
\mt, low \amt and dropping the \metSig requirement. Requirements similar to those in the \tWZeroL\ SRs
are also imposed on the presence of a $W$-tagged large-$R$ jet and on \dRWb\ and \mWb\ to ensure that
this control region scans a topology similar to that in the SRs.
The \tWOneL\ \ttbar CR is enriched in dileptonic \ttbar\ events by inverting the constraints on
\amt\ and \ptbtwo. Since this region is used to estimate the \ttbar\ background in both \tWOneL\ SRs,
if a variable has different requirements in the two SRs, the requirement is either dropped
in the CR or chosen to be the looser one.
 
The \Zjets\ background, dominated in the \tWZeroL\ signal region by $\Zboson(\rightarrow\nu\nu)$+jets,
is estimated by selecting a large high-purity sample of events with two
same-flavour opposite-sign (SF-OS) leptons, as presented in Figure~\ref{fig:scheme_0L1Lchan}.
The leptons from the $Z$-boson decay are treated as invisible particles and
added to the \met of the event, now denoted by \METprime, to mimic the behaviour of the
\Zjets\ background in the \tWZeroL\ SR, where this background is dominant. The CR is defined by following the
selection criteria of the \tWZeroL\ SR, but variables built with \met\ in the SR are built with \METprime\ instead.
The \Wjets background is estimated in a CR selecting events with exactly one lepton and \mtw\ in the
\Wboson-boson mass range $[40, 100]$~\GeV\ (as presented in Figure~\ref{fig:scheme_0L1Lchan}),
high \metSig and low \mWHad to ensure high acceptance for \Wjets events.

The  estimation of the \ttZ\ background is
performed in a selection requiring exactly three leptons as described
in Ref.~\cite{SUSY-2018-08}. A \Zboson-boson candidate is
reconstructed from the SF-OS lepton
pair with invariant mass closest to the \Zboson-boson mass. The resulting
lepton pair is treated as invisible. The contribution from jets
misidentified as leptons in this control region is estimated using MC samples and
amounts to less than $10\%$~\cite{SUSY-2018-08}.
 
Finally, to correctly estimate the single-top-quark background and, therefore, to reduce
the systematic uncertainties arising from its modelling,
a dedicated CR with two leptons and high \met\ is
constructed. Most \Zjets events are removed by requiring \mll\ to be outside of
the \Zboson-boson mass range i.e.\ $\notin [71, 111]$~\GeV. Events with a leptonically decaying
\Wboson\ boson are selected by means of a low
\mttwo\ requirement built using both leptons in the event and \met. The variables \minmbl\ and \mblt~\cite{TOPQ-2017-05,EXOT-2018-43} are built by combining
the leptons and jets in the event and present an endpoint in the range 153--170~\GeV, close to the mass of the top quark.
They are highly efficient in separating the single-top-quark, \ttbar\ and \ttZ\ backgrounds and are used to increase this CR's purity
in single-top-quark events.
 
A summary of all control region definitions can be found in Tables \ref{tab:CRs1} and \ref{tab:CRs2}.
Normalization factors for all of the aforementioned SM backgrounds are
fitted simultaneously in these regions using the background-only fit configuration.
Their values are  $\mu_{\ttbar}^{\tWZeroL} = 1.00 \pm 0.12 $, $\mu_{\ttbar}^{\tWOneL} = 0.92  \pm 0.06$, $\mu_{\Zjets} = 0.98  \pm 0.07$,
$\mu_{\Wjets}=1.08  \pm 0.09$, $\mu_{\mathrm{single top}} = 0.43  \pm 0.13$ and $\mu_{\ttZ} = 1.18  \pm 0.19$.
There is a large discrepancy between the fitted value of the single-top normalization parameter and the Monte Carlo predicted value.
This discrepancy is driven by the dominant contribution of the $tW$ process to the single-top channel and related to the modelling of the
interference between single-resonant and double-resonant top-quark production.
It is found that the default scheme used to model this interference (diagram removal~\cite{Frixione:2008yi}) and the alternative scheme used
to estimate the associated uncertainty (diagram subtraction, see Section~\ref{sec:unc} for details) bracket the observed number of events
in the single-top CR data, with a large difference between the two predictions. The single-top CR allows $\mu_{\mathrm{single top}}$
to be constrained by data independently of the choice of default interference scheme. Residual shape differences between the two schemes are
assigned as systematic uncertainties as described in Section~\ref{sec:unc}.
 
\begin{figure}[htb]
\center
\subfloat[]{
\includegraphics[width=0.45\textwidth]{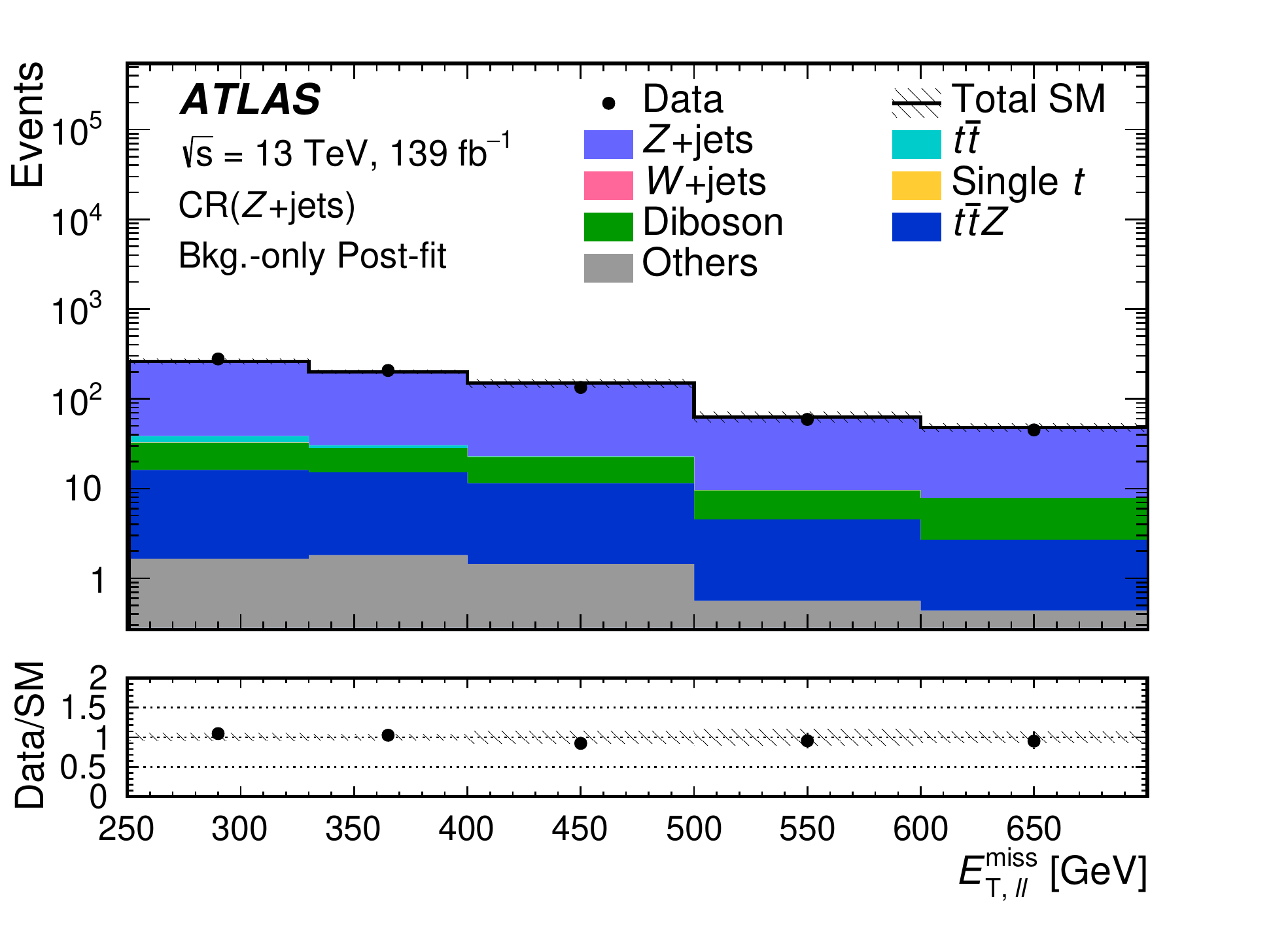}}
\subfloat[]{
\includegraphics[width=0.45\textwidth]{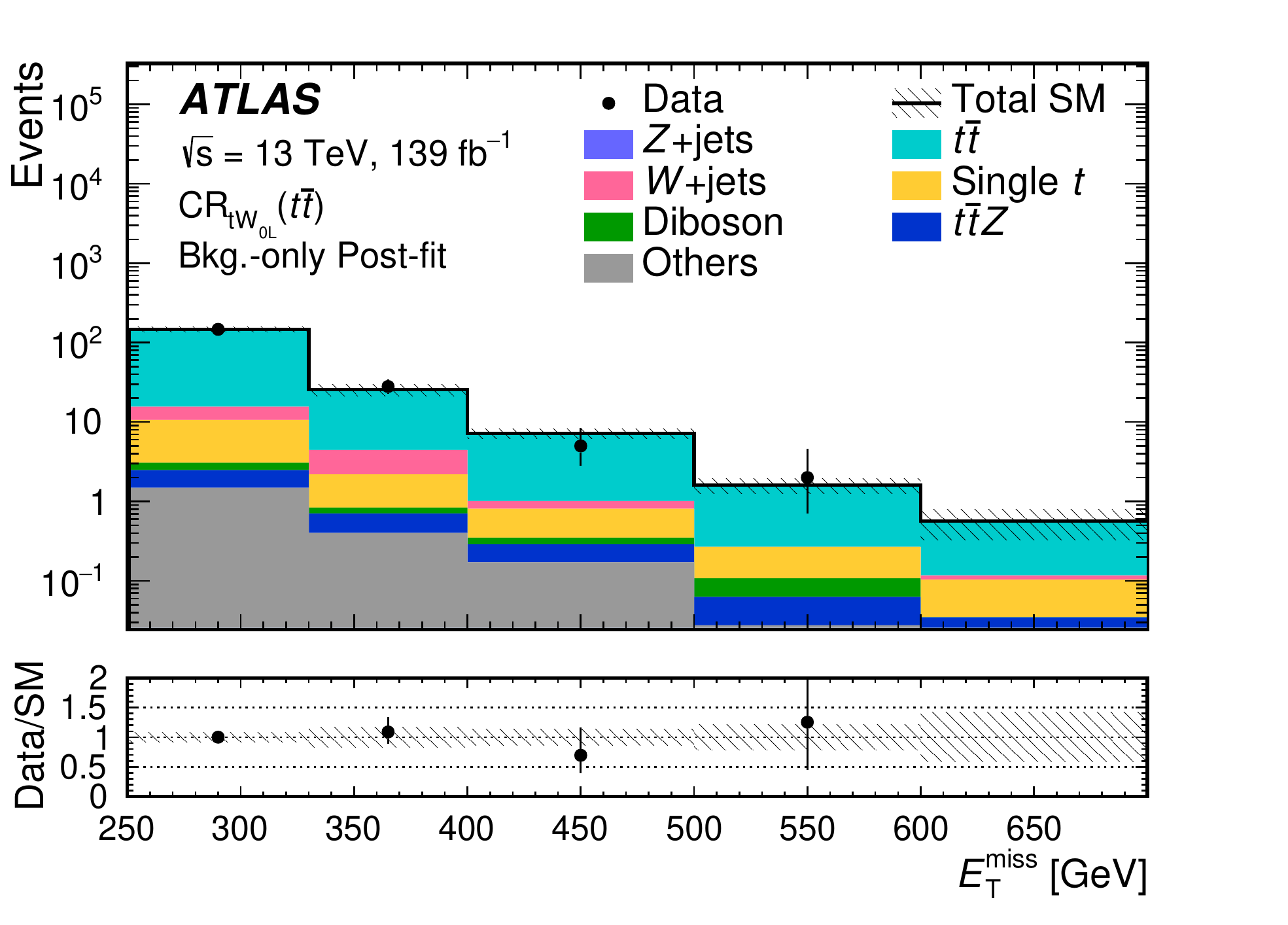}} \\
\subfloat[]{
\includegraphics[width=0.45\textwidth]{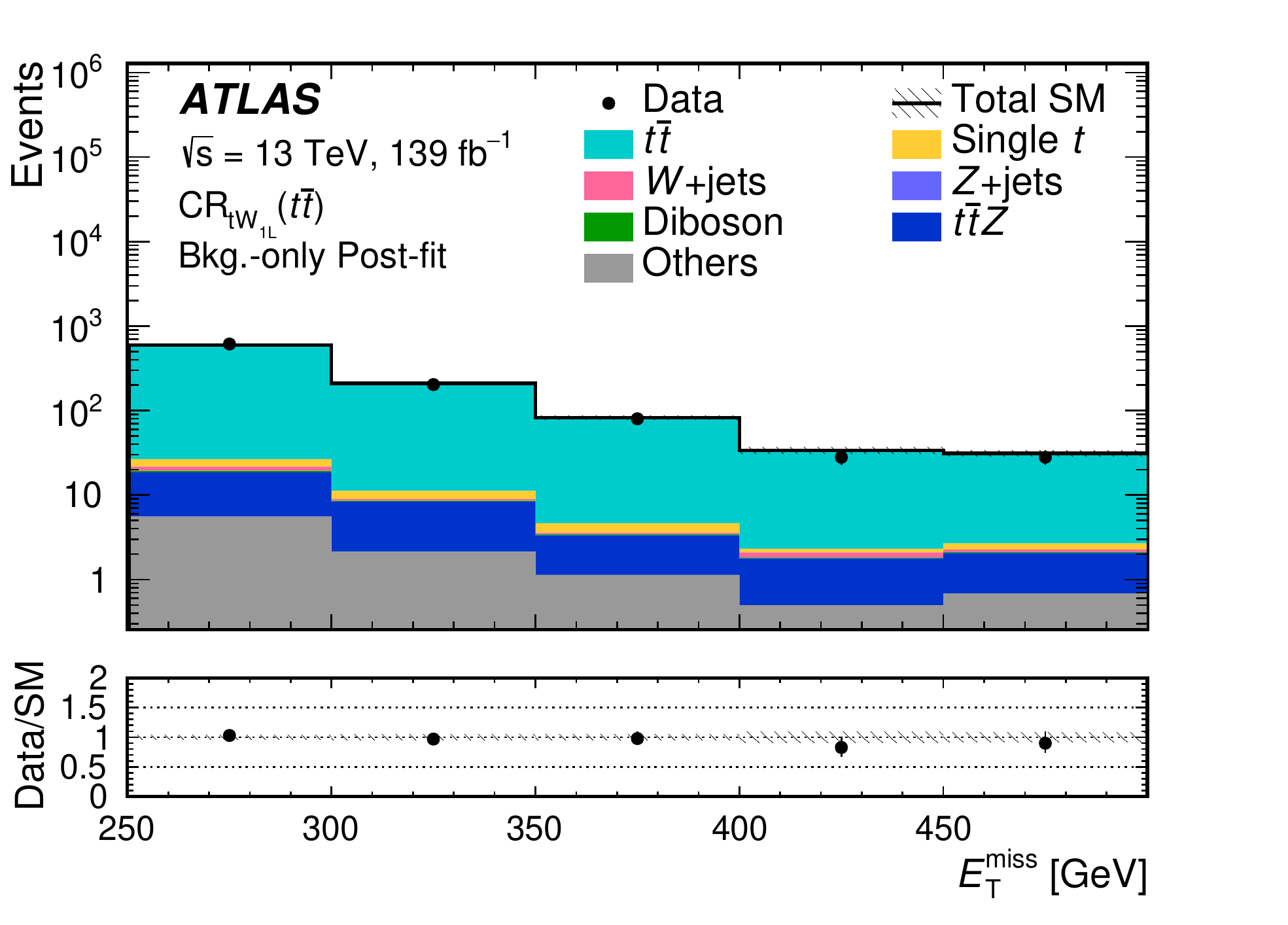}}
\subfloat[]{
\includegraphics[width=0.45\textwidth]{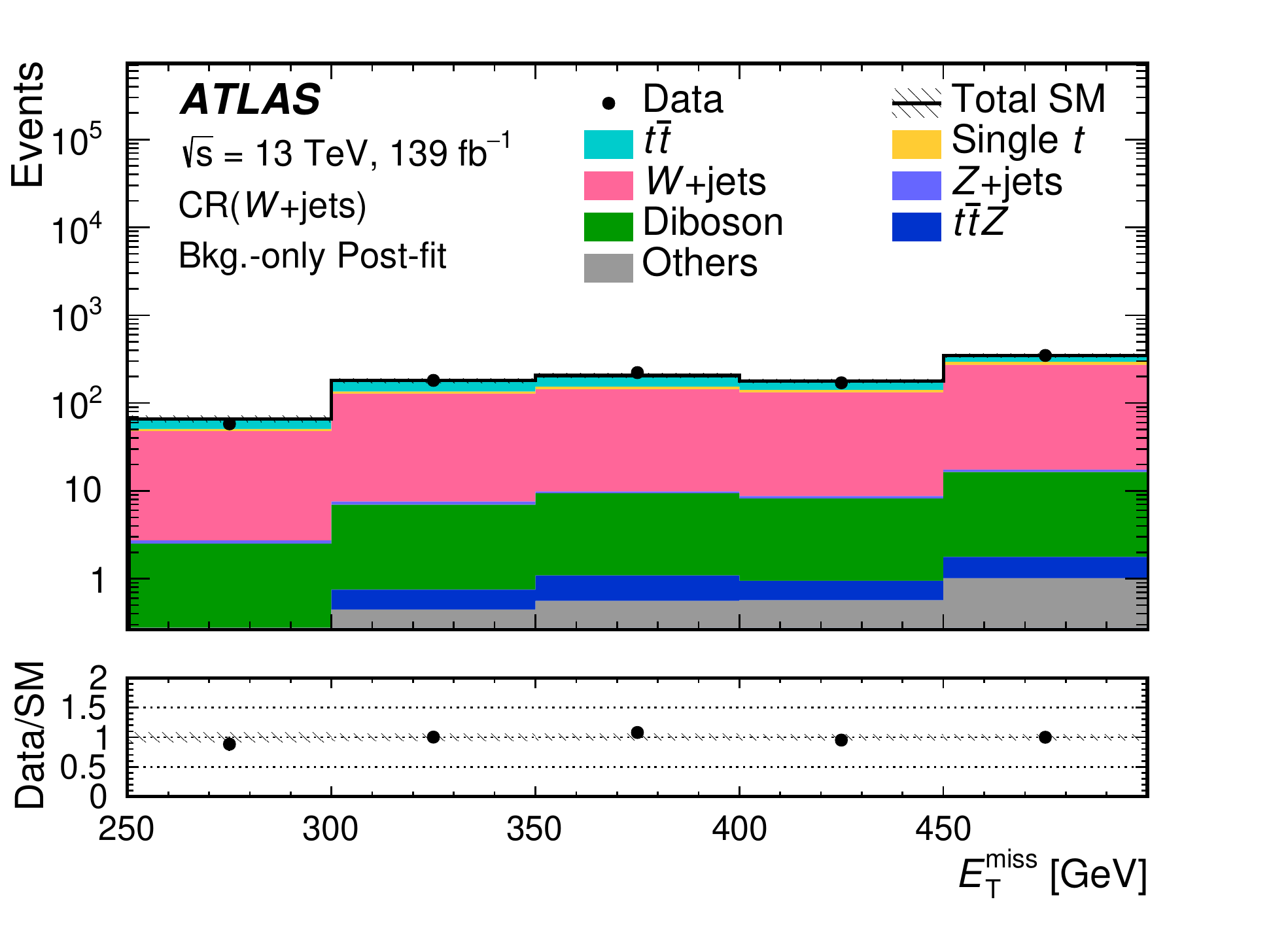}} \\
\subfloat[]{
\includegraphics[width=0.45\textwidth]{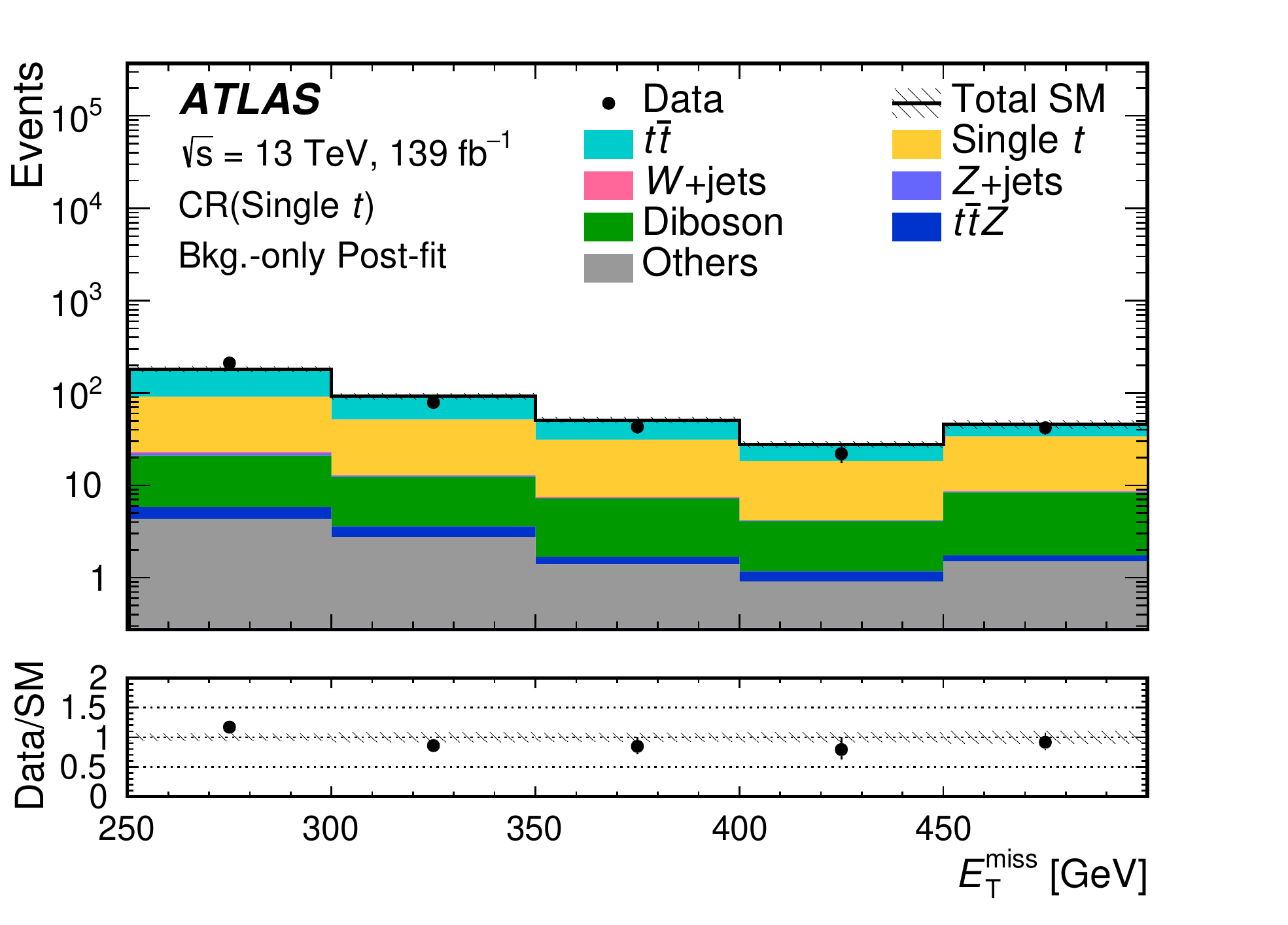}}
\subfloat[]{
\includegraphics[width=0.45\textwidth]{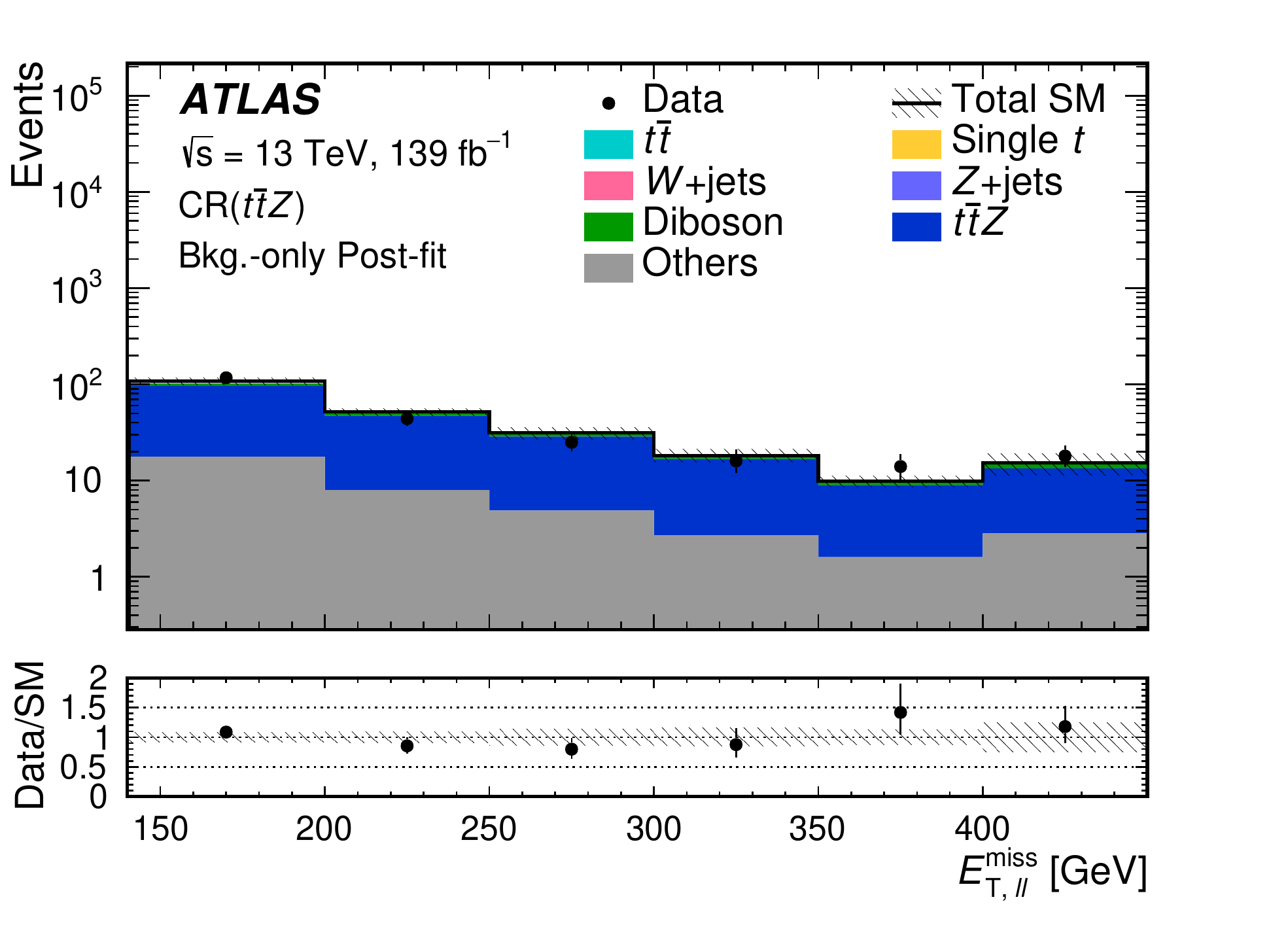}}
\caption{
\label{fig:CRs}
The post-fit \met and \METprime distributions in the (a) \Zjets, (b) \tWZeroL\ \ttbar, (c) \tWOneL\ \ttbar, (d) \Wjets, (e) single-top, and (f) \ttZ control regions. The last bin in the histogram includes the overflow events. The bottom panel shows the ratio of data to total SM background.
The uncertainties shown are the sum of the statistical and post-fit systematic uncertainties as detailed in Section~\ref{sec:unc}. The fit set-up corresponds to the background-only fit configuration.
The `Others' category includes contributions from rare processes such as $tWZ$, $tZ$, triboson, $ttt$, $t\bar{t}t\bar{t}$, $ttW$ and $ttH$.
}
\end{figure}

Figure~\ref{fig:CRs} shows the post-fit \met and \METprime distributions for all CRs, where good agreement between data and fitted predictions can be observed.
\def\arraystretch{1.5}
\begin{table}
\centering
\caption{Summary of \Wjets, \Zjets, and both \ttbar control regions. All variables used in \Zjets CR are calculated using a corrected version of the \met, denoted by \METprime, treating all leptons in the event as invisible. `SF-OS' indicates that the selected leptons are required to have the same flavour and  opposite-sign electric charges, such that they are compatible with the decay of a $Z$ boson.}
\label{tab:CRs1}
\begin{tabular}{l|c|c|c|c}
\hline
Variable  & CR$_\tWZeroL$ (\ttbar)  & CR$_\tWOneL$ (\ttbar) & CR (\Wjets) & CR (\Zjets)\\
\addlinespace[-3ex] & & & & \\
\hline
Trigger & \met\  & \met\  & \met  & Single-lepton \\
\hline
\met\ [\GeV]& $\geq 250$ & $\geq 250$ & $\geq 250$ & $\leq 120$ \\
\METprime[\GeV] & -- & -- & -- & $\geq 250$ \\
\metSig & -- & -- & $\geq 15$ & -- \\
\metSigPrime & -- & -- & -- & $\geq 14$\\
\dphiminfour & $\geq 0.5 $ & $\geq 0.5 $ & $\geq 0.5 $& -- \\
\dphiminfourprime & --  & -- & -- & $\geq 0.5 $ \\
\hline
Number of baseline leptons & 1 & 1 & 1 & 2 \\
Number of signal leptons & 1 & 1 & 1 & 2 (SF-OS) \\
\ptLone[\GeV] & $\geq 30$ & $\geq 30$ & $\geq 30$ & $\geq 30$  \\
\ptLtwo[\GeV] & -- & -- & -- & $\geq 20$  \\
\hline
Number of signal jets & $\geq 4$ & $\geq 3$ & $\geq 3$ & $\geq 4$ \\
\ptjone[\GeV] & $\geq 100$~~ & $\geq 30$ & $\geq 30$ & $\geq 100$~~ \\
\ptjtwo[\GeV] & $\geq 60$ & $\geq 30$ & $\geq 30$ & $\geq 60$ \\
\ptjthree[\GeV] & $\geq 60$ & $\geq 30$ & $\geq 30$ & $\geq 60$ \\
\ptjfour[\GeV] & $\geq 40$ & -- & -- & $\geq 40$ \\
Number of $b$-tagged jets & $\geq 1$ & $\geq 2$ & $\geq 1$ & $\geq 1$ \\
\ptbone[\GeV] & $\geq 50$ & $\geq 50$ & $\geq 50$ & $\geq 50$ \\
\ptbtwo[\GeV] & $\leq 50$ & $\geq 50$ & $\leq 50$ & $\leq 50$ \\
\hline
Number of $W$-tagged jets (\nwtagged)  & $\geq 1$ & -- & $= 0$ &  $\geq 0$  \\
\dRWb & $\geq 1.0$ & -- & -- & -- \\
\mWb\ [\GeV] & $\geq 220$ & -- & -- & -- \\
\hline
\mll[\GeV]& -- & -- & -- & $\in [81,101]$ \\
\hline
$m_\mathrm{T}$($\text{b}_1$, \METprime) [\GeV] & -- & -- & -- & $\geq 180$ \\
\mtw[\GeV] & $\in [30,130]$ & $\geq 130$ & $\in [40,100]$ & -- \\
\amt[\GeV] & $< 180$ & $< 180$ & $\geq 180$ & -- \\
\mWHad[\GeV] & -- & -- & ~~$< 60$ & -- \\
\hline
\end{tabular}
\end{table}
 
\def\arraystretch{1.5}
\begin{table}
\centering
\caption{Summary of the single-top-quark and \ttZ control regions. `SF' and `OS' indicate that the two leptons are required to have the same flavour and  opposite-sign electric charges, respectively. For the \ttV CR, the leptons treated as invisible in \METprime are the SF-OS pair with invariant mass closest to the $Z$-boson mass.}
\label{tab:CRs2}
\begin{tabular}{l|c|c}
\hline
Variable & CR (Single $t$) & CR (\ttZ) \\
\hline
Trigger & \met\ & Single-lepton   \\
\hline
\met\ [\GeV]&  $\geq 250$ & -- \\
\METprime[\GeV] & -- & $\geq 140$ \\
\dphiminfour & $\geq 0.5 $& -- \\
\hline
Number of baseline leptons & 2 & 3 \\
Number of signal leptons &  2 (OS) & 3 (at least one SF-OS pair) \\
\ptLone[\GeV] & $\geq 25$ & $\geq 30$ \\
\ptLtwo[\GeV] & $\geq 20$ & $\geq 20$ \\
\ptLthree[\GeV] & -- & $\geq 20$ \\
\hline
Number of signal jets & $\geq 1$ & $\geq 3$ \\
\ptjone[\GeV] & $\geq 50$ & $\geq 30$ \\
\ptjtwo[\GeV] & -- & $\geq 30$ \\
\ptjthree[\GeV] & -- & $\geq 30$ \\
Number of $b$-tagged jets & $\geq 1$ & $\geq 2$ \\
\ptbone[\GeV] & $\geq 50$ & $\geq 30$ \\
\ptbtwo[\GeV] & --  & $\geq 30$ \\
\hline
\mll[\GeV]& $\geq 40$, $\not\in [71,111]$ if SF & $\in [71,111]$ for at least one SF-OS pair \\
\hline
\mtw[\GeV] & $ > 30$ &  -- \\
\mttwo[\GeV] & $< 100$ & -- \\
\minmbl[\GeV] & $> 170$ & -- \\
\mblt[\GeV] & $> 150$ & -- \\
\hline
\end{tabular}
\end{table}


 
Validation regions are defined in order to verify that the background estimation strategy is robust.
One or more VRs are designed to validate each background estimate from the CRs.
The \tWZeroL\ \ttbar background estimate is validated using a
selection with zero reconstructed leptons.
In order to ensure orthogonality to the SRs and a high
\ttbar\ background purity, the \mWb\
and \metSig selection requirements are inverted. A similar
strategy is used for the \Zjets and \Wjets VRs, where
the normalization factors are extrapolated
from two-lepton and one-lepton control regions to a
zero-lepton selection. However, since the definition of a \Wjets-enriched region
using an event selection with no leptons poses a challenge due to its similarity to \Zjets,
a combined \Vjets validation region is defined for \Wjets and \Zjets with the goal of high
acceptance for the sum of the two processes. In the \Vjets validation region, the selection requirement on \dRWb\
is also inverted to be orthogonal to the signal region, but the \metSig selection requirement is kept the same, as this
ensures orthogonality to the \tWZeroL\ \ttbar VR. Furthermore, in order to decrease statistical uncertainties,
the \nwtagged selection requirement is relaxed, as shown in Figure~\ref{fig:scheme_0L1Lchan}, and requirements on
$\min[\Delta\phi(\mathrm{j^{all}},\met)]$ and $\Delta R_\mathrm{j1,j2}$ are
imposed to increase the \Zjets and \Wjets purity.
 
\def\arraystretch{1.5}
\begin{table}[p]
\centering
\caption{Summary of the \tWZeroL\ \ttbar and the \Vjets validation regions. }
\label{tab:0LVR}
\begin{tabular}{l|c|c}
\hline
Variable & VR$_\tWZeroL$ (\ttbar) & VR$_\tWZeroL$ (\Vjets) \\
\addlinespace[-3ex] & & \\
\hline
Trigger & \met\  & \met\  \\
\hline
\met\ [\GeV]& $\geq 250$ & $\geq 250$ \\\
\metsigObj & $\in$ [10, 14] & $\geq 14$ \\
\dphiminfour & $\geq 0.9 $ & $\geq 0.5 $\\
\hline
Number of baseline leptons & 0 & 0\\
Number of signal leptons & 0 & 0\\
\hline
Number of signal jets & $\geq 4$ & $\geq 4$	\\
\ptjone[\GeV] & $\geq 100$ & $\geq 100$ \\
\ptjtwo[\GeV] & $\geq 60$ & $\geq 60$ \\
\ptjthree[\GeV] & $\geq 60$ & $\geq 60$ \\
\ptjfour[\GeV] & $\geq 40$ & $\geq 40$ \\
Number of $b$-tagged jets & $\geq 1$ & $\geq 1$ \\
\ptbone[\GeV] & $\geq 50$ & $\geq 50$ \\
\ptbtwo[\GeV]& $\leq 50$ & $\leq 50$ \\
\hline
Number of $W$-tagged jets (\nwtagged)  & $\geq 1$ & $\geq 0$ \\
\dRWb & - & < 1.0\\
\mWb\ [\GeV] & $ < 220$ & - \\
\hline
\mtbmet[\GeV] & $\geq 180$ & $\geq 180$ \\
\hline
$\min[\Delta\phi(\mathrm{j^{all}},\met)]$ & -- & $\geq$ 1.2\\
$\Delta R_\mathrm{j1,j2}$ & -- & $\geq$ 1.2 \\
\hline
\end{tabular}
\end{table}
\def\arraystretch{1.}
 
To validate the \ttbar prediction in the \tWOneL channel, one validation region per
SR is constructed. In both VRs,  low \amt\ is required, both to ensure orthogonality to
the signal regions and to enhance the \ttbar fraction. To
increase the acceptance in the regions, the \Wboson-tagging requirement is dropped in the
\lepTop\ \ttbar validation region and the \mWHad requirement is dropped in
the \hadTop\ \ttbar validation region. The \Wjets VR is kinematically close to \hadTop.
The acceptance of \Wjets events is increased by
constraining \mtw to be in the \Wboson-boson mass range and demanding high
\metSig. The resulting region has large \ttbar and single-top-quark
contributions and can be considered a simultaneous validation region for all three
backgrounds. Good agreement between data and the \ttZ background predicted
by its CR was reported in Ref.~\cite{SUSY-2018-08}, so no dedicated \ttZ VR is considered in this analysis.
\def\arraystretch{1.5}
\begin{table}[p]
\caption{Summary of the \tWOneL\ \ttbar, \Wjets and single-top-quark validation regions.}
\label{tab:1LVRs}
\centering
\begin{tabular}{l|c|c|c|c}
\hline
Variable & VR$_\tWOneL^{\mathrm{lep.top}}$ (\ttbar) & VR$_\tWOneL^{\mathrm{had.top}}$ (\ttbar) & VR$_\tWOneL$ (\Wjets) & VR (Single $t$) \\
\addlinespace[-3ex] & & & & \\
\hline
Trigger & \met\  & \met\  & \met  & \met\ \\
\hline
\met\ [\GeV]& $\geq 250$ & $\geq 250$ & $\geq 250$ & $\geq 250$ \\
\metSig & $\geq 15$ & -- & $\geq 15$ & $\geq 15$\\\hline
\dphiminfour & $\geq 0.5 $ & $\geq 0.5 $ & $\geq 0.5 $& $\geq 0.5 $ \\
\hline
Number of baseline leptons & 1 & 1 & 1 & 1\\
Number of signal leptons & 1 & 1 & 1 & 1 \\
\ptLone[\GeV] &  $\geq 30$ & $\geq 30$ & $\geq 30$ &$\geq 30$\\
\hline
Number of signal jets & $\geq 2$ & $\geq 3$ & $\geq 3$ & $\geq 3$\\
Number of $b$-tagged jets & $\geq 1$ & $\geq 1$ & $\geq 1$ & $\geq 2$ \\
\ptbone[\GeV] & $\geq 50$ & $\geq 50$ & $\geq 50$ & $\geq 50$ \\
\ptbtwo[\GeV]& $\leq 50$ & $\leq 50$ & $\leq 50$ & $\geq 50$ \\
\hline
\mtw[\GeV] & $\geq 130$ & $\geq 200$ & $\in [40,100]$ & $\in [30,100]$ \\
\hline
\mbj[\GeV] & $\geq 200$ & $< 200$ & $< 200$ & -- \\
\amt[\GeV] & $< 180$ & $< 180$ & $\geq 180$ & $\geq 180$ \\
\hline
\mWHad[\GeV] & -- & -- & $\geq 60$ & -- \\
\hline
\end{tabular}
\end{table}
\def\arraystretch{1.}
 
\begin{figure}[p]
\center
\subfloat[]{    
\includegraphics[width=0.45\textwidth]{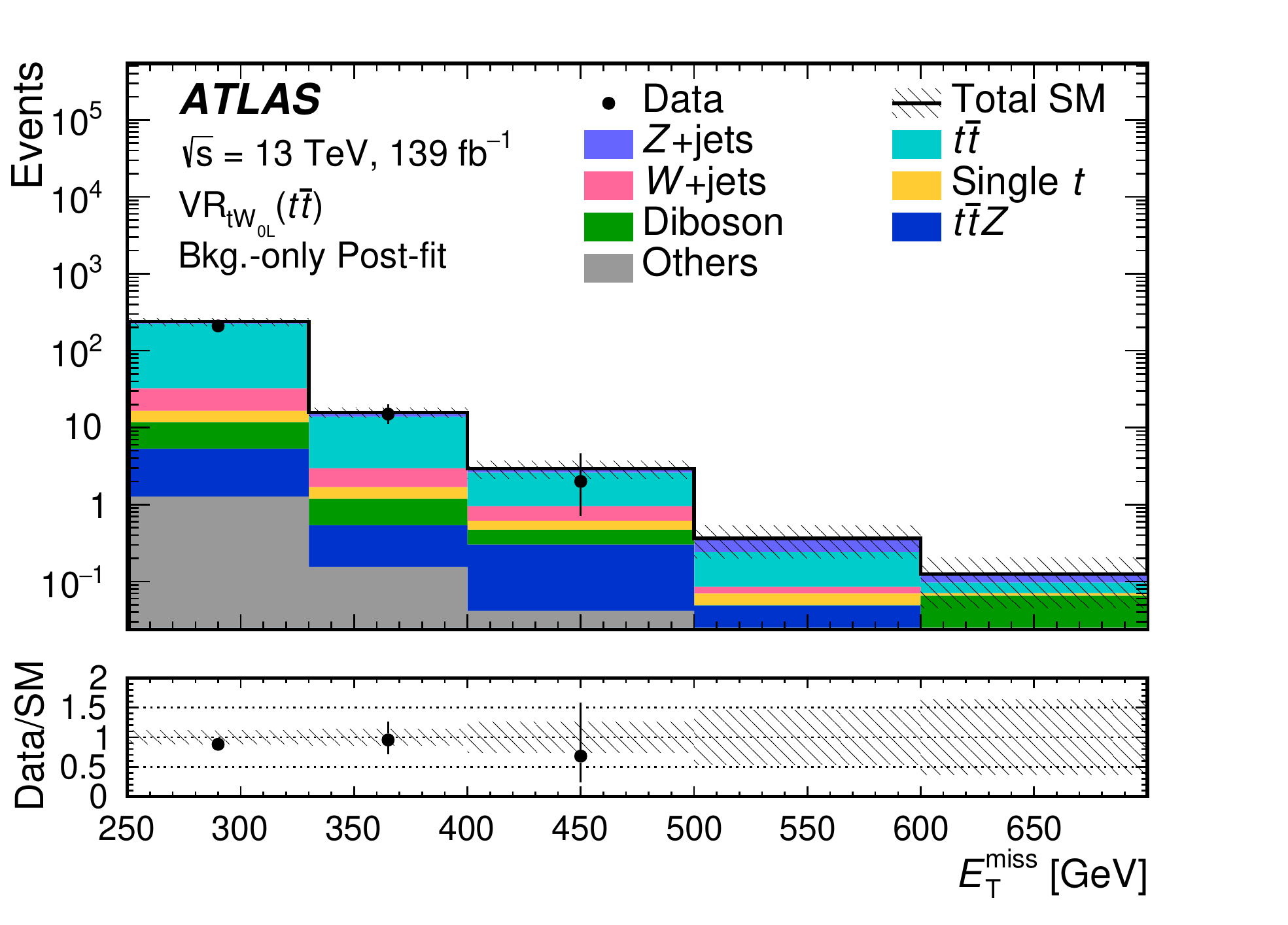}}
\subfloat[]{    
\includegraphics[width=0.45\textwidth]{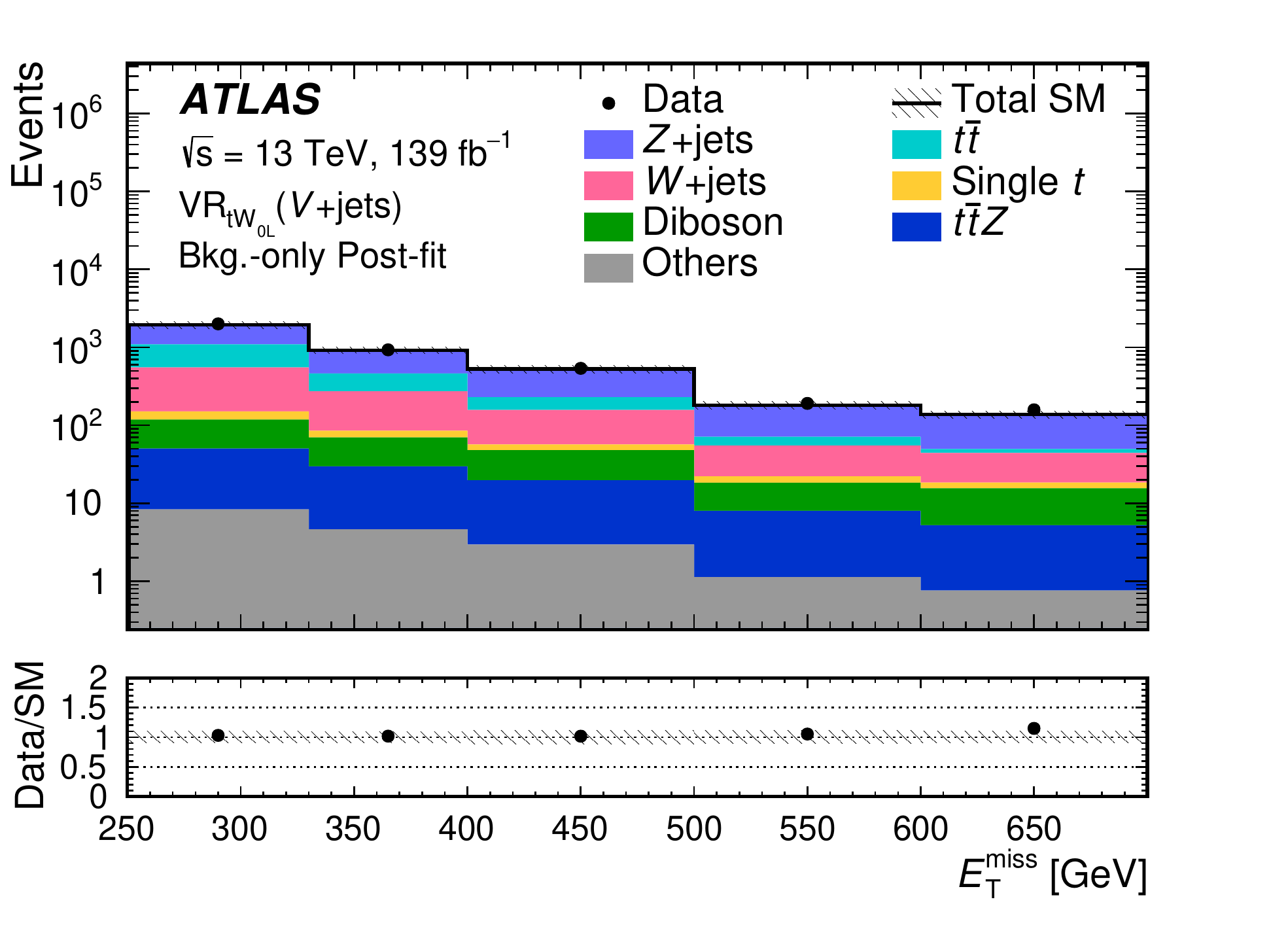}}\\
\subfloat[]{    
\includegraphics[width=0.45\textwidth]{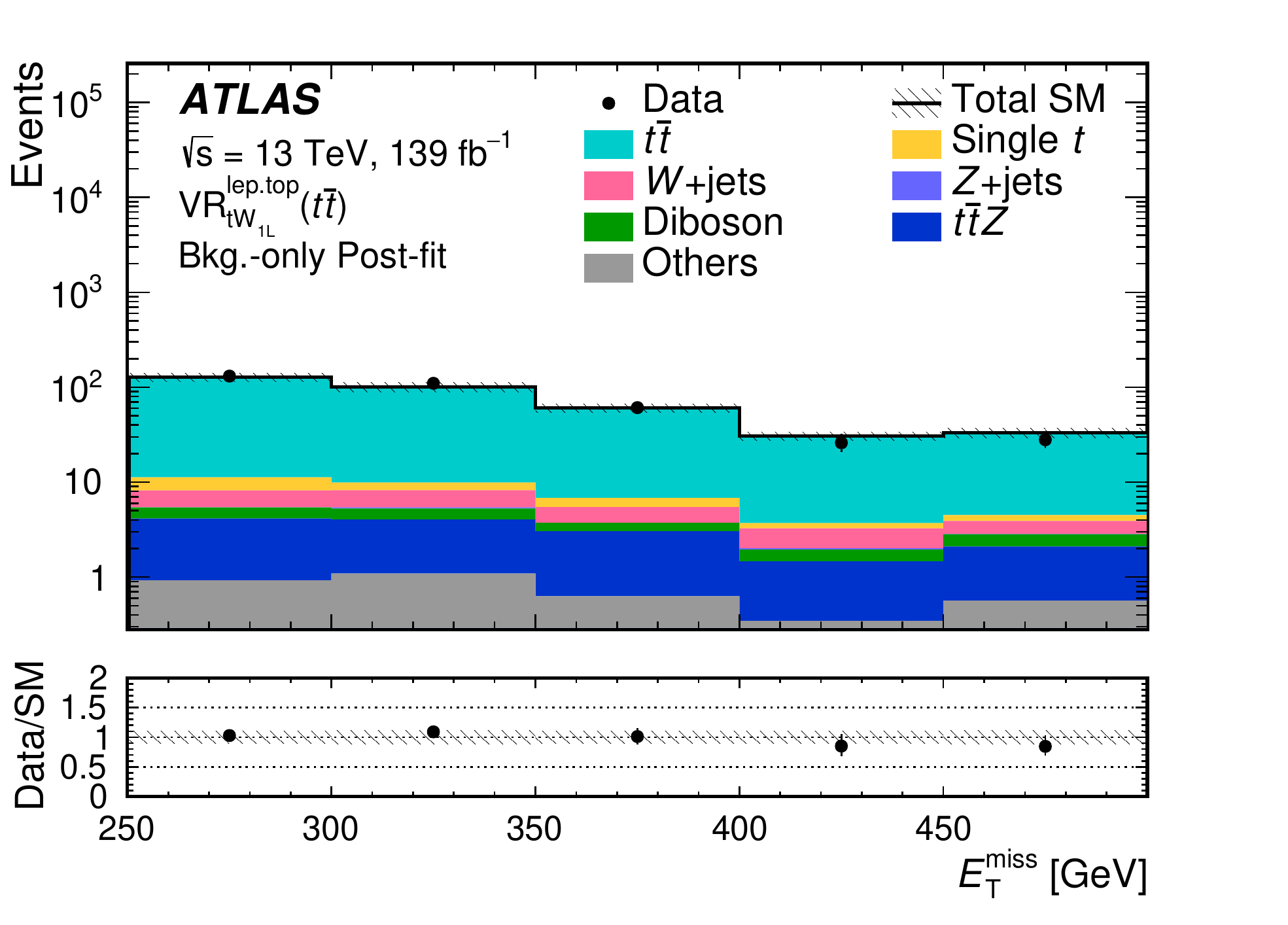}}
\subfloat[]{    
\includegraphics[width=0.45\textwidth]{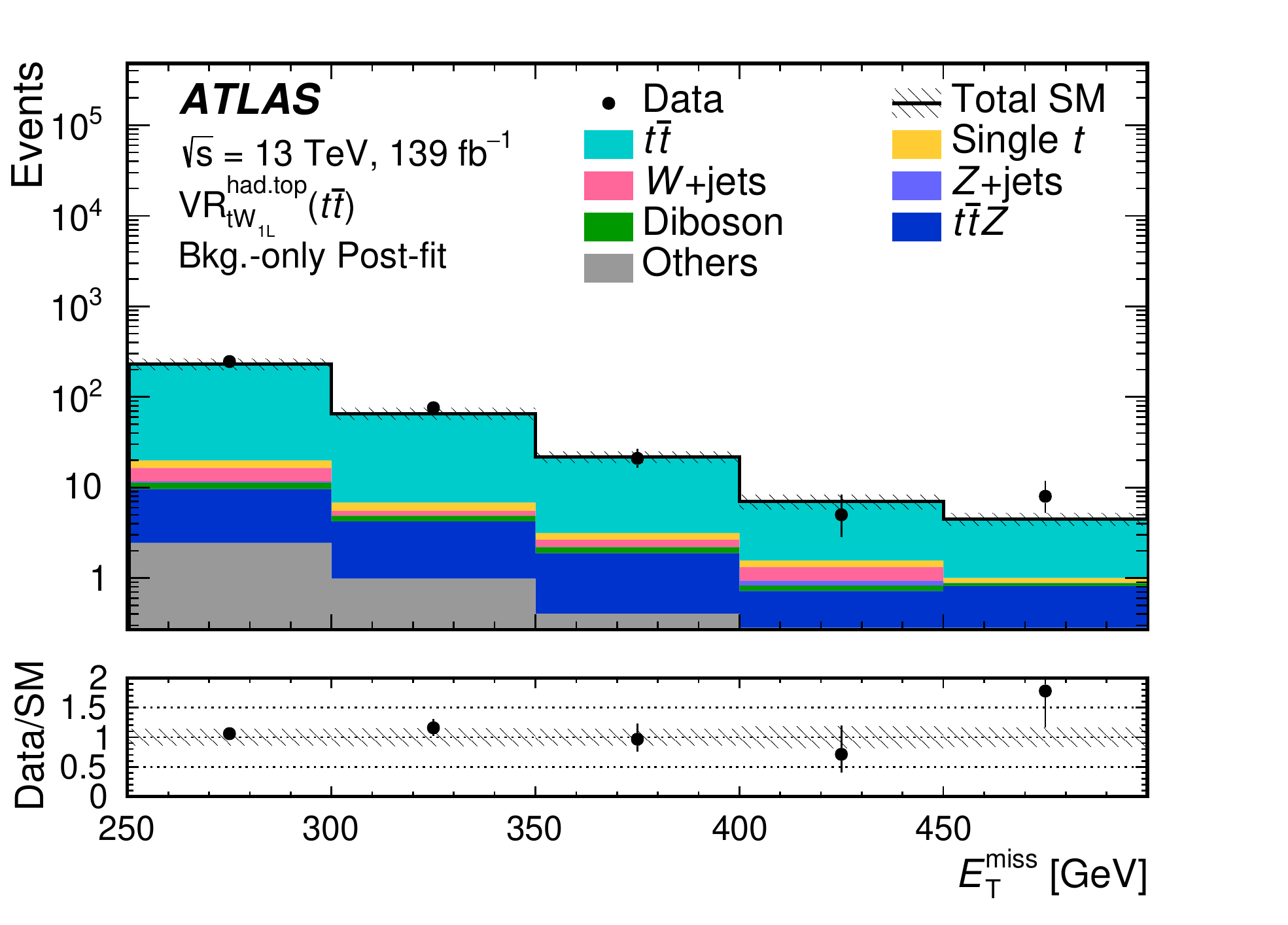}} \\
\subfloat[]{    
\includegraphics[width=0.45\textwidth]{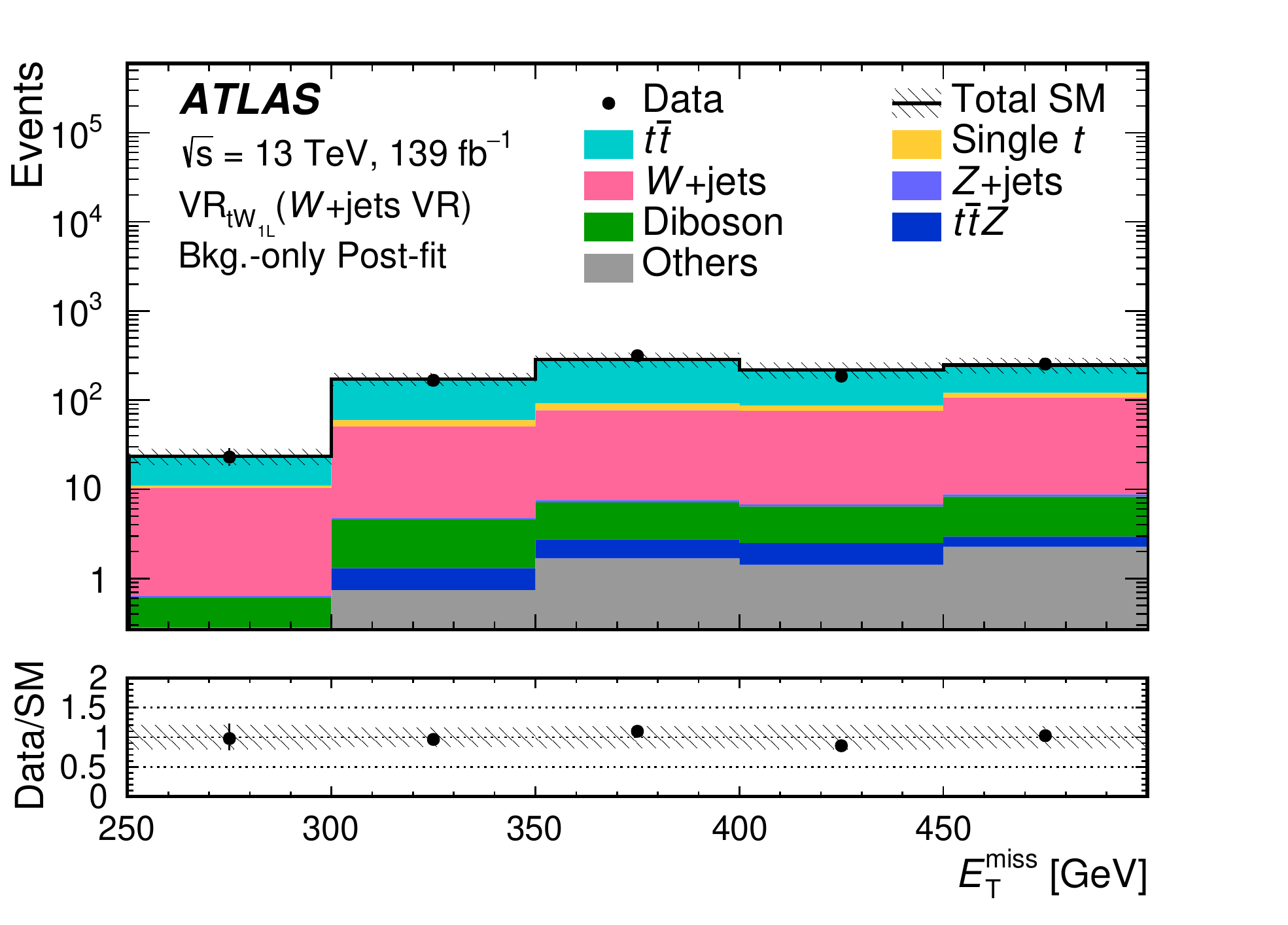}}
\subfloat[]{    
\includegraphics[width=0.45\textwidth]{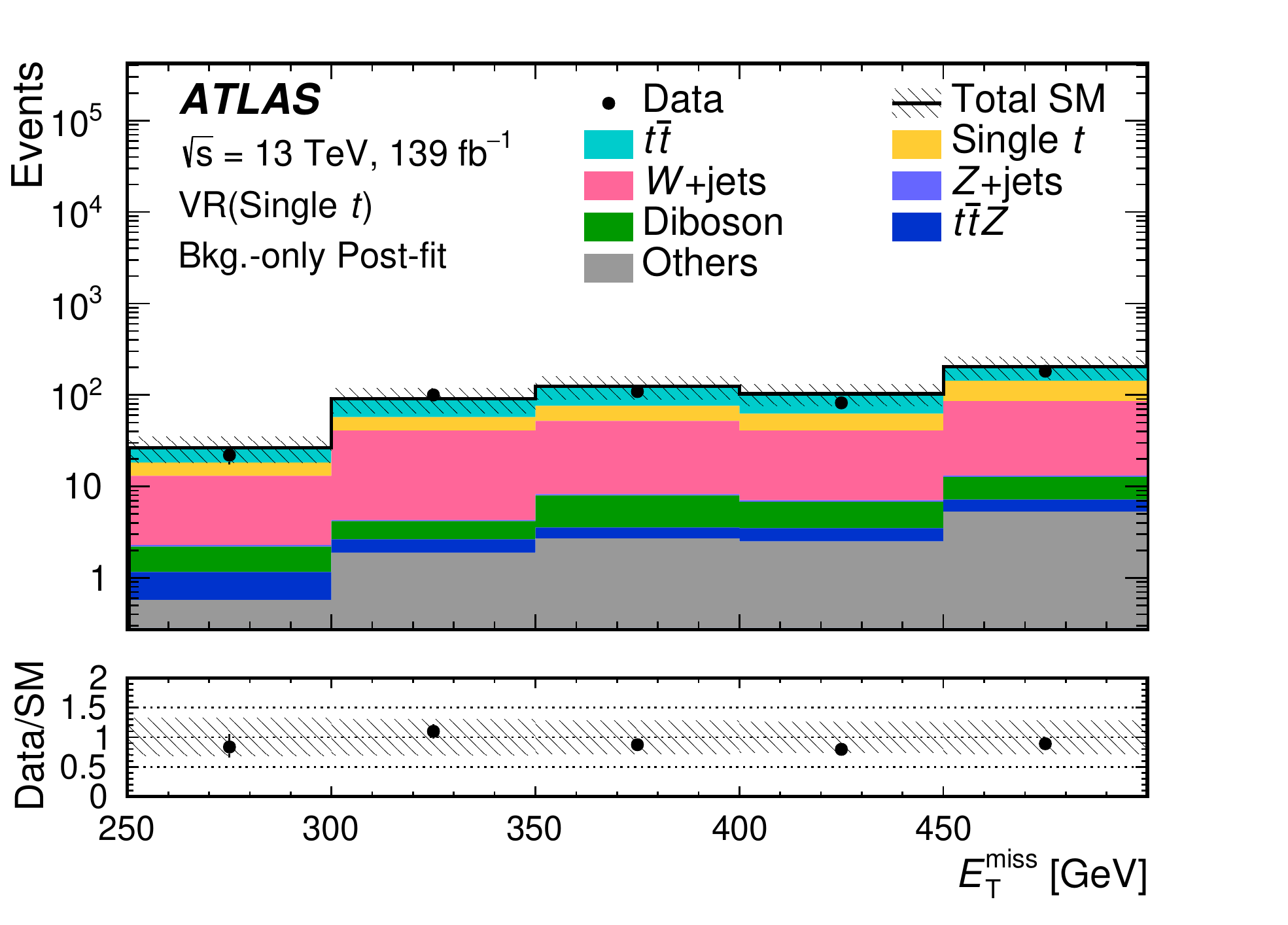}}
\caption{
\label{fig:1LVR}
The post-fit \met distributions in the (a) \tWZeroL\ \ttbar, (b) \tWZeroL\ \Vjets (c) \lepTop\ \ttbar, (d) \hadTop \ttbar, (e) \Wjets and (f) single-top-quark validation regions. The last bin in the histogram includes the overflow events. The bottom panel shows the ratio of data to the total SM background.
The uncertainties shown are the sum of the statistical and post-fit systematic uncertainties  as detailed in Section~\ref{sec:unc}. The fit set-up corresponds to the background-only fit configuration.
The `Others' category includes contributions from rare processes such as $tWZ$, $tZ$, triboson, $ttt$, $t\bar{t}t\bar{t}$, $ttW$ and $ttH$.
}
\end{figure}
 
Finally, the single-top-quark prediction is validated in
a one-lepton region. The single-top-quark acceptance is enhanced by
applying a low \mt requirement. The \ttbar events are suppressed by
demanding high \amt and \metSig. The \Wjets contribution is reduced by
selecting events with high sub-leading $b$-jet transverse momentum.
\begin{figure}[p]
\center
\includegraphics[width=0.7\textwidth]{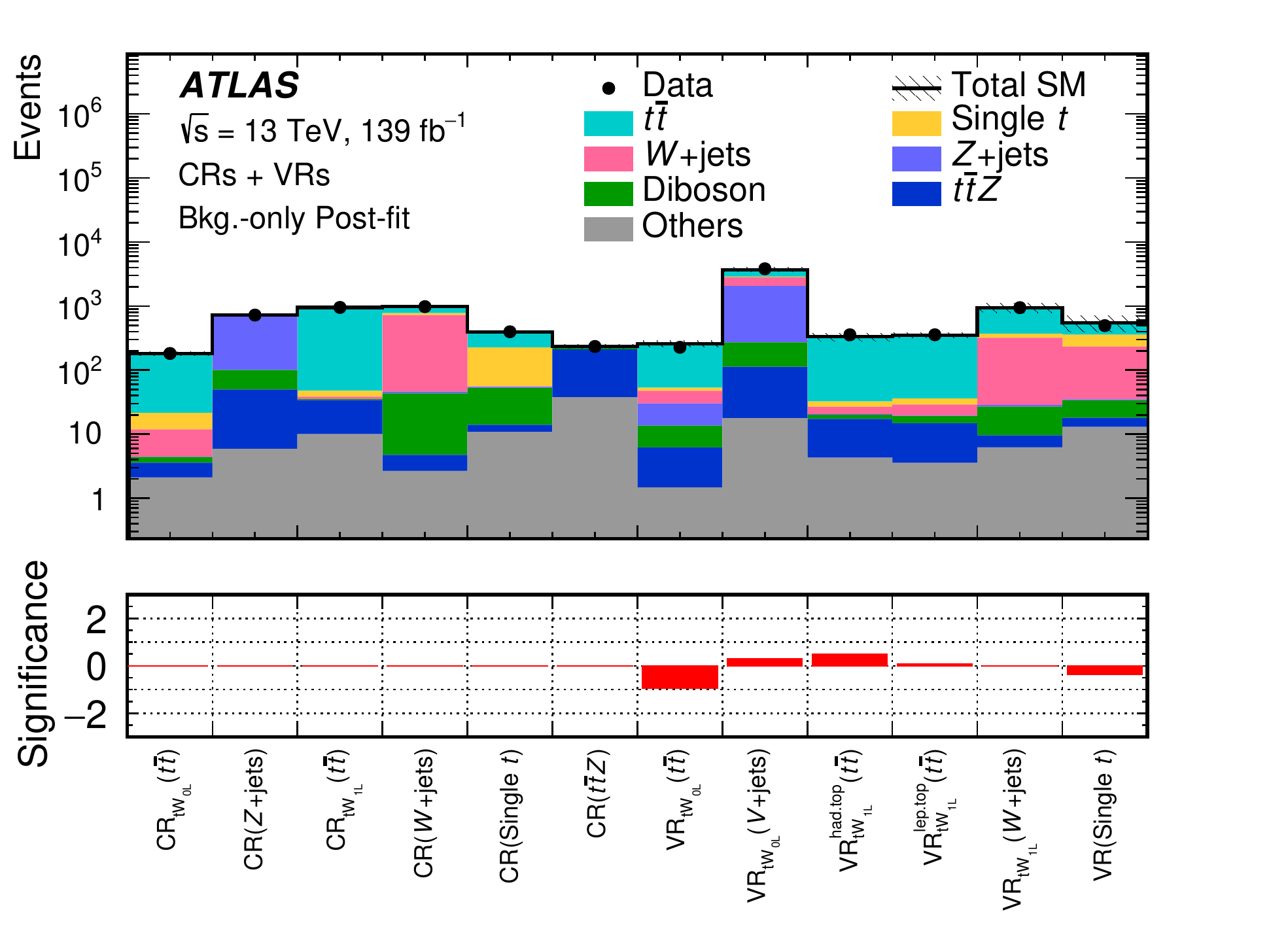}
\caption{
\label{fig:CRVR}
Summary of all control and validation regions comparing the post-fit predicted SM background with the observed number of events.
The normalization parameters were extracted from the control regions. The fit set-up corresponds to the background-only
fit configuration. Statistical and systematic uncertainties are included in the shaded region of the top panel  as detailed in Section~\ref{sec:unc}. The bottom panel shows the statistical significance~\cite{ATL-PHYS-PUB-2020-025} of the excesses and deficits of data relative to the predicted SM background.
}
\end{figure}
 
A summary of all validation regions can be found in Tables~\ref{tab:0LVR}~and~\ref{tab:1LVRs}.
Figure~\ref{fig:1LVR} shows the post-fit \met distribution in each VR. In addition, observed data and
predicted background yields in all control and validation regions are presented in Figure~\ref{fig:CRVR}, together with the
ratio of their difference to the estimated background uncertainty. Good agreement between data and the
expected background predictions can be observed in both figures, thus validating the background estimation strategy of the analysis.


\FloatBarrier
\subsection{Statistical combination}
The SRs of the \tWZeroL\ and \tWOneL\ channels are
constructed to be statistically independent and they are combined to
derive the final results in Section~\ref{sec:results}. The CRs are
constructed to be in common for the two channels, with the exception of
the \ttbar\ CRs which are disjoint and the \ttbar\ background is
estimated in each channel with a separate normalization parameter.
These two channels are also statistically independent of the
\tWTwoL\ channel in Ref.~\cite{SUSY-2018-08}. For this reason the
results are also derived using the statistical combination of the
\tWZeroL\ and \tWOneL\ channels with the \tWTwoL\ channel, in order to
provide the most stringent constraints on the model considered in this
paper. The dominant SM background in the \tWTwoL\ channel is
\ttZ\ production and it is estimated in Ref.~\cite{SUSY-2018-08}
using a CR which is a subset of the \ttZ\ CR in this paper. In the
combination of the three channels, the \ttZ background is estimated
using a common normalization parameter fitted in the common
\tWZeroL\ and \tWOneL\ CR (Section~\ref{ssec:crs}). As the \ttZ\ CR
in this paper has less contamination from diboson processes than
the \tWTwoL\ \ttZ CR, the diboson CR of Ref.~\cite{SUSY-2018-08}, which
provides an estimate of the diboson processes in the \ttZ CR, is
not used. All other SM backgrounds in the \tWTwoL\ channel are
estimated directly from the MC simulation, as in
Ref.~\cite{SUSY-2018-08}.
These CR orthogonalization choices impact the final \tWTwoL\ background estimate by up to
10\%--15\% because the normalization factor for the \ttZ\ background in this channel
changes from $0.8\pm0.2$ to $1.2\pm0.2$.


\section{Systematic uncertainties}
\label{sec:unc}
 
\begin{figure}[!b]
\begin{center}
\includegraphics[width=0.85\textwidth]{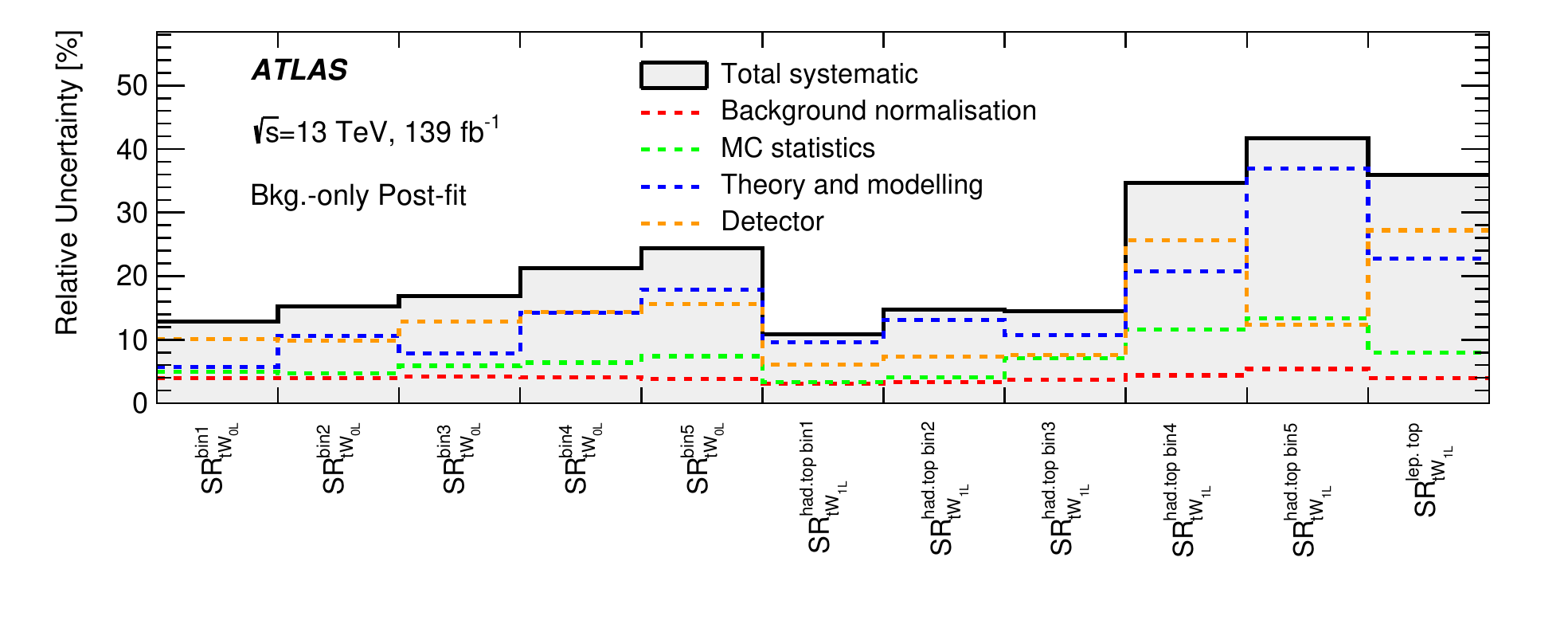}
\end{center}
\caption{Relative uncertainties (in percent) in the total background yield in each signal region of the two analysis channels, including the contributions
from the different sources of uncertainty. The `Detector' category contains all detector-related systematic uncertainties.
The `Background normalization' represents the uncertainty in the
fitted normalization factors, including the available data event counts in the CRs.
Individual uncertainties can be correlated, and do not necessarily add up in quadrature to the total background uncertainty. The fit configuration
used to estimate these uncertainties corresponds to the background-only fit explained in Section~\ref{sec:results}.}
\label{fig:SRsyst}
\end{figure}

This analysis considers several sources of uncertainty, of both
experimental and theoretical nature, that affect the prediction of the
SM background and the DM signal in all channels. Figure~\ref{fig:SRsyst}
provides an overview of the size of the \tWZeroL\ and \tWOneL\ systematic uncertainties,
estimated in a combined fit of the two channels.
 
The uncertainties related to the limited measurement precision of
reconstructed objects, the estimate of the dataset luminosity
and the modelling of the pile-up are broadly referred to as `detector
systematic uncertainties'. The dominant contributions to these uncertainties
arise from the small-$R$ jet energy scale and resolution and the
large-$R$ jet \Wtagging. The small-$R$ jet energy scale and resolution uncertainties have
a large impact on the high \met\ bins of the \hadTop\ and
\lepTop\ region, respectively. In addition, small-$R$ jet energy resolution uncertainties
are the source of the second-largest experimental uncertainty in the \tWZeroL\ SRs.
The \Wtagging\ uncertainties dominate across the \tWZeroL\ SRs, being the dominant
experimental uncertainty in this channel. The uncertainties associated with trigger requirements, pile-up
modelling, lepton reconstruction and energy measurements have a small
or negligible impact on the final results. The lepton, photon and jet-related uncertainties are propagated to
the calculation of the \etmiss, along with additional uncertainties due to the energy scale and resolution
of the soft term. These \met\ soft-term uncertainties are found to be small or negligible.
Finally, as mentioned in Section~\ref{sec:datamc},
a 1.7\% uncertainty in the combined 2015--2018 integrated luminosity
is included.

Theoretical uncertainties are estimated for the modelling of SM background processes in the MC simulation. Their
theoretical cross-section uncertainties are also taken into account. Modelling
uncertainties are important for both channels. In the \tWZeroL\ channel, single-top-quark
uncertainties are dominant in the bins of the SRs with highest \met requirements, while \zjets\ theory uncertainties
contribute significantly in the lowest \met\ bins. The \ttbar\ and \wjets\ uncertainties are the dominant ones in the \tWOneL\ channel.
The \zjets\ and \wjets\ modelling uncertainties are evaluated by
varying the CKKW-L scale for matching of the matrix element and
parton shower, and the resummation, renormalization and factorization scales independently by factors
of 0.5 and 2. The \ttbar\ and single-top-quark uncertainties from the renormalization and factorization scales and
initial- and final-state radiation parameters are evaluated similarly.
In addition, uncertainties due to our choices of
hard-scattering generator and parton-shower and hadronization
models are estimated for these two processes.  The impact of the latter is
evaluated by comparing the nominal simulated sample with a sample
generated using the same matrix element generator, \POWHEGBOX, interfaced
to an alternative shower generator,
\HERWIG[7]~\cite{Bahr:2008pv,Bellm:2015jjp}. This sample uses the H7UE
set of tuned
parameters~\cite{Bellm:2015jjp}. To assess the uncertainty due to the
choice of hard-scattering generator
and matching scheme, an alternative generator set-up using
\MGNLO~\cite{Alwall:2014hca} interfaced to
\PYTHIA[8]~\cite{Sjostrand:2014zea} is employed.
An additional uncertainty is considered for the single-top-quark $tW$
channel: the impact of interference between single-resonant and
double-resonant top-quark production on the implementation of the $W$-boson lineshape in the
generator is estimated by comparing the nominal sample generated using
the diagram removal method with samples using the alternative diagram subtraction method~\cite{Frixione:2008yi}.
For the \ttZ background, uncertainties related to the choice of renormalization and factorization scales are
assessed by varying the corresponding event generator parameters by factors of 0.5 and 2 from their nominal values.
Overall, the total SM uncertainties vary from 11\% to 42\% across the \tWZeroL\ and \tWOneL\ signal regions.
 
Detector and modelling uncertainties are also evaluated for the DM signal processes.
Detector uncertainties are found to have an impact of 9\%--43\% on the expected signal yields across the $m_a$--$m_{H^{\pm}}$ and $m_a$--$\tan\beta$
planes for the signal regions of the \tWZeroL and \tWOneL analysis channels. The largest uncertainties are found to be concentrated in the highest \met\ bins
of the SRs for both channels. In all SRs, the dominant experimental uncertainties affecting signal
yields are found to be the uncertainties associated with the jet energy scale and resolution and with \Wtagging, as observed for the SM processes.
These uncertainties are assumed to be fully correlated with those affecting the SM background.
Modelling uncertainties include renormalization and factorization scale
uncertainties and uncertainties related to the modelling of the parton
shower.  For the signal regions of the \tWZeroL (\tWOneL) analysis channel, the average value of these modelling uncertainties lies
between 3\% and 30\% (3\% and 24\%) across the $m_a$--$m_{H^{\pm}}$ and
$m_a$--$\tan\beta$  planes, but can reach 50\% for certain benchmark signals
in the highest \met\ regions of the channel.
 
The effects of the various sources of systematic uncertainty on the signal and background estimates are introduced in the
likelihood fit (see Section~\ref{sec:analysis}) through nuisance
parameters (NPs) that affect the expectation values of the Poisson terms for
each CR and SR bin. The probability density function of each nuisance parameter is described by a Gaussian distribution whose standard
deviation corresponds to a specific experimental or theoretical
modelling uncertainty. The preferred value of each nuisance parameter
is determined as part of the likelihood fit and
none of them is significantly altered or constrained by the fit.
The uncertainties arising from the total number of data events in the CRs are also included in the fit for each region.
Since the number of CRs matches the number of fitted background normalization parameters, the systematic uncertainties are not
constrained in the background-only fit of this analysis.
 
All uncertainties arising from the same source, including background and signal modelling uncertainties, are treated as correlated across the \tWZeroL\ and \tWOneL\ channels.
For the statistical combination of the \tWZeroL\ and \tWOneL\ channels with the \tWTwoL\ channel, a simplified approach which considers uncorrelated experimental and theoretical systematic uncertainties is adopted.
This is supported by the large differences between the definitions of the physics objects, the selection and quality criteria, and uncertainty schemes which were used in the \tWTwoL\ channel and the analyses described in this paper.
Only the modelling uncertainties for the DM signal are treated as correlated across all channels.


\section{Results}
\label{sec:results}
\FloatBarrier
 
\begin{figure}[htb]
\centering
\includegraphics[width=0.6\textwidth]{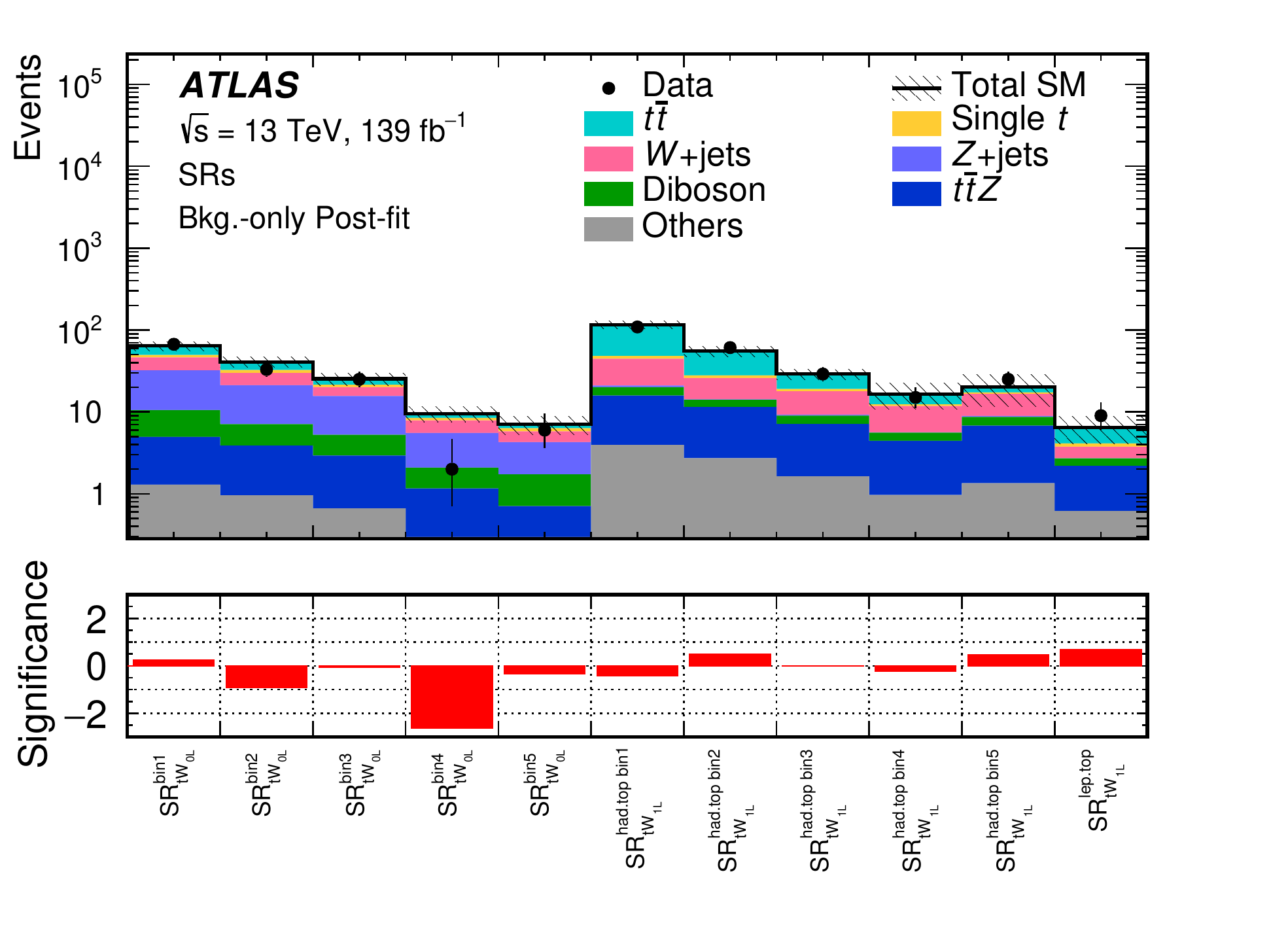}
\caption{Comparison of the background-only fit SM predictions extrapolated to all SRs with the observed data.
The normalization of the backgrounds is obtained from the fit to the CRs.
The upper panel shows the observed number of events and the predicted background yields.
The `Others' category includes contributions from rare processes such as $tWZ$, $tZ$, triboson, $ttt$, $t\bar{t}t\bar{t}$, $ttW$ and $ttH$.
All uncertainties defined in Section~\ref{sec:unc} are included in the uncertainty band.
The bottom panel shows the statistical significance \cite{ATL-PHYS-PUB-2020-025} of the excesses and deficits of data events relative to the predicted SM background.
}
\label{fig:SRpulls}
\end{figure}

The expected and observed numbers of events in the \tWZeroL\ and \tWOneL\ SRs are
shown in  Tables~\ref{tab:fit_results_zerolep} and~\ref{tab:fit_results_onelep}, respectively,
together with the SM prediction breakdown for
the background processes.
The expected yields are derived using the background-only fit configuration.
All systematic and statistical uncertainties described in
Section~\ref{sec:unc} are included in the predictions.
A graphical representation of the tables is given in Figure~\ref{fig:SRpulls},
where the bottom panel shows
the statistical significance~\cite{ATL-PHYS-PUB-2020-025} of the difference between the observation and prediction.
No significant deviation of the observed data from the SM prediction is found. The largest difference
appears in \srzerobinFour, corresponding to $500 < \met < 600$~\GeV, and amounts to
a data event deficit of around $2.5\sigma$ considering statistical and systematic uncertainties of the SM prediction.
Since the data and predictions agree well in the bins below and above, this deficit is considered to be a statistical fluctuation.
 
\begin{figure}[htb]
\centering
\subfloat[][]{
\includegraphics[width=0.4\textwidth]{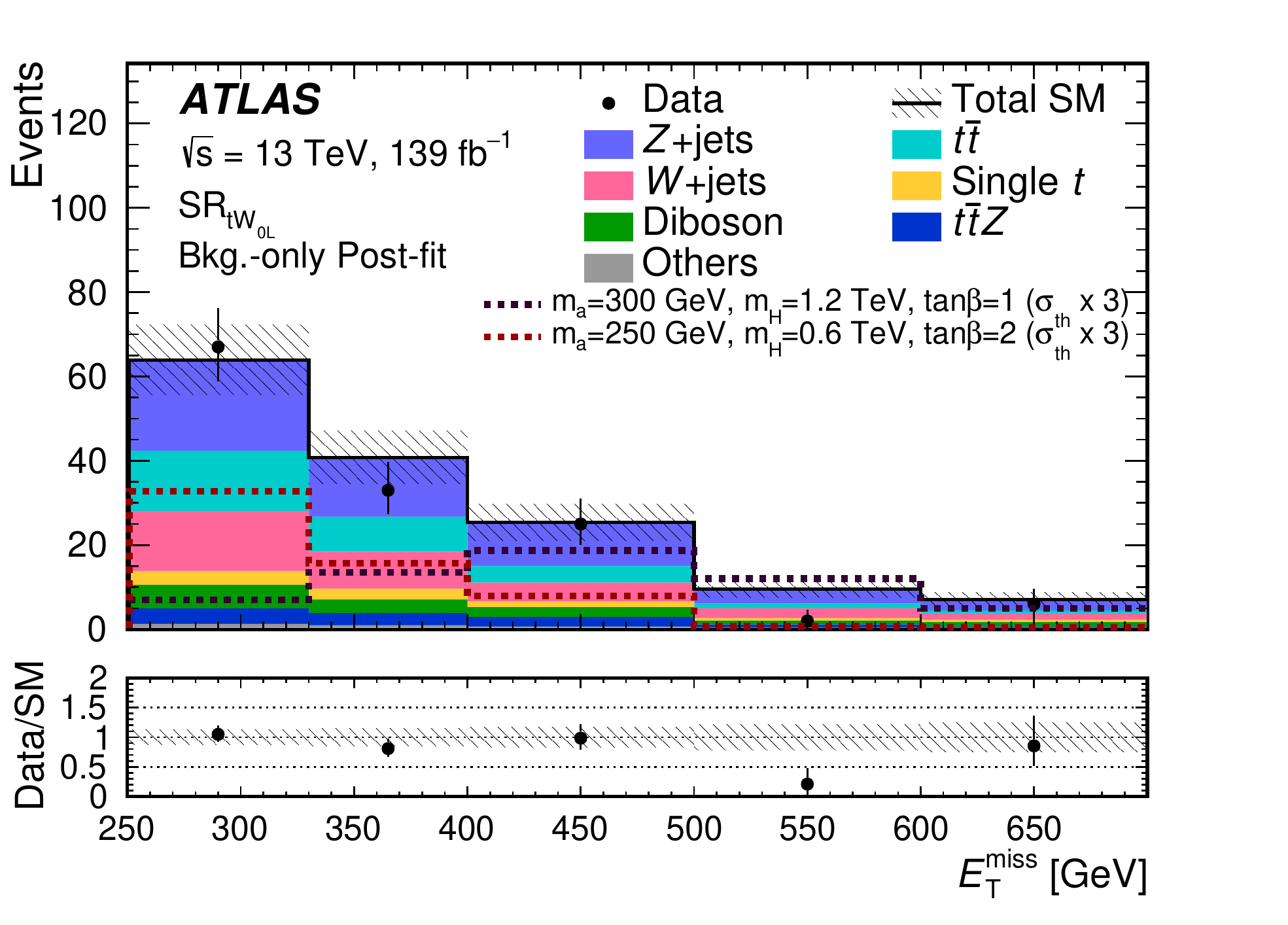}
\label{fig:tW0L_Nm1_met}
}
\subfloat[][]{
\includegraphics[width=0.4\textwidth]{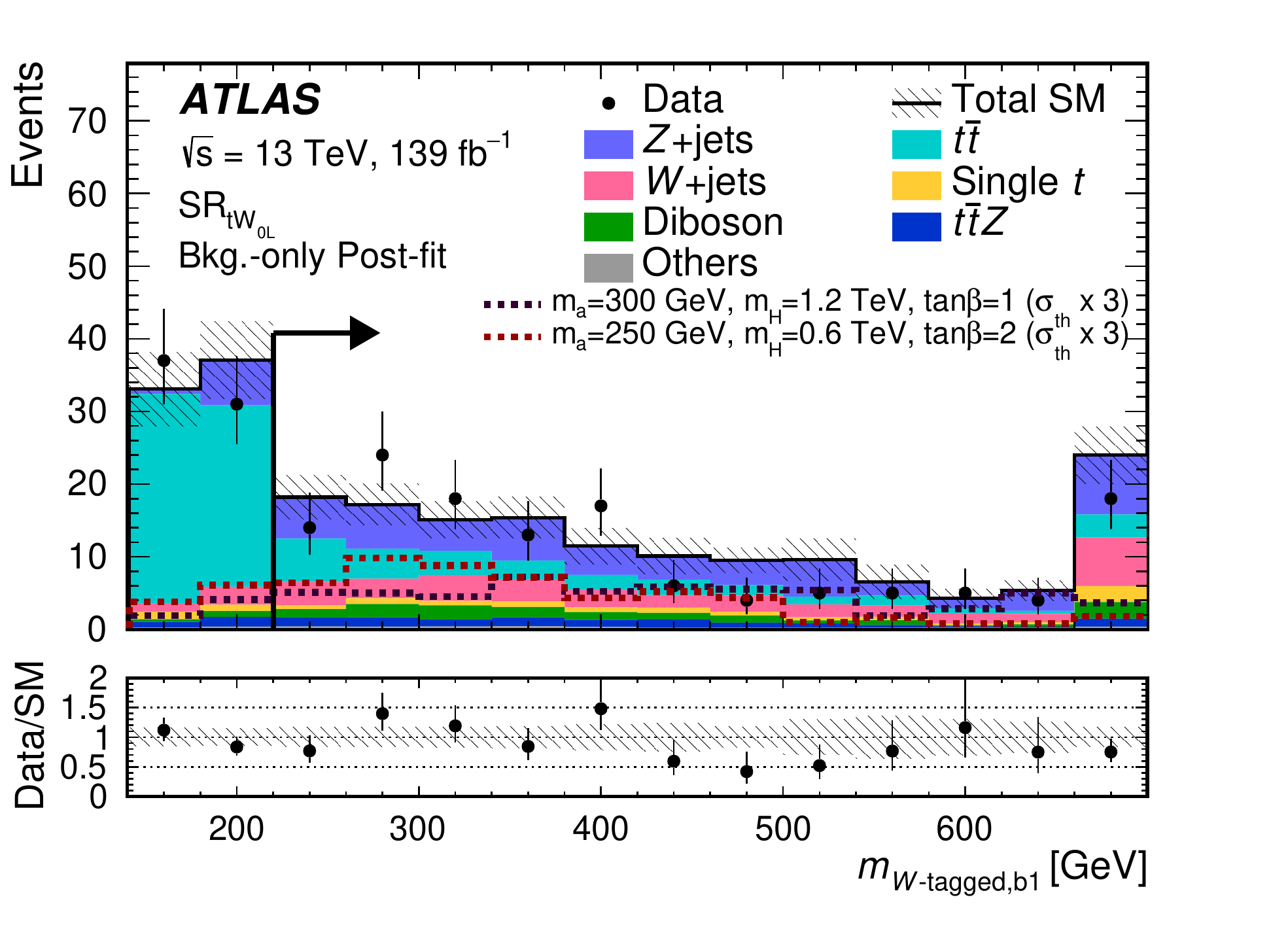}
\label{fig:tW0L_Nm1_wbmass}
}
 
\subfloat[][]{
\includegraphics[width=0.4\textwidth]{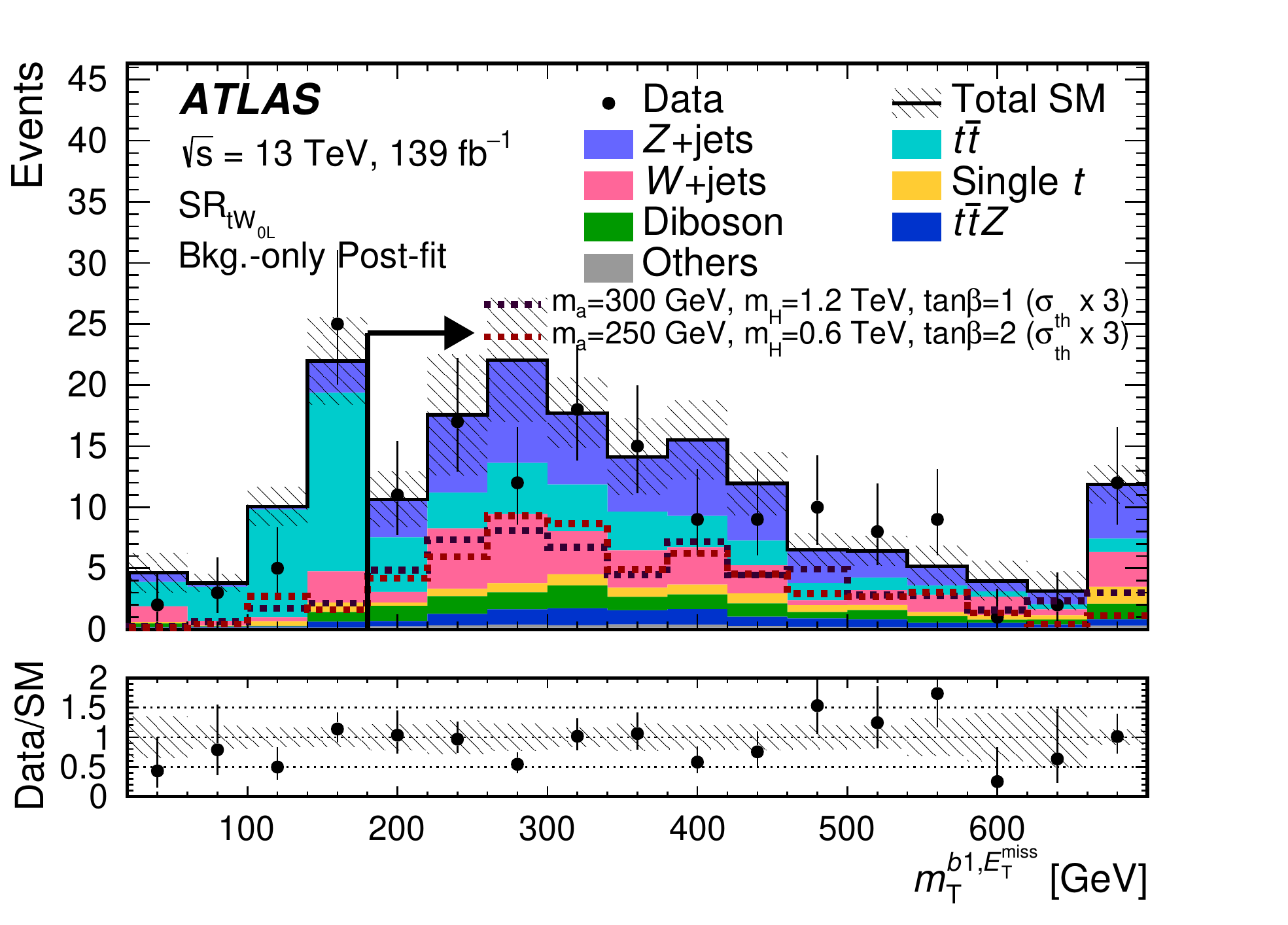}
\label{fig:tW0L_Nm1_mtb}
}
\subfloat[][]{
\includegraphics[width=0.4\textwidth]{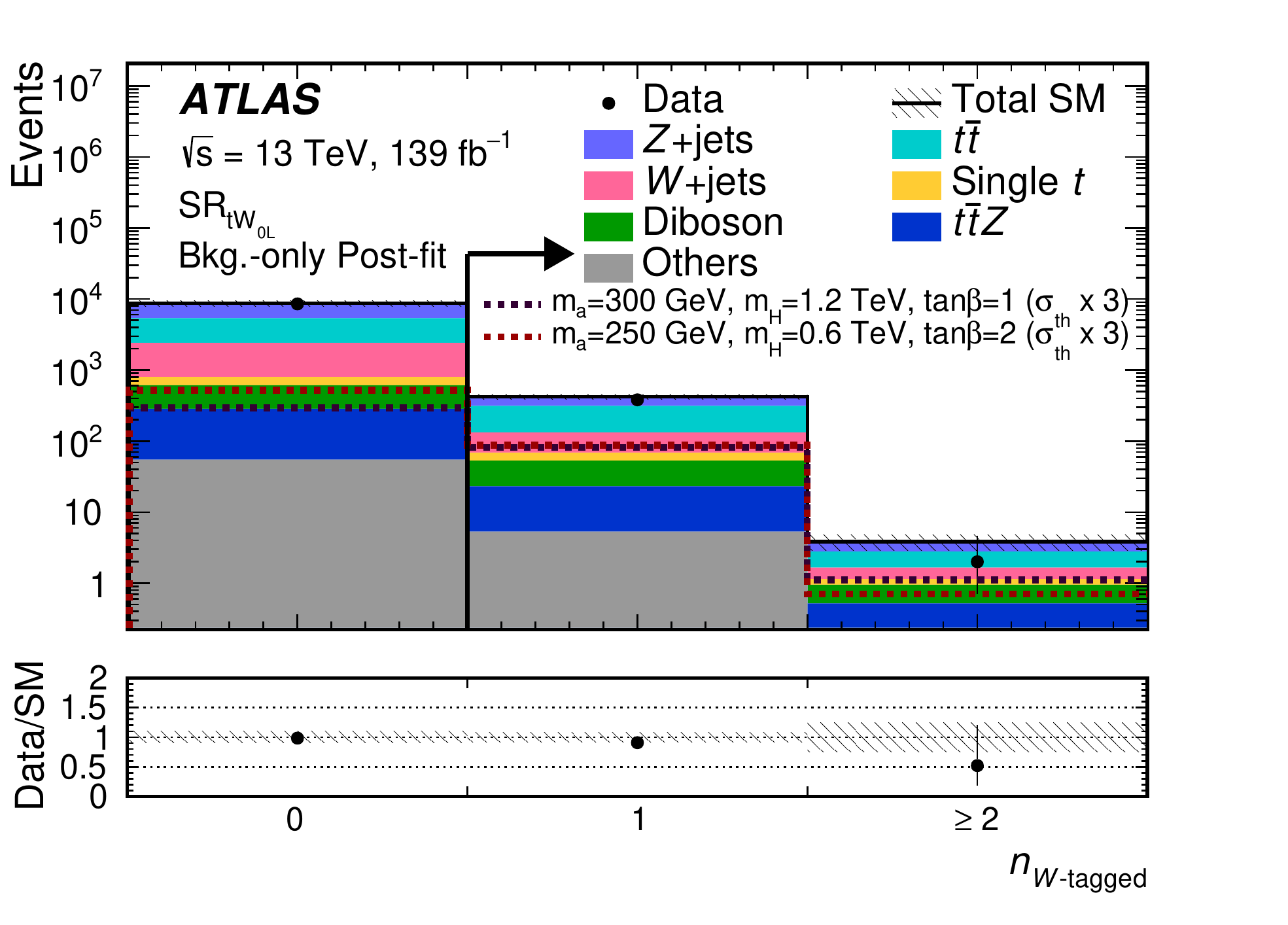}
\label{fig:tW0L_Nm1_nWtag}
}
\caption{
Representative distributions of \protect\subref{fig:tW0L_Nm1_met} \met,  \protect\subref{fig:tW0L_Nm1_wbmass} \mWb,  \protect\subref{fig:tW0L_Nm1_mtb} \mtbmet, and \protect\subref{fig:tW0L_Nm1_nWtag} the number of $W$-tagged large-$R$ jets in the \tWZeroL\ signal region. Observed data are compared with the SM background predictions extrapolated from the background-only fit.
The `Others' category includes contributions from rare processes such as $tWZ$, $tZ$, triboson, $ttt$, $t\bar{t}t\bar{t}$, $ttW$ and $ttH$.
The expected distributions for representative scenarios with different $m_a$,  $ m_{H^{\pm}}$, and $\tan\beta$ are
shown for illustrative purposes. All signal theory cross-sections ($\sigma_{\mathrm{th}}$) have been multiplied by three for better visibility.  The overflow events, where present, are included in the last bin.
The lower panels show the ratio of data to the background prediction.
The hatched error bands indicate the combined experimental, theoretical and MC statistical uncertainties of these background predictions. The arrows,
when present, indicate the value of the SR requirement on the quantity presented on the $x$-axis.
}
 
\label{fig:tW0L_SROpt_Nminus1}
\end{figure}

Figure~\ref{fig:tW0L_SROpt_Nminus1} shows the observed data and the SM prediction in the
\tWZeroL\ channel for the \met\ distribution in the SR using the binning of the final fit.
In the same figure, the $W$-tagged jet multiplicity, the \mWb\ observable and
the \mtbmet\ observable are shown in a region that contains all SR requirements with the
exception of the one on the variable shown in the plot. Small local deficits are seen
in the \mWb\ variable around 480~\GeV, although no significant trend is observed in any of the distributions
and overall, given the uncertainties, there is good agreement between data and predictions.
 
\begin{figure}[htb]
\centering
\subfloat[][]{
\includegraphics[width=0.4\textwidth]{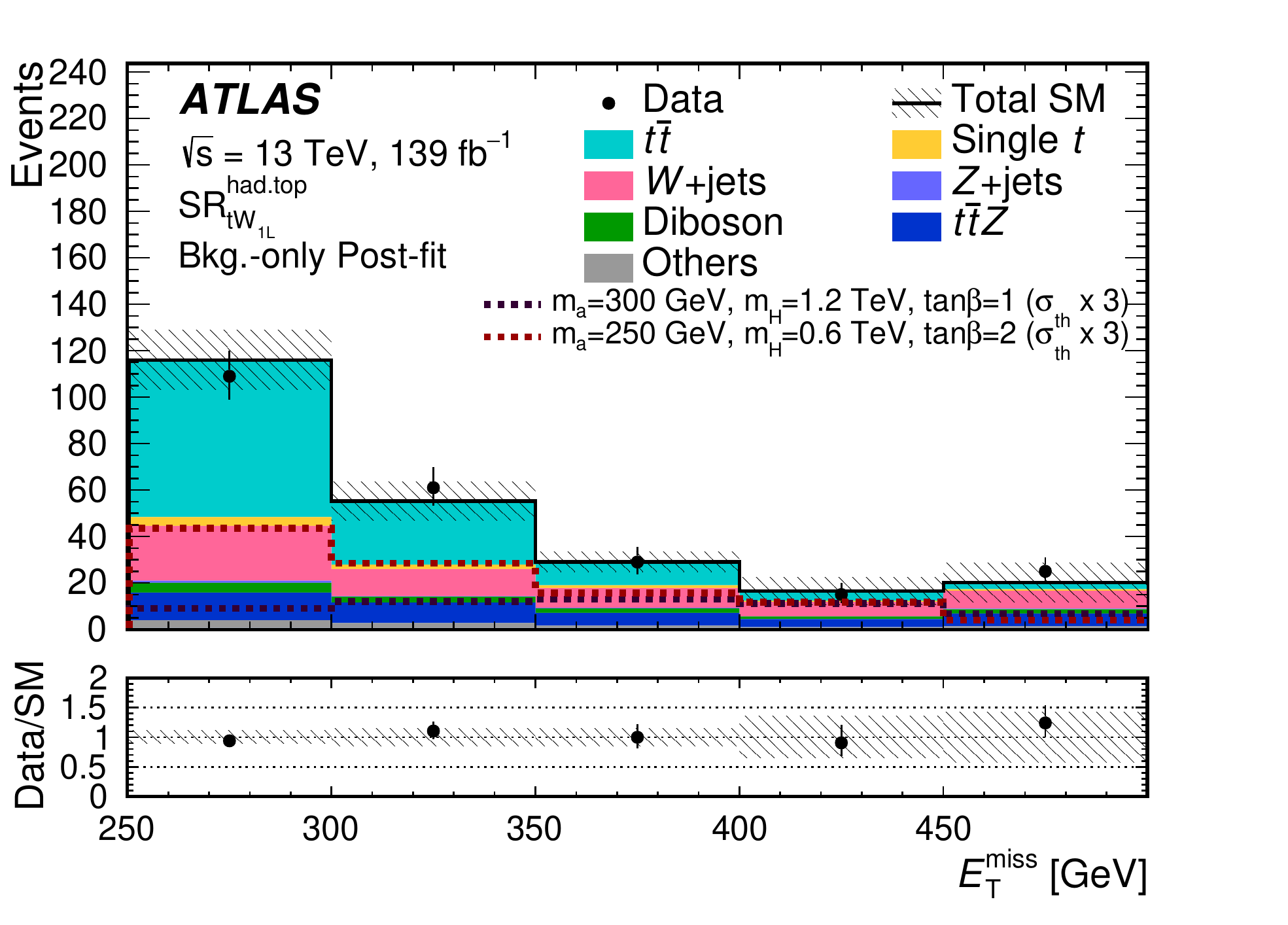}
\label{fig:tW1L_Had_Nm1_met}
}
\subfloat[][]{
\includegraphics[width=0.4\textwidth]{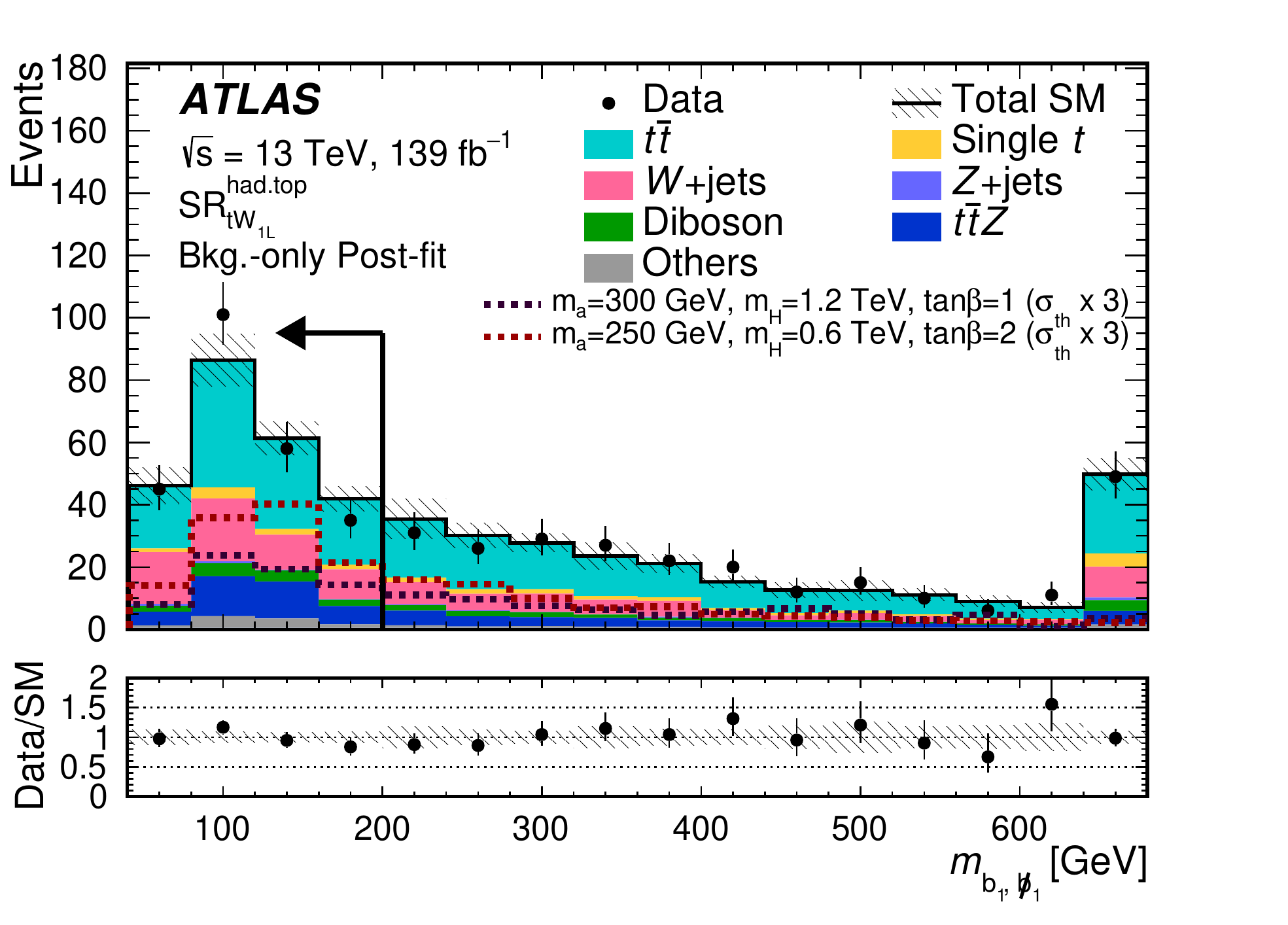}
\label{fig:tW1L_Had_Nm1_mbj1}
}
 
\subfloat[][]{
\includegraphics[width=0.4\textwidth]{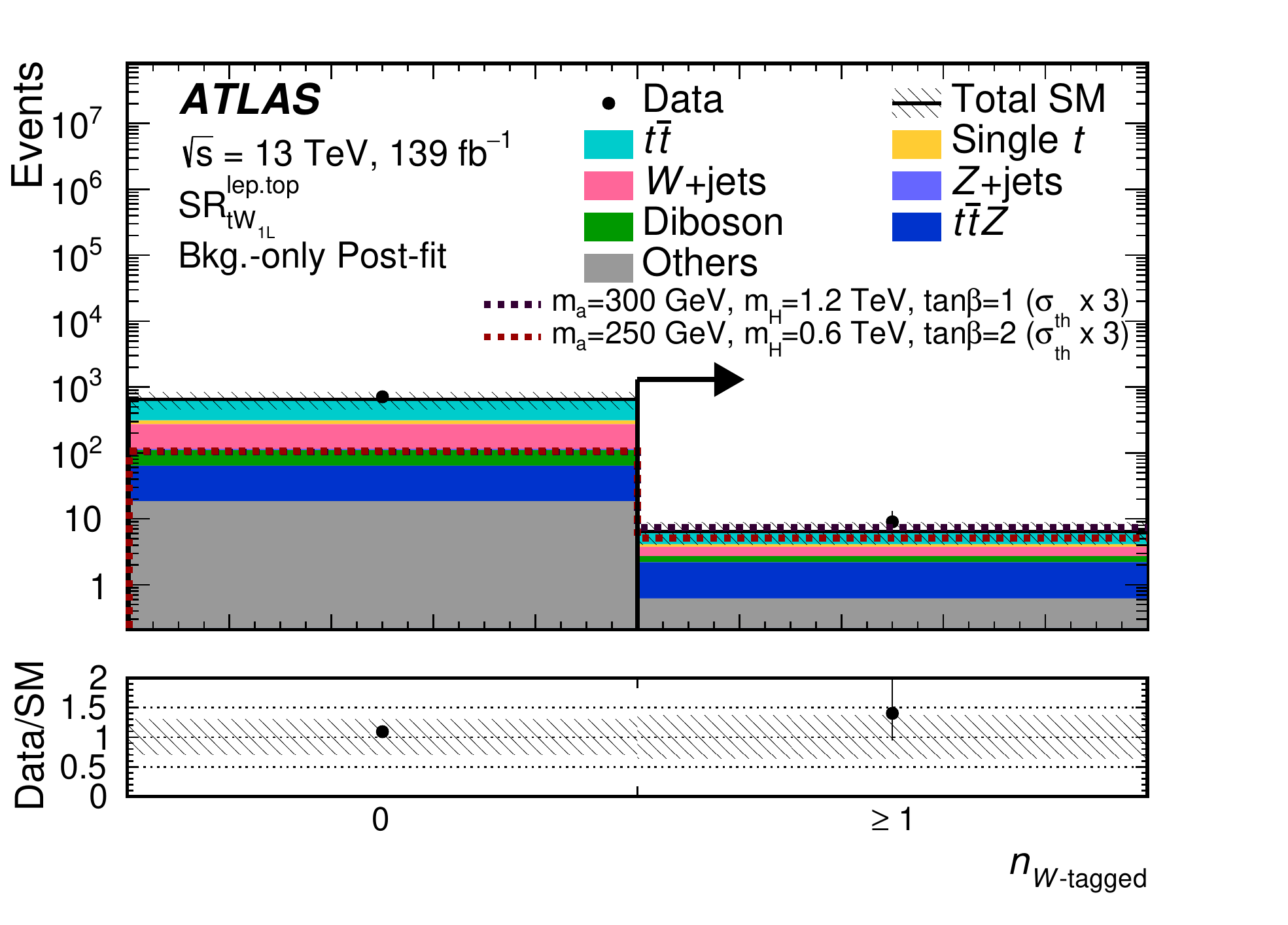}
\label{fig:tW1L_Lep_Nm1_nWtag}
}
\subfloat[][]{
\includegraphics[width=0.4\textwidth]{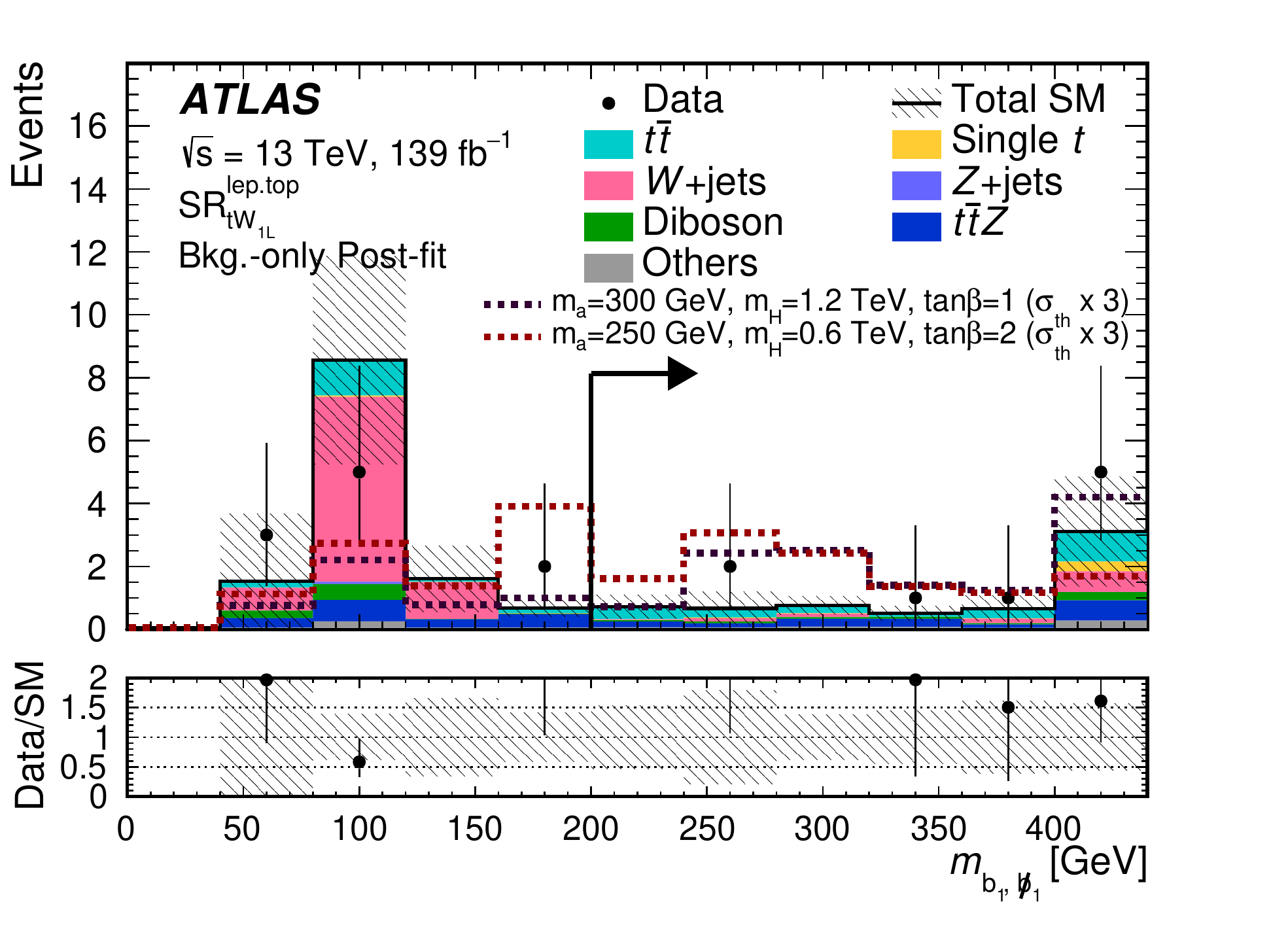}
\label{fig:tW1L_Lep_Nm1_mbj1}
}
\caption{
Representative distributions of \protect\subref{fig:tW1L_Had_Nm1_met} \met and  \protect\subref{fig:tW1L_Had_Nm1_mbj1} \mbj in \hadTop, as well as \protect\subref{fig:tW1L_Lep_Nm1_nWtag} the number of $W$-tagged large-$R$ jets and  \protect\subref{fig:tW1L_Lep_Nm1_mbj1} \mbj in \lepTop. Observed data are compared with the SM background predictions extrapolated from the background-only fit.
The `Others' category includes contributions from rare processes such as $tWZ$, $tZ$, triboson, $ttt$, $t\bar{t}t\bar{t}$, $ttW$ and $ttH$.
The expected distributions for representative scenarios with different $m_a$,  $ m_{H^{\pm}}$, and $\tan\beta$ values are
shown for illustrative purposes. All signal theory cross-sections ($\sigma_{\mathrm{th}}$) have been multiplied by three for better visibility. The overflow events, where present, are included in the last bin.
The lower panels  show the ratio of data to the background prediction. The hatched error bands indicate
the combined experimental, theoretical and MC statistical uncertainties of these background predictions. The arrows, when present, indicate the
value of the SR requirement on the quantity presented on the $x$-axis.
}
 
\label{fig:tW1L_SROpt_Nminus1}
\end{figure}

Figure~\ref{fig:tW1L_SROpt_Nminus1} shows the observed data and the SM prediction in \lepTop and \hadTop of the \tWOneL\ channel.
Panel (a) shows the \met distribution in \hadTop, using the
same bins as in the final fit. In the other panels, the $W$-tagged jet multiplicity in \lepTop\
and the \mbj\ observable in both \lepTop and \hadTop are shown. In all cases, all SR requirements
except the one on the shown quantity are applied. Similarly to the \tWZeroL\ channel, no significant trend is observed in the distributions
and overall there is good agreement between data and prediction. In addition, Figures~\ref{fig:tW1L_Lep_Nm1_nWtag}
and~\ref{fig:tW1L_Lep_Nm1_mbj1} show that \nwtagged\ and \mbj\ are powerful discriminating variables for rejecting
\ttbar\ and \Wjets\ backgrounds for the considered signal benchmark models.
 
\def\arraystretch{1.2}

\begin{table}[htb]
\begin{center}
\footnotesize
\caption{Event yields showing the observed data and the background-only fit SM predictions in the \tWZeroL\ signal regions. Signal regions are defined according to the
five \met\ bins presented in Section~\ref{ssec:srs} for the \tWZeroL\ channel, corresponding to increasing \met\ values in bins 1--5. SM predictions
are decomposed into the main backgrounds of the analysis. Smaller backgrounds ($tWZ$, $tZ$, triboson, $ttt$, $t\bar{t}t\bar{t}$, $ttW$ and $ttH$ events)
are grouped together and labelled as `Others'.
The quoted uncertainties in the fitted SM background include both the statistical and systematic uncertainties. }
\label{tab:fit_results_zerolep}
\resizebox{\textwidth}{!}{
\begin{tabular*}{\textwidth}{@{\extracolsep{\fill}}lccccc}
\toprule
& \srzerobinOne& \srzerobinTwo& \srzerobinThree& \srzerobinFour& \srzerobinFive\\
\midrule
Observed events          & $67$              & $33$              & $25$              & $2$              & $6$                    \\
\midrule
Fitted SM bkg. events         &$64 \pm 8$~~          & $41 \pm 6$~~          & $25 \pm 4$~~          & $9.6 \pm 2.1$          & $7.0 \pm 1.7$              \\
\midrule
\Zjets\         & $22 \pm 5$~~          & $14.1 \pm 2.9$~~          & $10.4 \pm 2.7$~~          & $3.4 \pm 0.9$          & $2.5 \pm 0.8$              \\
\Wjets\         &  $14.2 \pm 3.2$~~          & $8.8 \pm 1.8$          & $4.4 \pm 1.3$          & $2.4 \pm 0.8$          & $1.51 \pm 0.28$              \\
\ttbar\         & $14.2 \pm 2.6$~~          & $8.3 \pm 3.3$          & $3.9 \pm 1.2$          & $1.1 \pm 0.5$          & $0.7 \pm 0.6$              \\
\ttZ\         & $3.6 \pm 1.6$          & $2.9 \pm 1.1$          & $2.3 \pm 0.9$          & $1.0_{-1.0}^{+1.0}$\,          & $0.5 \pm 0.4$              \\
Single-top         & $3.3 \pm 2.5$          & $2.5_{-2.5}^{+3.1}$\,          & $1.4 \pm 1.1$          & $0.6_{-0.6}^{+0.8}$\,          & $0.5_{-0.5}^{+1.0}$\, \\
Diboson         & $5.6 \pm 1.2$          & $3.2 \pm 0.8$          & $2.3 \pm 0.8$          & $0.92 \pm 0.26$          & $1.02 \pm 0.31$              \\
Others         & $1.29 \pm 0.23$          & $0.96 \pm 0.17$          & $0.66 \pm 0.11$          & $0.20 \pm 0.04$          & $0.20 \pm 0.04$              \\
\bottomrule
\end{tabular*}
}
\end{center}
\end{table}


\begin{table}[htb]
\begin{center}
\footnotesize
\caption{Event yields showing the observed data and the background-only fit SM predictions in the \tWOneL\ signal regions. Bin numbers for SR$^{\mathrm{had.top}}_{\tWOneL}$ refer to  the
five \met\ bins presented in Section~\ref{ssec:srs} for the \tWOneL\ channel, corresponding to increasing \met\ values in bins 1--5. SM predictions
are decomposed into the main backgrounds of the analysis. Smaller backgrounds ($tWZ$, $tZ$, triboson, $ttt$, $t\bar{t}t\bar{t}$, $ttW$ and $ttH$ events)
are grouped together and labelled as `Others'.
The quoted uncertainties in the fitted SM background include both the statistical and systematic uncertainties.}
\label{tab:fit_results_onelep}
\resizebox{\textwidth}{!}{
\begin{tabular*}{\textwidth}{@{\extracolsep{\fill}}lcccccc}
\toprule
&\hadTopBinOne & \hadTopBinTwo& \hadTopBinThree& \hadTopBinFour& \hadTopBinFive &  \lepTop \\
\midrule
Observed events          & $109$              & $61$              & $29$              & $15$              & $25$              & $9$                    \\
\midrule
Fitted SM bkg. events        & $116 \pm 13$~~          & $55 \pm 8$~~          & $29 \pm 4$~~          & $17 \pm 6$~~          & $20 \pm 8$~~          & $6.4 \pm 2.3$ \\
\midrule
\Zjets\         & $0.9_{-0.9}^{+1.1}$\,          & $0.22 \pm 0.13$          & $0.25 \pm 0.25$          & $0.04_{-0.04}^{+0.09}$~          & $0.25 \pm 0.08$          & $0.01 \pm 0.00$  \\
\Wjets\         & $24 \pm 8$~~          & $12 \pm 5$~~          & $8.6 \pm 2.8$          & $6 \pm 5$          & $8 \pm 8$          & $1.1_{-1.1}^{+1.6}$\,              \\
\ttbar\         & $68 \pm 8$~~          & $27 \pm 6$~~          & $10.0 \pm 2.1$~~          & $4.2 \pm 1.2$          & $2.9 \pm 1.2$          & $2.3 \pm 1.4$ \\
\ttZ\         & $11.9 \pm 3.4$~~          & $8.8 \pm 1.9$          & $5.5 \pm 2.4$          & $3.5 \pm 1.5$          & $5.5 \pm 1.7$          & $1.6 \pm 0.7$   \\
Single-top         & $4_{-4}^{+4}$\,  & $1.9_{-1.9}^{+2.5}$\,  & $1.2_{-1.2}^{+1.5}$\,   & $0.6_{-0.6}^{+1.3}$\,   & $0.7 \pm 0.7$  & $0.4_{-0.4}^{+0.7}$\, \\
Diboson         & $4.2 \pm 0.4$          & $2.58 \pm 0.32$          & $1.9 \pm 0.4$          & $1.16 \pm 0.22$          & $1.88 \pm 0.31$          & $0.51 \pm 0.09$   \\
Others         & $3.95 \pm 0.28$          & $2.72 \pm 0.28$          & $1.63 \pm 0.17$          & $0.96 \pm 0.14$          & $1.35 \pm 0.32$          & $0.62 \pm 0.12$   \\
\bottomrule
\end{tabular*}
}
\end{center}
\end{table}


\def\arraystretch{1.}
 
\FloatBarrier
 
\subsection{Model-independent exclusion upper limits}
 
Model-independent upper limits exclude the presence of a larger generic signal independently for each discovery region
considered in this analysis. These limits are evaluated by extrapolating the SM background predictions obtained from the background-only fit configuration to the single-bin inclusive SR.
Table~\ref{table.results.exclxsec.pval.upperlimit} presents the results of this evaluation, provided in the form of
\textit{CL}$_\mathrm{B}$ representing the probability of the predicted SM background to fluctuate to at least the observed number of events.
In addition, 95\% CL upper limits are set on the observed ($S^{95}_\mathrm{obs}$) and expected ($S^{95}_\mathrm{exp}$) numbers of BSM events
as well as on the visible cross-section ($\sigma^{}_\textrm{vis}$) for all discovery regions.

\begin{table}
\centering
\footnotesize
\setlength{\tabcolsep}{0.1pc}
\caption[Breakdown of upper limits.]{The first column presents the name of the treated discovery region. The next two columns present the observed number of data events and expected SM contribution from the background-only fit in the model-independent regions.  The fourth and fifth columns present the 95\% CL upper limits on the visible cross-section ($\langle\epsilon\sigma\rangle_{\mathrm{obs}}^{95}$) and on the observed number of signal events ($S_{\mathrm{obs}}^{95}$), respectively. The sixth column ($S_{\mathrm{exp}}^{95}$) shows the 95\% CL upper limit on the expected number of signal events, given the expected number of background events and $\pm 1\sigma$ excursions of the expectation. The last two columns indicate the \textit{CL}$_{\text{B}}$ value, i.e. the confidence level observed for the background-only hypothesis, and the discovery $p$-value ($p(s = 0)$). These values are calculated using an asymptotic calculator based on a one-sided profile likelihood. Therefore, the $p$-values corresponding to regions with deficits are capped at 0.5.}
\begin{tabular*}{\textwidth}{@{\extracolsep{\fill}}l|cc|ccccc}
\noalign{\smallskip}\hline\noalign{\smallskip}{\textbf{ Signal channel}} & Obs. & \hspace{2mm} SM.\ exp \hspace{4mm} & $\langle\epsilon{\mathrm \sigma}\rangle_{\mathrm{obs}}^{95}$ [fb]  &  $S_{\text{obs}}^{95}$  & $S_{\mathrm{exp}}^{95}$ & \textit{CL}$_{\text{B}}$ & $p(s=0)$ ($Z$) \\
\noalign{\smallskip}\hline\noalign{\smallskip}
\addlinespace[-2ex] & & & & & \\
SR$_{\mathrm{tW}_{\mathrm{0L}}} (E_{\mathrm{T}}^{\mathrm{miss}} \geq 250 \;\GeV)$ & 133 & 147 $\pm$ 15~~~ & $0.21$ &  $29$~~ & $ { 36 }^{ +14 }_{ -10 }$~~ & $0.24$ & $ 0.50$~$(0.00)$ \\%
\addlinespace[-2ex] & & & & & \\
SR$_{\mathrm{tW}_{\mathrm{0L}}} (E_{\mathrm{T}}^{\mathrm{miss}} \geq 330 \;\GeV)$ & ~~66 & 83 $\pm$ 9~~ & $0.11$ &  $15.5$ & $ { 24 }^{ +10 }_{ -7 }$~~ & $0.09$ & $ 0.50$~$(0.00)$ \\%
\addlinespace[-2ex] & & & & & \\
SR$_{\mathrm{tW}_{\mathrm{0L}}} (E_{\mathrm{T}}^{\mathrm{miss}} \geq 400 \;\GeV)$ & ~~33 & 42 $\pm$ 6~~ & $0.08$ &  $11.7$ & $ { 16 }^{ +7 }_{ -5 }$~~~ & $0.15$ & $ 0.50$~$(0.00)$ \\%
\addlinespace[-2ex] & & & & & \\
SR$_{\mathrm{tW}_{\mathrm{0L}}} (E_{\mathrm{T}}^{\mathrm{miss}} \geq 500 \;\GeV)$ & ~~~~8 & ~~16.6 $\pm$ 2.3 & $0.04$ &  ~$5.4$ & ~~~${ 9.7 }^{ +4.3 }_{ -2.8 }$ & $0.03$ & $ 0.50$~$(0.00)$ \\%
\addlinespace[-2ex] & & & & & \\
SR$_{\mathrm{tW}_{\mathrm{0L}}} (E_{\mathrm{T}}^{\mathrm{miss}} \geq 600 \;\GeV)$ & ~~~~6 & ~~~~7.0 $\pm$ 1.7 & $0.05$ &  ~$6.5$ & ~~~$ { 7.4 }^{ +3.3 }_{ -2.1 }$ & $0.38$ & $ 0.50$~$(0.00)$ \\%
\addlinespace[-2ex] & & & & & \\
\hline
\addlinespace[-2ex] & & & & & \\
SR$_{\mathrm{tW}_{\mathrm{1L}}}^{\mathrm{had. top}} (E_{\mathrm{T}}^{\mathrm{miss}} \geq 250 \;\GeV)$\hspace{2mm} & 239 & 237 $\pm$ 25~~~ & $0.42$ &  $58$~~~ & $ { 57 }^{ +21 }_{ -15 }$~~ & $0.53$ & $ 0.47$~$(0.06)$ \\%
\addlinespace[-2ex] & & & & & \\
SR$_{\mathrm{tW}_{\mathrm{1L}}}^{\mathrm{had. top}} (E_{\mathrm{T}}^{\mathrm{miss}} \geq 300 \;\GeV)$ & 130 & 121 $\pm$ 17~~ & $0.33$ &  $46.4$ & $ { 40 }^{ +15 }_{ -11 }$~~ & $0.67$ & $ 0.33$~$(0.44)$ \\%
\addlinespace[-2ex] & & & & & \\
SR$_{\mathrm{tW}_{\mathrm{1L}}}^{\mathrm{had. top}} (E_{\mathrm{T}}^{\mathrm{miss}} \geq 350 \;\GeV)$ & ~~69 & 66 $\pm$ 9~~ & $0.19$ &  $26.3$ & $ { 24 }^{ +10 }_{ -7 }$~~ & $0.60$ & $ 0.39$~$(0.27)$ \\%
\addlinespace[-2ex] & & & & & \\
SR$_{\mathrm{tW}_{\mathrm{1L}}}^{\mathrm{had. top}} (E_{\mathrm{T}}^{\mathrm{miss}} \geq 400 \;\GeV)$ & ~~40 & 37 $\pm$ 9~~ & $0.17$ &  $23.7$ & $ { 22 }^{ +8 }_{ -6 }$~~ & $0.62$ & $ 0.38$~$(0.30)$ \\%
\addlinespace[-2ex] & & & & & \\
SR$_{\mathrm{tW}_{\mathrm{1L}}}^{\mathrm{had. top}} (E_{\mathrm{T}}^{\mathrm{miss}} \geq 450 \;\GeV)$ & ~~25 & 20 $\pm$ 9~~ & $0.16$ &  $22.0$ & $ { 19 }^{ +6 }_{ -5 }$~~ & $0.69$ & $ 0.30$~$(0.51)$ \\%
\addlinespace[-2ex] & & & & & \\
SR$_{\mathrm{tW}_{\mathrm{1L}}}^{\mathrm{lep. top}}$ & ~~~~9 &  ~~~~6.4 $\pm$ 2.3 & $0.07$ &  $10.2$ & ~~~~${ 8.0 }^{ +3.8 }_{ -2.3 }$ & $0.74$ & $ 0.24$~$(0.72)$ \\%
\noalign{\smallskip}\hline\noalign{\smallskip}
\end{tabular*}
\label{table.results.exclxsec.pval.upperlimit}
\end{table}


\subsection{Exclusion limits for the 2HDM$+a$ model}
 
The \tWZeroL\ and \tWOneL\ channels are
statistically combined with the \tWTwoL\
channel of Ref.~\cite{EXOT-2018-43} as described in Sections~\ref{sec:analysis}~and~\ref{sec:unc},
in order to provide the most stringent
constraints for 2HDM$+a$ models using the \twdm\ channel.
 
\begin{figure}[tbp]
\centering
\subfloat[]{
\includegraphics[width=0.45\textwidth]{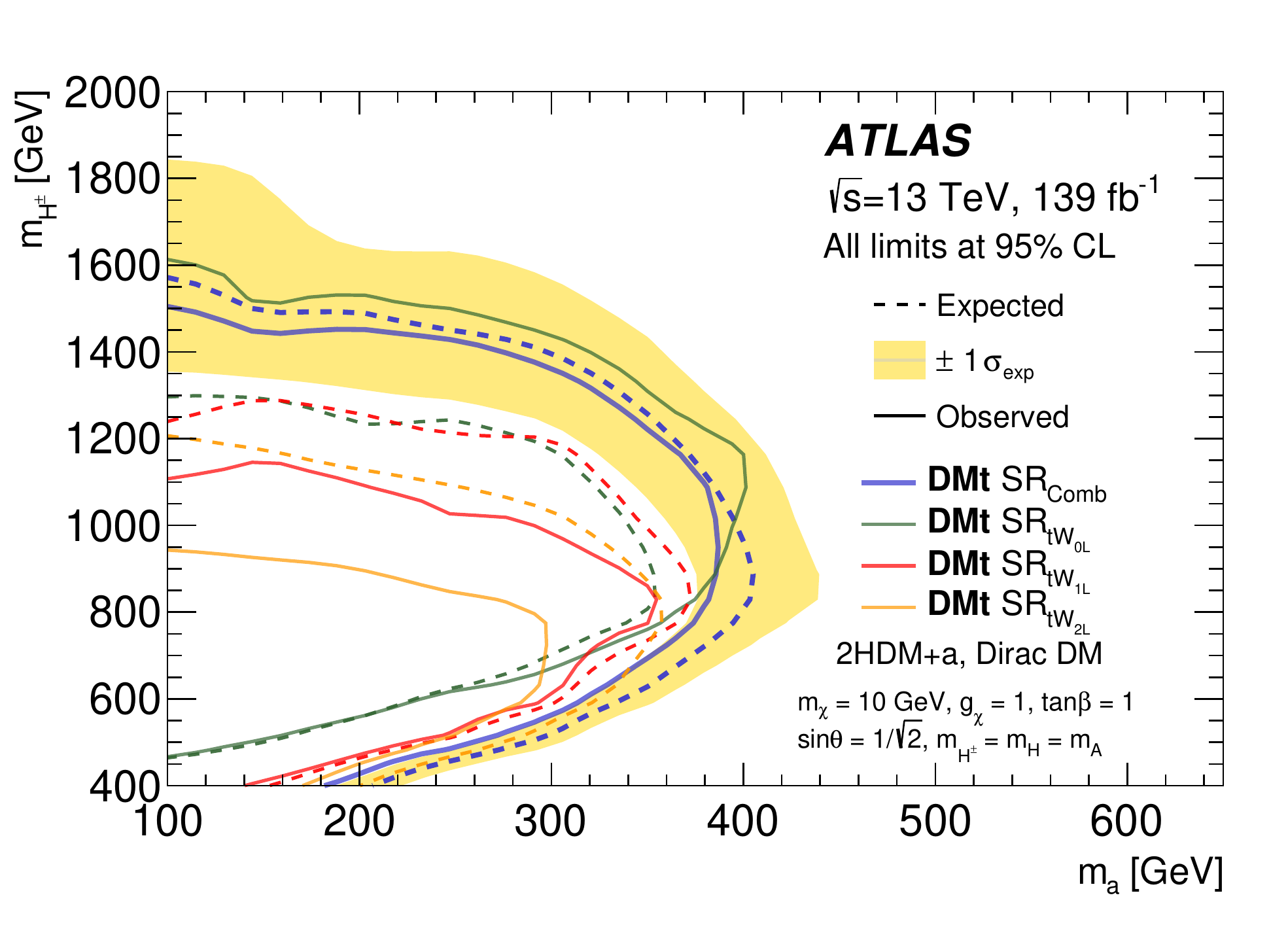}
\label{fig:comb_nominal_ma}
}
\subfloat[]{
\includegraphics[width=0.45\textwidth]{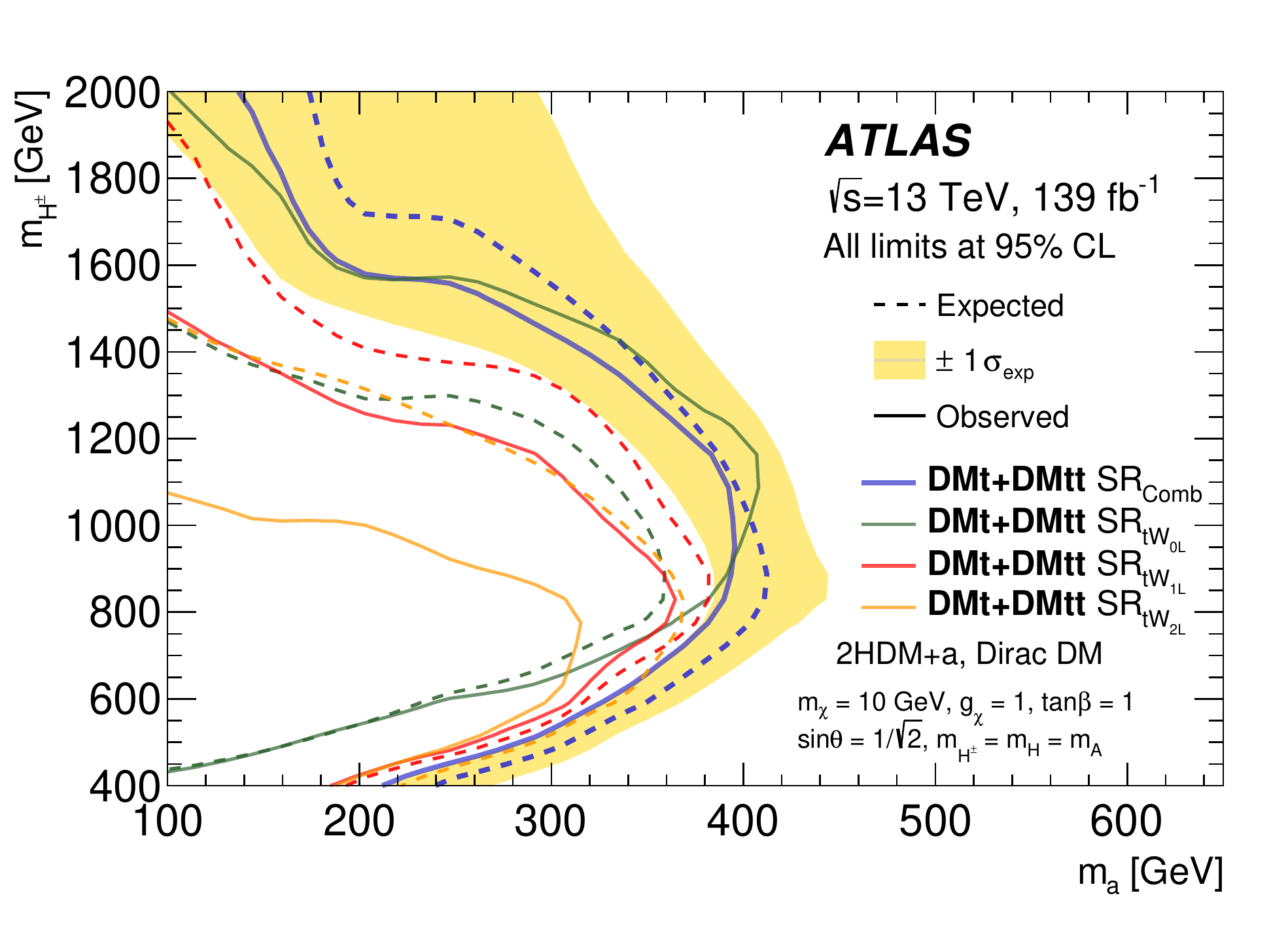}
\label{fig:comb_dmtt_ma}
}
\\
\subfloat[]{
\includegraphics[width=0.45\textwidth]{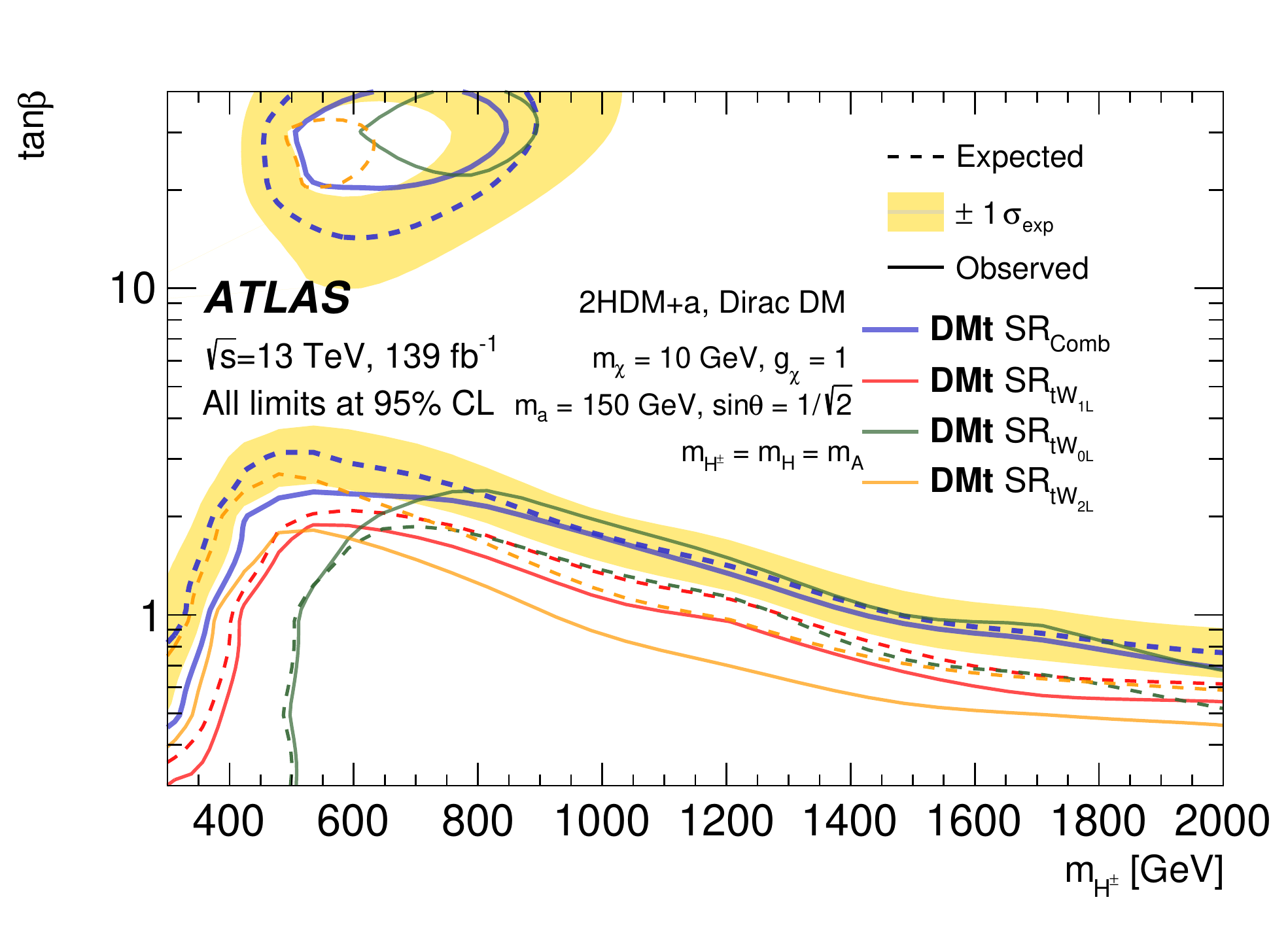}
\label{fig:comb_nominal_tanb}
}
\subfloat[]{
\includegraphics[width=0.45\textwidth]{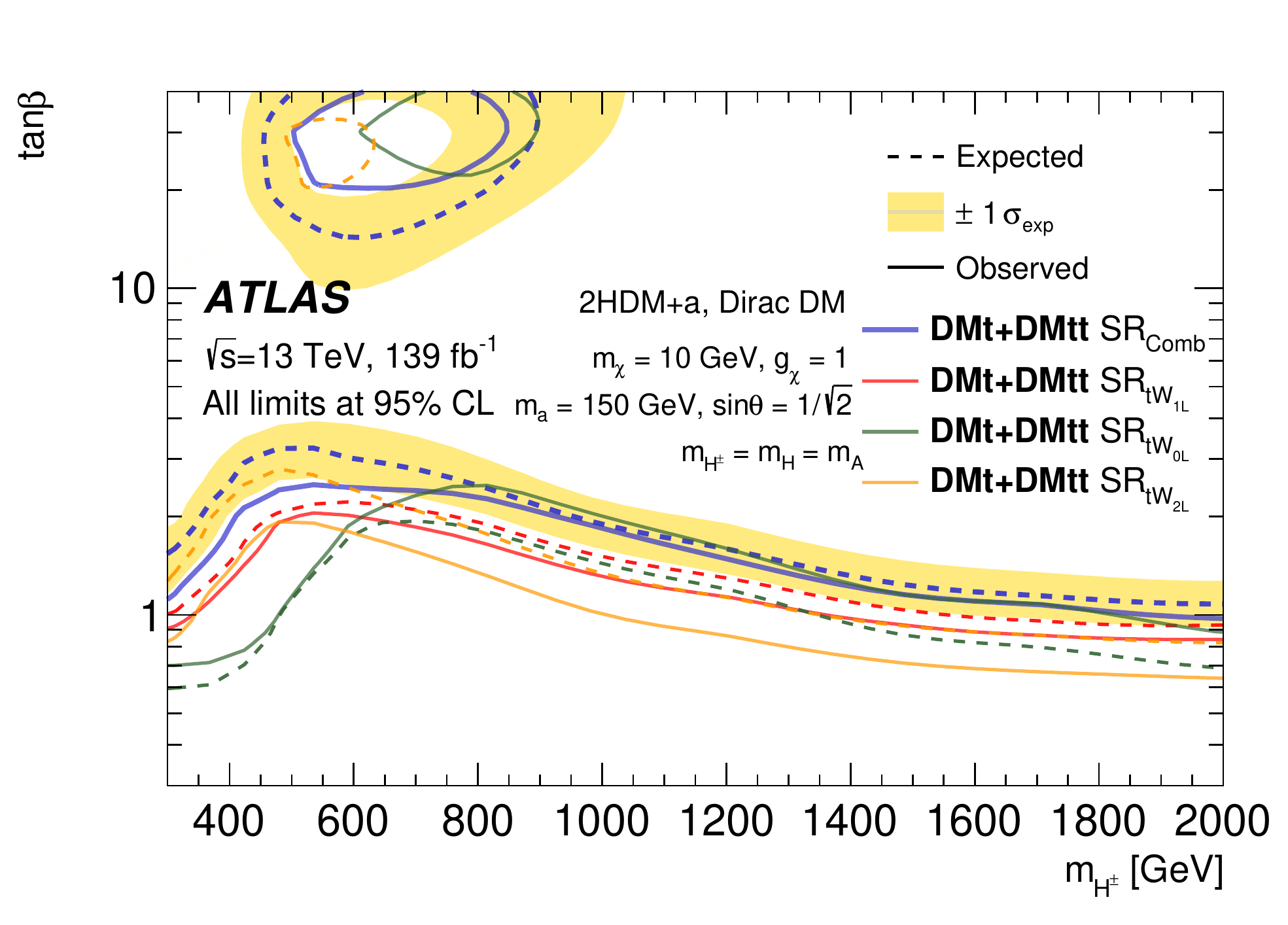}
\label{fig:comb_dmtt_tanb}
}
\\
\subfloat[]{
\includegraphics[width=0.45\textwidth]{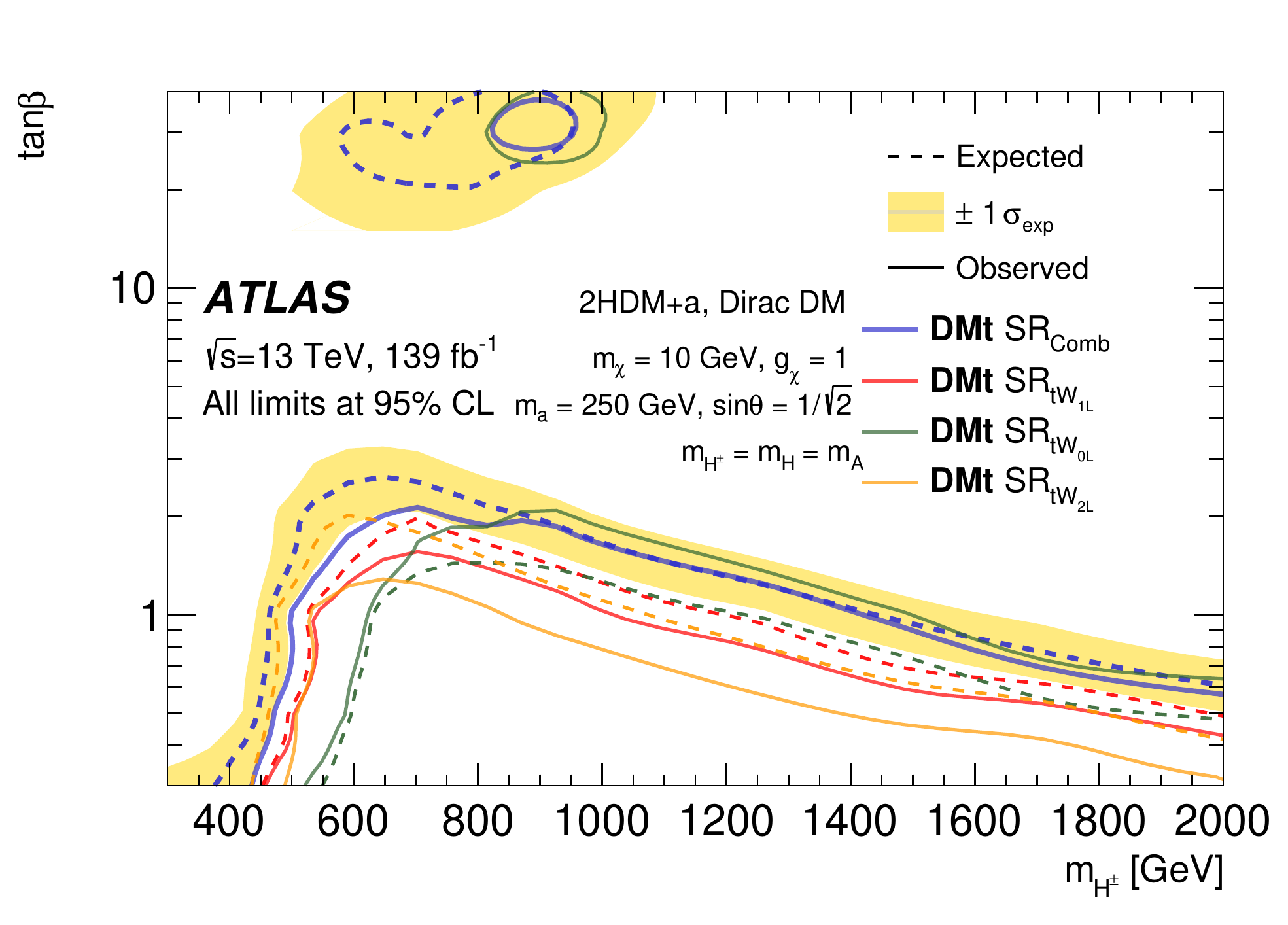}
\label{fig:comb_nominal_tanb_250}
}
\subfloat[]{
\includegraphics[width=0.45\textwidth]{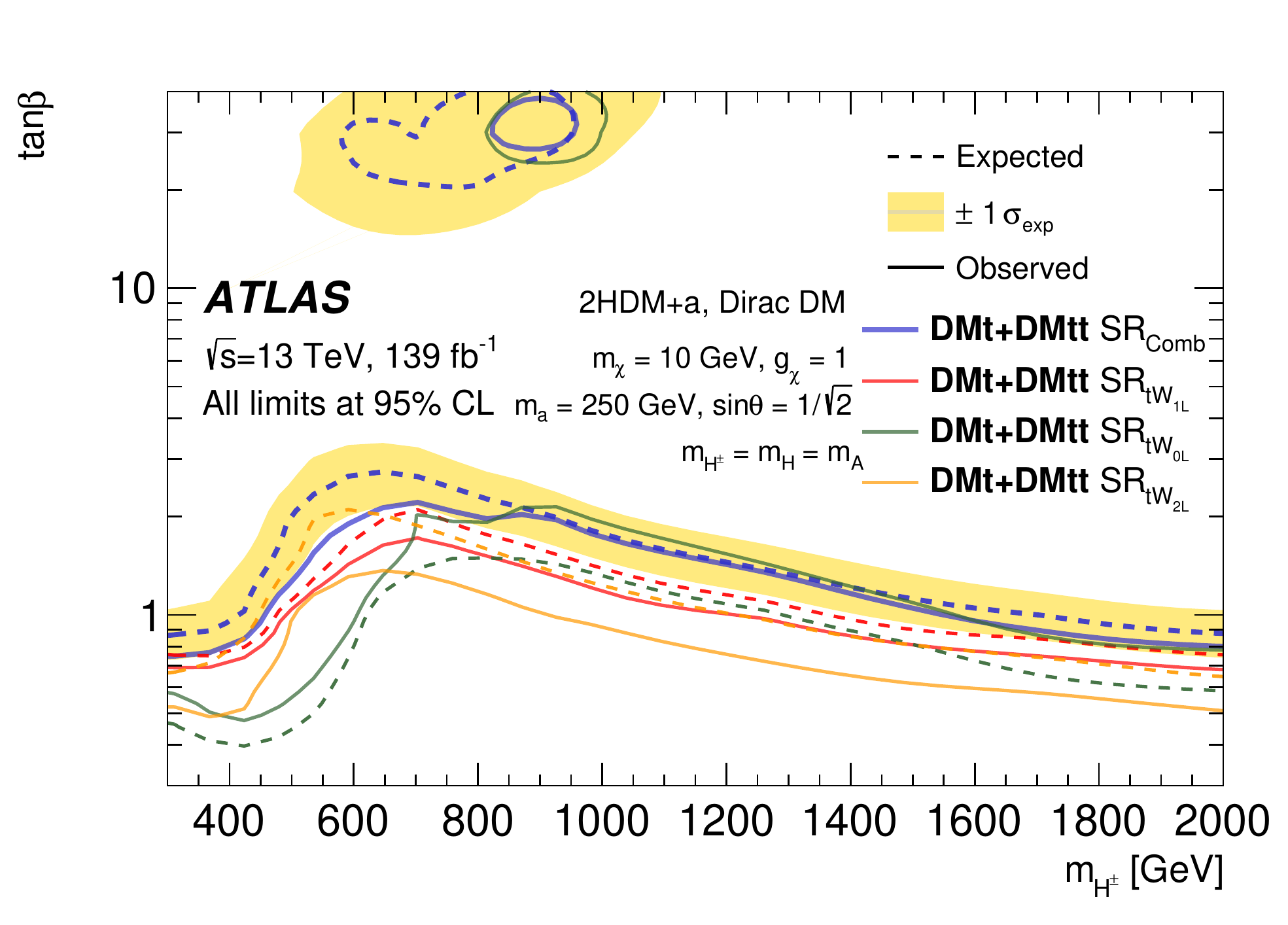}
\label{fig:comb_dmtt_tanb_250}
}\\
\caption{
The expected and observed exclusion contours as a function of
\protect\subref{fig:comb_nominal_ma}--\protect\subref{fig:comb_dmtt_ma} $(m_a, m_{H^{\pm}})$,
\protect\subref{fig:comb_nominal_tanb}--\protect\subref{fig:comb_dmtt_tanb} $( m_{H^{\pm}}, \tan\beta)$ assuming $m_a = 150\;\GeV$
and \protect\subref{fig:comb_nominal_tanb_250}--\protect\subref{fig:comb_dmtt_tanb_250} $( m_{H^{\pm}}, \tan\beta)$
assuming $m_a = 250\;\GeV$. The individual \tWZeroL\ (green line), \tWOneL\ (red line) and \tWTwoL\ (orange line) analysis
channels are shown together with their statistical combination (blue
line). Only \twdm\ contributions are considered in \protect\subref{fig:comb_nominal_ma}, \protect\subref{fig:comb_nominal_tanb} and
\protect\subref{fig:comb_nominal_tanb_250}, while \protect\subref{fig:comb_dmtt_ma}, \protect\subref{fig:comb_dmtt_tanb} and
\protect\subref{fig:comb_dmtt_tanb_250} consider \twdm\ and \ttdm\ contributions.
Experimental and theoretical systematic uncertainties, as
described in Section~\ref{sec:unc}, are applied to background
and signal samples and illustrated by the $\pm 1$
standard-deviation
yellow band and the blue dashed contour lines, respectively, for the
statistical combination.
}
 
\label{fig:results}
\end{figure}

Exclusion limits on the 2HDM$+a$ model are derived as a function of the parameters
$m_a$,  $ m_{H^{\pm}}$, and $\tan\beta$ in a combined likelihood fit to the events in all
CRs and SRs of the three channels and are shown in Figure~\ref{fig:results}.
The results are presented as a function of $(m_a,  m_{H^{\pm}})$ assuming $\tan\beta = 1$
and as a function of $( m_{H^{\pm}}, \tan\beta)$ assuming $m_a = 150\;\GeV$ or $m_a = 250\;\GeV$. Values of $\tan\beta$ up to 30 are considered in order to probe the
local maximum at $\tan\beta \in [20,30]$, as explained in Section~\ref{sec:intro}.
The 1$\sigma$ uncertainty bands are shown as shaded areas around the expected limit contour of the statistical combination. The typical acceptance,
i.e.\ percentage of events passing the selection requirements defined in Section~\ref{sec:analysis}, times detector efficiency
for the \twdm\ benchmark signals is 0.02\%--1.2\% for the inclusive \tWZeroL\ SRs, 0.001\%--0.4\% for \lepTop\ and 0.04\%--1.1\% for the \hadTop\ bins.
Figure~\ref{fig:results} also shows the sensitivity of each individual channel in both the $(m_a,  m_{H^{\pm}})$ and $( m_{H^{\pm}}, \tan\beta)$ planes.
For the fits in the individual channels, the non-\ttbar\ background estimates in the signal regions are derived using all control regions defined
in \Sec\ref{ssec:crs}, including the common \ttZ\ region for the \tWTwoL\ channel. For the \ttbar\
process, the \tWZeroL\ fit uses the \tWZeroL\ \ttbar\ CR, while the
\tWOneL\ \ttbar\ CR is used in the \tWOneL\ fit. The left panels in Figure~\ref{fig:results} consider only the \twdm process
as signal for the interpretation of the results, while the right panels in the same figure consider the contributions of both
the \twdm\ and \ttdm\ processes as predicted by the 2HDM$+a$ model.
 
The introduction of the \tWZeroL\ channel and the statistical combination performed in this paper extend the
sensitivity towards large $H^\pm$ boson masses. Exclusion limits are placed
in the high $\tan\beta$ parameter space for the first time in this final state.
Signal models assuming $H^{\pm}$ boson masses up to 1.5~\TeV\ and $a$-boson masses up to 350~\GeV\ can be excluded at 95\% CL for $\tan\beta = 1$.
For an $a$-boson mass of 150 (250)~\GeV, $\tan\beta$ values up to 2 are excluded for $H^{\pm}$ masses between 300 (400)~\GeV\ and 1.5~\TeV.
Signals with $\tan\beta$ values between 20 and 30 are also excluded for $H^{\pm}$ masses between 500 and 800~\GeV (900~\GeV) and a $a$-boson mass of 150 (250)~\GeV.
If \ttdm\ contributions are considered together with \twdm, $a$-boson masses up to 250~\GeV\ can be excluded at 95\% CL for an $H^\pm$ mass of
1.5~\TeV\ assuming $\tan\beta =1$. For low $H^\pm$ boson masses, the lower limit on $m_a$ is 20--40~\GeV\ higher than when considering only the \twdm\ contribution at the same $\tan\beta$ value.
Assuming an $m_a$ value of 150~\GeV\ or 250~\GeV, $H^\pm$ boson masses below 400~\GeV\ can be excluded for $\tan\beta$
values lower than 1. No additional constraints are observed at $\tan\beta$ > 10 when adding the \ttdm\ contribution to the \twdm\ contribution since,
as discussed in Refs.~\cite{Abe:2018bpo,EXOT-2017-32}, the \ttdm\ cross-section in the 2HDM+$a$ model is proportional to $1/\tan^2\beta$ and is expected to be subdominant at high $\tan\beta$ values.

\FloatBarrier
 
\section{Conclusions}
\label{sec:conclusion}
 
A search for dark matter in final states with a single top quark and an energetic \Wboson\ boson using 139 fb$^{-1}$ of $pp$ collisions delivered by the LHC
at a centre-of-mass energy of 13~\TeV\ and collected by the ATLAS detector is presented.
The search focuses on a two-Higgs-doublet model together with an additional pseudoscalar mediator, $a$,
which decays into dark-matter particles.
Final states which include either zero or one charged lepton (electron or muon) and a
significant amount of missing transverse momentum are considered.
No significant excess relative to Standard Model predictions was found and
95\% confidence-level limits are set on the 2HDM$+a$ signal models considered.
These limits exclude $a$-boson mediator masses up to 350~\GeV\ and $H^{\pm}$ boson masses up to
1.5~\TeV\ for $\tan\beta = 1$ in comparison with the current 1.3~\TeV\ bound, and are the most stringent limits on \twdm\ signal models obtained so far at the LHC.
This analysis also provides the first limits for a 2HDM$+a$ signal model assuming $\tan\beta \geq 10$ and using the single-top-quark
production signature.

\section*{Acknowledgements}


We thank CERN for the very successful operation of the LHC, as well as the
support staff from our institutions without whom ATLAS could not be
operated efficiently.
 
We acknowledge the support of
ANPCyT, Argentina;
YerPhI, Armenia;
ARC, Australia;
BMWFW and FWF, Austria;
ANAS, Azerbaijan;
CNPq and FAPESP, Brazil;
NSERC, NRC and CFI, Canada;
CERN;
ANID, Chile;
CAS, MOST and NSFC, China;
Minciencias, Colombia;
MEYS CR, Czech Republic;
DNRF and DNSRC, Denmark;
IN2P3-CNRS and CEA-DRF/IRFU, France;
SRNSFG, Georgia;
BMBF, HGF and MPG, Germany;
GSRI, Greece;
RGC and Hong Kong SAR, China;
ISF and Benoziyo Center, Israel;
INFN, Italy;
MEXT and JSPS, Japan;
CNRST, Morocco;
NWO, Netherlands;
RCN, Norway;
MEiN, Poland;
FCT, Portugal;
MNE/IFA, Romania;
MESTD, Serbia;
MSSR, Slovakia;
ARRS and MIZ\v{S}, Slovenia;
DSI/NRF, South Africa;
MICINN, Spain;
SRC and Wallenberg Foundation, Sweden;
SERI, SNSF and Cantons of Bern and Geneva, Switzerland;
MOST, Taiwan;
TENMAK, T\"urkiye;
STFC, United Kingdom;
DOE and NSF, United States of America.
In addition, individual groups and members have received support from
BCKDF, CANARIE, Compute Canada and CRC, Canada;
PRIMUS 21/SCI/017 and UNCE SCI/013, Czech Republic;
COST, ERC, ERDF, Horizon 2020 and Marie Sk{\l}odowska-Curie Actions, European Union;
Investissements d'Avenir Labex, Investissements d'Avenir Idex and ANR, France;
DFG and AvH Foundation, Germany;
Herakleitos, Thales and Aristeia programmes co-financed by EU-ESF and the Greek NSRF, Greece;
BSF-NSF and MINERVA, Israel;
Norwegian Financial Mechanism 2014-2021, Norway;
NCN and NAWA, Poland;
La Caixa Banking Foundation, CERCA Programme Generalitat de Catalunya and PROMETEO and GenT Programmes Generalitat Valenciana, Spain;
G\"{o}ran Gustafssons Stiftelse, Sweden;
The Royal Society and Leverhulme Trust, United Kingdom.
 
The crucial computing support from all WLCG partners is acknowledged gratefully, in particular from CERN, the ATLAS Tier-1 facilities at TRIUMF (Canada), NDGF (Denmark, Norway, Sweden), CC-IN2P3 (France), KIT/GridKA (Germany), INFN-CNAF (Italy), NL-T1 (Netherlands), PIC (Spain), ASGC (Taiwan), RAL (UK) and BNL (USA), the Tier-2 facilities worldwide and large non-WLCG resource providers. Major contributors of computing resources are listed in Ref.~\cite{ATL-SOFT-PUB-2021-003}.


\clearpage
 
\printbibliography

\clearpage
 
\begin{flushleft}
\hypersetup{urlcolor=black}
{\Large The ATLAS Collaboration}

\bigskip

\AtlasOrcid[0000-0002-6665-4934]{G.~Aad}$^\textrm{\scriptsize 101}$,
\AtlasOrcid[0000-0002-5888-2734]{B.~Abbott}$^\textrm{\scriptsize 119}$,
\AtlasOrcid[0000-0002-7248-3203]{D.C.~Abbott}$^\textrm{\scriptsize 102}$,
\AtlasOrcid[0000-0002-1002-1652]{K.~Abeling}$^\textrm{\scriptsize 55}$,
\AtlasOrcid[0000-0002-8496-9294]{S.H.~Abidi}$^\textrm{\scriptsize 29}$,
\AtlasOrcid[0000-0002-9987-2292]{A.~Aboulhorma}$^\textrm{\scriptsize 35e}$,
\AtlasOrcid[0000-0001-5329-6640]{H.~Abramowicz}$^\textrm{\scriptsize 150}$,
\AtlasOrcid[0000-0002-1599-2896]{H.~Abreu}$^\textrm{\scriptsize 149}$,
\AtlasOrcid[0000-0003-0403-3697]{Y.~Abulaiti}$^\textrm{\scriptsize 116}$,
\AtlasOrcid[0000-0003-0762-7204]{A.C.~Abusleme~Hoffman}$^\textrm{\scriptsize 136a}$
% ...
\bigskip
\\

% ...
$^{29}$Physics Department, Brookhaven National Laboratory, Upton NY; United States of America.\\
% ...
$^{35}$$^{(a)}$Facult\'e des Sciences Ain Chock, R\'eseau Universitaire de Physique des Hautes Energies - Universit\'e Hassan II, Casablanca;$^{(b)}$Facult\'{e} des Sciences, Universit\'{e} Ibn-Tofail, K\'{e}nitra;$^{(c)}$Facult\'e des Sciences Semlalia, Universit\'e Cadi Ayyad, LPHEA-Marrakech;$^{(d)}$LPMR, Facult\'e des Sciences, Universit\'e Mohamed Premier, Oujda;$^{(e)}$Facult\'e des sciences, Universit\'e Mohammed V, Rabat;$^{(f)}$Institute of Applied Physics, Mohammed VI Polytechnic University, Ben Guerir; Morocco.\\
% ...
$^{55}$II. Physikalisches Institut, Georg-August-Universit\"{a}t G\"ottingen, G\"ottingen; Germany.\\
% ...
$^{101}$CPPM, Aix-Marseille Universit\'e, CNRS/IN2P3, Marseille; France.\\
$^{102}$Department of Physics, University of Massachusetts, Amherst MA; United States of America.\\
% ...
$^{116}$Department of Physics, New York University, New York NY; United States of America.\\
% ...
$^{119}$Homer L. Dodge Department of Physics and Astronomy, University of Oklahoma, Norman OK; United States of America.\\
% ...
$^{136}$$^{(a)}$Departamento de F\'isica, Pontificia Universidad Cat\'olica de Chile, Santiago;$^{(b)}$Millennium Institute for Subatomic physics at high energy frontier (SAPHIR), Santiago;$^{(c)}$Instituto de Investigaci\'on Multidisciplinario en Ciencia y Tecnolog\'ia, y Departamento de F\'isica, Universidad de La Serena;$^{(d)}$Universidad Andres Bello, Department of Physics, Santiago;$^{(e)}$Instituto de Alta Investigaci\'on, Universidad de Tarapac\'a, Arica;$^{(f)}$Departamento de F\'isica, Universidad T\'ecnica Federico Santa Mar\'ia, Valpara\'iso; Chile.\\
% ...
$^{149}$Department of Physics, Technion, Israel Institute of Technology, Haifa; Israel.\\
$^{150}$Raymond and Beverly Sackler School of Physics and Astronomy, Tel Aviv University, Tel Aviv; Israel.\\
% ...
$^{a}$ Also Affiliated with an institute covered by a cooperation agreement with CERN.\\
$^{b}$ Also at Borough of Manhattan Community College, City University of New York, New York NY; United States of America.\\
$^{c}$ Also at Bruno Kessler Foundation, Trento; Italy.\\
$^{d}$ Also at Center for High Energy Physics, Peking University; China.\\
$^{e}$ Also at Centro Studi e Ricerche Enrico Fermi; Italy.\\
$^{f}$ Also at CERN, Geneva; Switzerland.\\
% ...

\end{flushleft}


\end{document}